\documentclass{article}
\usepackage{authblk}
\usepackage{amsfonts}
\usepackage[english]{babel}
\usepackage[utf8]{inputenc} 
\usepackage[T1]{fontenc}
\usepackage[letterpaper,top=2cm,bottom=2cm,left=3cm,right=3cm,marginparwidth=1.75cm]{geometry}

\usepackage{array}
\usepackage[table]{xcolor}
\definecolor{tablegray}{gray}{0.94}
\usepackage{hhline}
\usepackage{amsmath}
\usepackage{amssymb}
\usepackage{graphicx,enumitem}
\usepackage{xcolor}
\usepackage{soul}
\usepackage{mathtools}
\usepackage{cancel}
\usepackage{natbib}
\usepackage{amsthm}
\usepackage[colorlinks=true, allcolors=blue]{hyperref}
\usepackage[capitalise, sort, noabbrev]{cleveref}

\newtheorem{theorem}{Theorem}[section]
\newtheorem{proposition}[theorem]{Proposition}
\newtheorem{lemma}[theorem]{Lemma}
\newtheorem{corollary}[theorem]{Corollary}

\theoremstyle{definition}
\newtheorem{definition}{Definition}[section]

\theoremstyle{remark}
\newtheorem{remark}[theorem]{Remark}

\title{Rapid Mixing of Stereographic MCMC \\
for Heavy-Tailed Sampling}
\date{}

\author[1]{Federica Milinanni}
\author[2]{Tyler Farghly}
\author[3]{Jun Yang}

\affil[1]{Department of Industrial Engineering and Management Sciences, Northwestern University, \href{mailto:federica.milinanni@northwestern.edu}{federica.milinanni@northwestern.edu}}
\affil[2]{Inria, CNRS, École Normale Supérieure, PSL Research University, \href{mailto:tyler.farghly@inria.fr}{tyler.farghly@inria.fr}}
\affil[3]{Department of Mathematical Sciences, University of Copenhagen, \href{mailto:jy@math.ku.dk}{jy@math.ku.dk}}

\begin{document}
\maketitle

\begin{abstract}
Sampling from high-dimensional, heavy-tailed distributions is a fundamental challenge in computational statistics, as many standard Markov chain Monte Carlo (MCMC) methods mix poorly in such settings. Recently, Stereographic MCMC [Yang et al., 2024] and the Sub-Cauchy Projection Sampler [Grazzi et al., 2026] have been shown to perform successfully on such tasks. However, establishing their non-asymptotic convergence properties remains an important open problem. In this work, we fill this gap by establishing non-asymptotic upper bounds on the mixing time of the stereographic projection and sub-Cauchy projection samplers. Our results demonstrate that, under certain conditions on the target and initial distributions, the mixing time is polynomial in dimension for a broad class of distributions, including light- and heavy-tailed cases.
Motivated by the theoretical analysis, we further establish a new weighted isoperimetric inequality that extends the classical version for log-concave distributions to the heavy-tailed setting with optimal dimension dependence.
The proof techniques provide new insights into the geometric properties of heavy-tailed distributions that govern rapid mixing in high dimensions.
\end{abstract}

\vfill
\section*{AI use disclosure}
We explicitly affirm that all theoretical frameworks, mathematical formulations, and logical proofs in this manuscript represent original human work conducted entirely by the authors without the assistance of generative AI or automated theorem-proving tools. The core results were established and publicly presented at BayesComp 2025 (Singapore, June 2025), Advances in MCMC Methods (Eindhoven, December 2025), and the IMS International Conference on Data Science (ICSDS 2025, Seville, December 2025), establishing clear historical provenance before the proliferation of LLM-based mathematical tools. Generative AI was strictly limited to assistance with literature survey (including two examples in Bayesian statistics), final language copyediting, and grammatical polishing.

\newpage

\tableofcontents

\newpage

\section{Introduction}
\label{sec:introduction}

Markov chain Monte Carlo (MCMC) methods are a popular
class of algorithms widely used to sample from probability
distributions that are known only up to a normalizing
constant \citep{robert2004monte, asmussen2007stochastic, andrieu2003introduction}. These methods generate a Markov chain whose invariant distribution
is the target distribution. MCMC methods play a central role in Bayesian posterior
computation and have applications in computational biology,
statistical mechanics, Bayesian inverse problems, and beyond
\citep{robert2004monte,brooks2011handbook,stuart2010inverse,kramer2023uqsa}.
Sampling is often challenging in high dimensions,
particularly when the target distribution is heavy-tailed.
Such distributions are not uncommon in applications. For instance, they are used in robust regression
\citep{geweke1993bayesian}, extreme value modeling
\citep{coles2001introduction}, and
weakly informative priors \citep{gelman2008weakly}.
In Bayesian inference, heavy-tailed posterior distributions can also arise
through the marginalization of uncertain scale parameters
\citep{christen2024bayesian}.
Heavy-tailed distributions assign substantial probability to
regions far from their center, making reliable exploration
of the tails an essential part of the sampling problem.

Many widely used MCMC methods belong to the
Metropolis--Hastings family~\citep{metropolis1953equation,hastings1970monte}, in which a proposed move is
accepted or rejected according to a rule that preserves
the target distribution.
A standard example is the Random Walk Metropolis (RWM)
algorithm, which proposes a new state of the chain by adding a random
increment, such as Gaussian noise, to the current state.
With such local proposals, the chain can get ``stuck'' in the tails, and thus it requires many
iterations to return from a distant tail region to the central part of the distribution. 
This difficulty is reflected in the lack of fast convergence properties, such as geometric ergodicity.
This is the case for the RWM algorithm used to sample from polynomially decaying target densities~\citep{jarner2000geometric}.
These limitations motivate the development of sampling methods that can effectively explore the tails while still allowing the chain to escape from them easily.

One source of difficulty for sampling is the non-compactness of Euclidean state spaces. This suggests that mapping the non-compact space to a (pre-)compact one could facilitate sampling. The Stereographic Projection Sampler (SPS), introduced by~\citet{yang2022stereographic}, follows this idea by mapping the Euclidean state space onto the sphere, excluding the north pole, through inverse stereographic projection. A Markov chain is then generated on the sphere via RWM to sample from the transformed target distribution, and the resulting samples are mapped back to Euclidean space. 

The SPS depends on two algorithmic hyperparameters: the projection scale $R>0$ and the step size $h>0$ of the random walk proposal on the sphere. Their precise definitions are given in \Cref{sec:SPS-and-SCPS}. Under stereographic projection, the south pole corresponds to the origin of the Euclidean space, while points approaching the north pole correspond to points with diverging norm. When the target is centered at the origin, these regions therefore represent its center and tails, respectively. Consequently, displacements of the same geodesic length can correspond to Euclidean displacements of very different magnitude, with movements near the north pole being magnified much more strongly than those near the south pole. Thus, the stereographic projection induces a position-dependent rescaling of the chain's movements in Euclidean space.

The projection strategy in fact helps the sampling process. As proved in~\cite[Theorem~2.1]{yang2022stereographic}, SPS can exhibit stronger convergence properties than RWM. For example, for certain polynomially decaying targets, the SPS chain is uniformly ergodic (a property stronger than geometric ergodicity) even though RWM is not even geometrically ergodic. The same work~\cite[Section~5]{yang2022stereographic} also provides a high-dimensional analysis of SPS for product-form targets.
With the projection scale and proposal step size chosen
as $R\asymp\sqrt d$ and $h\asymp d^{-1}$, respectively, the authors study the limiting acceptance probability and expected squared jumping distance, derive an optimal acceptance probability, and establish a diffusion limit for a modified version of the algorithm.

The sampling strategy of SPS is generalized by the Sub-Cauchy Projection Sampler (SCPS), introduced by~\citet{grazzi2026sub}, to accommodate a broader class of heavy-tailed targets. The projection used in SCPS differs slightly from the stereographic projection. First, note that we identify Euclidean space with the hyperplane tangent to the sphere at its south pole. In stereographic projection, an observer is placed at the north pole. Each point on the sphere is mapped to Euclidean space by extending the line joining it to the observer until it intersects this hyperplane. In the Sub-Cauchy projection, the observer is instead moved into the interior of the sphere. As a result, the transformed state space becomes a subset of the sphere, called the \emph{bright side}, whose boundary corresponds to Euclidean points with diverging norm, whereas under stereographic projection all such points approach the north pole. As proved in~\cite[Theorem 1]{grazzi2026sub}, this construction yields a uniformly ergodic sampler for targets with tails no heavier than those of a multivariate Cauchy distribution. We describe both projection maps and the corresponding sampling algorithms in \Cref{sec:SPS-and-SCPS}.

\textbf{The main objective of this work is to quantify the
mixing behavior of SPS and SCPS in high dimensions} by bounding the number
of iterations required to approximate the target
distribution to a prescribed accuracy. We formalize this concept via the Markov chain's total variation mixing time: for a Euclidean transition kernel $\bar T$ with target
distribution $\bar\pi$ and initial distribution
$\bar\pi_0$, we define the total variation mixing time by
\begin{equation}
\label{eq:mixing-time-definition}
n_{\rm mix}(\varepsilon,\bar\pi_0)
\coloneqq
\min\left\{
n\in\mathbb N_0:
\|\bar\pi_0\bar T^n-\bar\pi\|_{\rm TV}\le\varepsilon
\right\},
\end{equation}
where $\varepsilon\in(0,1)$, $\bar T^n$ denotes the $n$-step
transition kernel, and $\bar\pi_0\bar T^n$ is the distribution
of the chain after $n$ transitions from $\bar\pi_0$. Here, $\|\cdot\|_{\rm TV}$ denotes total variation distance,
and the minimum is understood to be $+\infty$ if the
set is empty. Even when a chain is uniformly ergodic for every fixed state space dimension $d$, its mixing time may grow rapidly with $d$.
\textbf{Our focus is the dependence of this mixing time on
the dimension $d$ of the state space.}
Under the assumptions stated precisely in
\Cref{sec:main_results}, we obtain upper bounds for the mixing time
that grow at most polynomially in $d$.
These results complement the existing ergodicity and
optimal-scaling analyses of the Stereographic and Sub-Cauchy samplers by providing quantitative
mixing-time guarantees.

Our mixing bounds are stated for warm initial
distributions. An initial distribution $\bar\pi_0$
is $M$-warm with respect to the target $\bar\pi$ if $\bar\pi_0(A)\le M\cdot \bar\pi(A)$, for all measurable sets $A$. Equivalently,
\[
\bar\pi_0\ll\bar\pi,
\qquad
\tfrac{d\bar\pi_0}{d\bar\pi}\le M
\quad \bar\pi\text{-almost everywhere}.
\]
Thus, when both distributions admit densities, the ratio between the initial density and the target density is bounded by $M$. Observe that, necessarily, $M\ge1$.

Our analysis concerns $\kappa$-concave distributions, which generalize log-concave distributions and include many heavy-tailed families~\citep{borell1974convex,borell1975convex}. For $\kappa<0$, this class is strictly broader than the log-concave case and allows for polynomially decaying tails, while still containing every log-concave distribution. For $\kappa<0$, $\kappa$-concave densities have the form $V^{1/\kappa-d}$, up to normalization, on a convex support, where $V$ is positive and convex. This representation allows polynomial tail decay while retaining convex upper level sets. The definition and basic properties of $\kappa$-concavity are reviewed in \Cref{sec:kappa-concave-def}.

\paragraph{Motivating examples from Bayesian statistics.}
A familiar example of a $\kappa$-concave distribution is the multivariate Student's $t$ distribution with $\nu$ degrees of freedom, which is $\kappa$-concave with $\kappa=-1/\nu$. Student's $t$ distributions arise naturally in Bayesian analysis, for instance as marginal posteriors for regression coefficients in Gaussian linear models with unknown variance~\citep{tiao1964bayesian,tiao1964bayes}. In \Cref{sec:example-student}, we provide explicit mixing-time guarantees for SPS and SCPS on this canonical family of heavy-tailed $\kappa$-concave targets.

Beyond Student's $t$ distributions, $\kappa$-concave posteriors also arise in Bayesian models with more complex likelihoods and latent structure. We consider, in particular, two Bayesian regression settings in which integrating out unknown scale parameters leads to $\kappa$-concave posteriors with polynomial tails: exponential-power regression with unknown scale and Gaussian regression with unknown variance and censored responses~\citep{ferreira2014bayesian,karling2024conjugacy}. The resulting posteriors can be nonelliptical or asymmetric, providing Bayesian applications of our mixing-time analysis beyond the Student's $t$ family. \Cref{sec:example-quartic,sec:example-censored} present these models, verify their $\kappa$-concavity, and establish mixing-time guarantees under explicit conditions.

\paragraph{Main contributions.} 
Our main results, stated precisely in
\Cref{sec:main_results}, address two regimes
of the concavity parameter.
\begin{enumerate}
\item When $-c_1/d\le\kappa\le0$, we establish a total variation mixing-time bound of $\mathcal O(d^2\log d)$ for both SPS and SCPS. The algorithms' hyperparameters satisfy $R\asymp\sqrt d$ and $h\asymp d^{-1}$, with a fixed observer latitude $l_o$ for SCPS. The result holds for every fixed accuracy $\varepsilon\in(0,1)$, under covariance and local regularity assumptions, and for initial distributions $\bar\pi_0$ whose warmness parameter $M$ grows at most polynomially in $d$. This regime includes Student's $t$ targets with degrees of freedom scaling linearly in the dimension. \Cref{cor:SPS_mixing_kappa_concave} states the result for SPS, and \Cref{rmk:vanishing-kappa-SCPS} extends it to SCPS. These results are summarized in the first row of \Cref{table:main-mixing-bounds}.

\item When $\kappa\in(-1,0)$ is fixed independently of dimension, we establish a total variation mixing-time bound of $\mathcal O(d^{2\beta+2})$ for SCPS, where
\[
R\asymp d^\beta,\qquad h\asymp d^{-(\beta+1)},\qquad
\beta>\tfrac{1}{2}+\tfrac{3|\kappa|}{2(1-|\kappa|)}.
\]
The result holds for every fixed accuracy $\varepsilon\in(0,1)$, under a quantile condition on the Euclidean norm and local regularity assumptions, and for initial distributions $\bar\pi_0$ whose warmness parameter $M$ is uniformly bounded in $d$. This regime includes Student's $t$ targets with fixed degrees of freedom $\nu>1$, including cases with infinite second moments. \Cref{cor:SCPS_mixing_kappa_concave} states the precise result. The bound is summarized in the second row of \Cref{table:main-mixing-bounds}.

\item For radially symmetric targets satisfying a curvature condition (see \Cref{prop:sharper_isoperimetry} for details), we establish sharper weighted isoperimetric inequalities, improving the lower bound on the Cheeger constant from $\Phi=\Omega(1)$ to $\Phi=\Omega(\sqrt d)$. These inequalities may be of independent interest. This refinement follows by interpreting a weighted Poincar\'e inequality as an unweighted one in a conformally changed geometry and applying a Buser-type argument under the associated curvature-dimension condition. Using this sharpened isoperimetric inequality, the mixing-time bounds improve to $\mathcal O(d\log d)$ for both SPS and SCPS in the regime $-c_1/d\le\kappa\le0$, and to $\mathcal O(d^{2\beta+1})$ for SCPS in regime with fixed $\kappa\in(-1,0)$. In the latter case, the condition on $\beta$ is relaxed to
\[
\beta>\tfrac{1}{2}+\tfrac{|\kappa|}{1-|\kappa|}.
\]
Under these additional assumptions, the exponent $2\beta+1$ in the mixing time upper bound can be chosen arbitrarily close to $2$ as $|\kappa|\to0$. The precise results are given in \Cref{cor:SPS_mixing_CD,cor:SCPS_mixing_CD}, with the SCPS extension in the moderately heavy-tailed regime provided by \Cref{rmk:vanishing-kappa-SCPS}. These bounds are summarized in the third and fourth rows of \Cref{table:main-mixing-bounds}.

\end{enumerate}

We emphasize that the results in the first two points apply without assuming that the target is a product distribution or that its density is spherically symmetric.

 \begin{table}[ht]
\begin{center}

\renewcommand{\arraystretch}{1.5}

\begin{tabular}{@{}c@{\hspace{0.05cm}}c@{\hspace{0.05cm}}c@{}}

\begin{tabular}{|c|c|c|c|}
\hhline{~|---|}
\multicolumn{1}{c|}{}
& \cellcolor{tablegray}$R$
& \cellcolor{tablegray}$l_o$
& \cellcolor{tablegray}$h$
\\
\hline

\cellcolor{tablegray} $-c_1/d\le\kappa\le0$
& $\asymp\sqrt d$
& $\mathrm{const}$
& $\asymp d^{-1}$
\\
\hline

\cellcolor{tablegray} $\kappa\in(0,1)$ fixed
& $\asymp d^\beta$
& $\mathrm{const}$
& $\asymp d^{-(\beta+1)}$
\\
\hline

\cellcolor{tablegray} $-c_1/d\le\kappa\le0$ $+$ curvature
& $\asymp\sqrt d$
& $\mathrm{const}$
& $\asymp d^{-1}$
\\
\hline

\cellcolor{tablegray} $\kappa\in(0,1)$ fixed $+$ curvature
& $\asymp d^\beta$
& $\mathrm{const}$
& $\asymp d^{-(\beta+1)}$
\\
\hline
\end{tabular}

&

\raisebox{0.1ex}{
    \scalebox{1.5}{$\Rightarrow$}
}

&

\begin{tabular}{|c|}
\hline
\cellcolor{tablegray}$n_{\rm mix}(\varepsilon,\bar\pi_0)$
\\
\hline

$\mathcal O(d^2\log d)$
\\
\hline

$\mathcal O(d^{2\beta+2})$
\\
\hline

$\mathcal O(d\log d)$
\\
\hline

$\mathcal O(d^{2\beta+1})$
\\
\hline
\end{tabular}
\end{tabular}
\end{center}
\caption{Summary of the parameter scalings and mixing-time upper bounds under the assumptions of the corresponding results in \Cref{sec:main_results}. Rows 1 and 3 apply to both SPS and SCPS, by \Cref{cor:SPS_mixing_kappa_concave,cor:SPS_mixing_CD} and \Cref{rmk:vanishing-kappa-SCPS}. Rows 2 and 4 apply to SCPS, by \Cref{cor:SCPS_mixing_kappa_concave,cor:SCPS_mixing_CD}. Rows 3 and 4 additionally require radial symmetry of the target distribution and the hypotheses of \Cref{prop:sharper_isoperimetry}. In rows 1 and 3, the initial distribution is $M$-warm with $M=\mathcal O(d^q)$ for fixed $q\ge0$. In rows 2 and 4, $M=\mathcal O(1)$. The conditions on $\beta$ in rows 2 and 4 are $\beta>1/2+3|\kappa|/(2(1-|\kappa|))$ and $\beta>1/2+|\kappa|/(1-|\kappa|)$, respectively. The observer latitude $l_o\in[1,2)$ is relevant only to SCPS and is fixed independently of $d$. All bounds hold for fixed accuracy $\varepsilon\in(0,1)$ and sufficiently small step-size prefactors $c_h$.
}
\label{table:main-mixing-bounds}
\end{table}

\paragraph{Proof techniques.}
The proofs use an argument based on
$s$-conductance, which relates the mixing time to the
probability of moving between regions of the state
space \citep{lovasz1993random}.
Building on the approach used
in~\citep{dwivedi2019log,wu2022minimax}, we combine
three ingredients: an isoperimetric inequality,
a $(\delta,\rho)$-coupling condition, and a tail bound. As described in \Cref{sec:proof-sketch-s-cond}, these ingredients allow us to determine a bound for the $s$-conductance and its mixing-time consequences.
The three ingredients are developed in
\Cref{sec:isoperimetric_inequalities,sec:coupling,sec:tail_bounds}, respectively, with sharper isoperimetric inequalities established under additional curvature assumptions in \Cref{sec:shaper_iso}.

A central and novel contribution of our analysis, essential to
our mixing-time results, is the transfer of weighted isoperimetric inequalities on $\mathbb{R}^d$ into unweighted
isoperimetric inequalities on the transformed state spaces.
Starting from inequalities for $\kappa$-concave measures
\citep{bobkov2009weighted,bobkov2009weighted2},
we choose weight functions adapted to the differentials of
the stereographic and Sub-Cauchy projection maps.
These weights account for the distortion of lengths
and boundaries under projection.
This transfer provides the isoperimetric estimates
for the transformed targets on the sphere needed in the conductance
argument (see
\Cref{sec:isoperimetric_inequalities}).

For radially symmetric targets, we further sharpen these inequalities under an additional curvature condition. Viewing the weighted Poincar\'e inequality as a standard Poincar\'e inequality in a conformally changed geometry, in which our condition plays the role of log-concavity, we adapt the Buser-type argument used for log-concave targets \citep{Ledoux2004-ry}. This improves the Cheeger constant from order one to order $\sqrt d$, and the mixing-time bound from $\mathcal O(d^2\log d)$ to $\mathcal O(d\log d)$.

The $(\delta,\rho)$-coupling condition requires the
one-step transition distributions from nearby starting
points to be close in total variation.
Establishing this condition requires controlling the
distortion induced by the projection, which becomes
particularly pronounced near the north pole for SPS
and near the boundary of the bright side for SCPS.
We therefore establish coupling outside a neighborhood
$\mathcal C$ of these regions
(see \Cref{sec:coupling}).

Under the projection maps, $\mathcal C$ corresponds
to a Euclidean tail region, whose probability we
control in \Cref{sec:tail_bounds}.
When $-c_1/d\le\kappa\le0$, the tail bound permits
the choice $R\asymp\sqrt d$ for either sampler.
For fixed $\kappa\in(-1,0)$, our SCPS analysis uses
a larger projection scale $R$ to make the excluded region $\mathcal{C}$
sufficiently unlikely despite the weaker tail bound.

\paragraph{Outline.}
The remainder of the paper is organized as follows. We first state our main mixing-time results in \Cref{sec:main_results}, and then describe the stereographic and Sub-Cauchy projection maps and the corresponding sampling algorithms in \Cref{sec:SPS-and-SCPS}. \Cref{sec:kappa-concave} is dedicated to $\kappa$-concave distributions: in \Cref{sec:kappa-concave-def}, we introduce this family of distributions and their basic properties, and in \Cref{sec:applications-results}, we apply our mixing-time results to several $\kappa$-concave Bayesian targets, including Student's $t$ posteriors, exponential-power regression posteriors, and Gaussian regression posteriors with censored responses. \Cref{sec:proof-sketch} provides an overview of the conductance argument underlying our mixing-time bounds (\Cref{sec:proof-sketch-s-cond}) and shows how the analysis reduces to three main ingredients: isoperimetric, coupling, and tail estimates. These are developed in \Cref{sec:isoperimetric_inequalities,sec:coupling,sec:tail_bounds}, respectively, with sharper isoperimetric inequalities established under additional assumptions in \Cref{sec:shaper_iso}. We conclude with a discussion in \Cref{sec:discussion}, while detailed proofs are collected in \Cref{sec:appendix-proofs} and the sharpened weighted isoperimetric inequalities corresponding to the additional curvature condition are developed in \Cref{sec:cheeger}.

\paragraph{Notation.}
Throughout the paper, we use the following conventions.
We write $|x|$ for the Euclidean norm,
$\|\cdot\|_{\rm op}$ for the operator norm, and
$\mathcal P(\mathcal X)$ for the probability measures
on a state space $\mathcal X$. We denote by $\{e_i\}_{i=1}^{d+1}$ the canonical orthonormal basis of $\mathbb{R}^{d+1}$.
For probability measures $\mu,\nu$ on the same space, the total variation distance between $\mu$ and $\nu$ is defined as $\|\mu-\nu\|_{\rm TV}
\coloneqq\sup_A|\mu(A)-\nu(A)|$, where the supremum is over measurable sets.
For a Markov kernel $T$, $\mu T^n$ denotes the distribution
after $n$ transitions from initial distribution $\mu$.
For a measurable map $F$, $F_{\#}\mu$ denotes the
pushforward measure, defined by
$(F_{\#}\mu)(A)=\mu(F^{-1}(A))$, for every measurable set $A$.
We use $\bar\pi$ for the Euclidean target and $\pi$
for its transformed counterpart, using the same symbols
for their densities when these exist.
Asymptotic notation refers to $d\to\infty$.
For positive sequences $a_d,b_d$, we write
$a_d=\mathcal O(b_d)$ if $a_d\le Cb_d$ for all sufficiently
large $d$, and $a_d=\Omega(b_d)$ if $a_d\ge cb_d$ for all
sufficiently large $d$, where $c,C>0$ are independent
of $d$.
We write $a_d\asymp b_d$ when both bounds hold.
\section{Main Results}
\label{sec:main_results}

The main objective of this work is to establish upper bounds on the total variation mixing times of the Stereographic Projection Sampler (SPS) and the Sub-Cauchy Projection Sampler (SCPS). Both algorithms transform a target distribution $\bar\pi$ on $\mathbb R^d$ into a distribution on a sphere of unit radius, implement a Random Walk Metropolis algorithm on this Riemannian manifold, and then map the samples back to $\mathbb R^d$. The SPS and SCPS, which are described in more detail in \Cref{sec:SPS-and-SCPS}, depend on the following algorithmic hyperparameters: $R>0$ controls the projection scale, $h>0$ is the proposal step size on the sphere, and the SCPS additionally introduces an observer latitude $l_o\in[1,2)$.

We recall that the total variation mixing time, defined in~\eqref{eq:mixing-time-definition}, corresponds to the number of iterations required to reach total variation distance at most $\varepsilon$ from the target distribution $\bar\pi\in\mathcal{P}(\mathbb{R}^d)$, starting from an initial distribution $\bar\pi_0 \in\mathcal{P}(\mathbb{R}^d)$. We consider in particular $M$-warm initial distributions with respect to $\bar\pi$, meaning that $ \bar\pi_0 (A)\le M\cdot \bar\pi(A)$ for every measurable $A\subseteq \mathbb{R}^d$.

Our results concern the class of $\kappa$-concave distributions (see \Cref{sec:kappa-concave-def}), of which we consider two regimes. In the first regime of $-c_1/d\le\kappa\le0$, we obtain an $\mathcal O(d^2\log d)$ mixing-time bound from polynomially warm initial distributions. In the second regime, $\kappa\in(-1,0)$ is fixed independent of dimension, and we obtain a polynomial mixing-time bound $\mathcal{O}(d^{2\beta+2})$, with $\beta>\tfrac{1}{2}+\tfrac{3|\kappa|}{2(1-|\kappa|)}$, from initial distributions with uniformly bounded warmness. We refer to these as the \textit{moderately heavy-tailed} and \textit{heavier-tailed} regimes, respectively, reflecting the increasing degree of tail heaviness allowed by their assumptions. These terms are used only to distinguish the two regimes considered here and do not exclude lighter-tailed targets, including log-concave distributions. Indeed, we remark that log-concave distributions are $\kappa$-concave for every $\kappa\le 0$.

\subsection{SPS and SCPS for moderately heavy-tailed targets}
\label{sec:SPS_results}

Our first result applies when the $\kappa$-concavity parameter satisfies $-{c_1}/{d}\le\kappa\le0$. This includes log-concave targets and allows polynomial tails whose decay becomes lighter as the dimension increases. For example, this is the case of multivariate Student's $t$ distributions with degrees of freedom $\nu_d\asymp d$.

\begin{theorem}[Mixing for moderately heavy-tailed targets]
\label{cor:SPS_mixing_kappa_concave}
Let $\bar\pi$ be a centered $\kappa$-concave probability distribution on $\mathbb R^d$, with a positive twice continuously differentiable density $\bar\pi$, where $-\frac{c_1}{d}\le\kappa\le0$ for a constant $c_1>0$. Assume that its covariance matrix $\Sigma_d$ satisfies
\begin{equation}
    \label{eq:cov-mat-eig-bounds-SPS-results}
     0<\lambda_L\le\lambda_{\min}(\Sigma_d)\le\lambda_{\max}(\Sigma_d)\le\lambda_U<\infty,
\end{equation}
with $\lambda_L$ and $\lambda_U$ constant in $d$. Suppose that, for every fixed $C>0$,
\begin{equation}
    \label{eq:grad-Hess-bounds-SPS-results}
    \sup_{|x|\le C\sqrt d}|\nabla_x\log\bar\pi(x)|=\mathcal O(\sqrt d),\qquad \sup_{|x|\le C\sqrt d}\|\nabla_x^2\log\bar\pi(x)\|_{\rm op}=\mathcal O(1),
\end{equation}
where the constants in the $\mathcal{O}(\cdot)$ may depend on $C$. Fix $c_R>0$ and set $R=c_R\sqrt d$. Then, there exists a sufficiently small constant $c_h>0$ such that the SPS with step size $h=c_hd^{-1}$ satisfies
\begin{align}
\label{eq:mixing_time_UB_SPS}
    n_{\rm mix}(\varepsilon,\bar\pi_0 )=\mathcal O(d^2\log d)
\end{align}
for every fixed $\varepsilon\in(0,1)$ and every $M$-warm initial distribution with $M=\mathcal O(d^q)$ for a fixed $q\ge0$.
\end{theorem}

\begin{proof}
    The proof of \Cref{cor:SPS_mixing_kappa_concave} is provided in \Cref{sec:proofs-main-results}.
\end{proof}

The assumption~\eqref{eq:cov-mat-eig-bounds-SPS-results} on the covariance matrix yields $\mathbb E[|X|^2]=\mathcal O(d)$, which implies a concentration of the mass of the Euclidean norm at scale $\sqrt{d}$. This motivates the choice $R\asymp\sqrt d$. As proved in \Cref{lem:tail_bound_kappa_concave_light_tail}, in this tail regime, the probability outside a sufficiently large fixed multiple of $\sqrt{d}$ decreases exponentially in $\sqrt{d}$.
This tail bound allows us to apply the result with an initial distribution with warm start parameter $M$ that grows polynomially in dimension.

Key to the proof of \Cref{cor:SPS_mixing_kappa_concave} is a Cheeger isoperimetric inequality which we develop for generic $\kappa$-concave targets, with a resulting Cheeger constant of order one. This can be sharpened for radially symmetric targets that satisfy an additional curvature condition. These additional assumptions yield a Cheeger constant of order $\sqrt d$ instead of order one, which improves the mixing time upper bound as in the theorem below.

\begin{theorem}[Mixing for moderately heavy-tailed targets under a curvature condition]
\label{cor:SPS_mixing_CD}
Consider the setting of \Cref{cor:SPS_mixing_kappa_concave}, and assume in addition that $\bar\pi$ is radially symmetric and satisfies the additional curvature condition of \Cref{prop:sharper_isoperimetry}. Then, there exists a sufficiently small constant $c_h>0$ such that the SPS with step size $h=c_hd^{-1}$ satisfies
\begin{align}
\label{eq:mixing_time_UB_SPS_CD}
    n_{\rm mix}(\varepsilon,\bar\pi_0)=\mathcal O(d\log d)
\end{align}
for every fixed $\varepsilon\in(0,1)$ and every $M$-warm initial distribution with $M=\mathcal O(d^q)$ for a fixed $q\ge0$.
\end{theorem}

\begin{proof}
    The proof of \Cref{cor:SPS_mixing_CD} is provided in \Cref{sec:proofs-main-results}.
\end{proof}

As explained in \Cref{sec:shaper_iso}, the condition of \Cref{prop:sharper_isoperimetry} relates to the curvature--dimension condition of the celebrated Bakry--Emery calculus. In \Cref{sec:shaper_iso}, we show that these conditions hold for the Student's $t$ distributions with $\nu_d\asymp d$ degrees of freedom and isotropic scale matrix.

\begin{remark}[Extension of the results to SCPS]
\label{rmk:vanishing-kappa-SCPS}
The conclusions of \Cref{cor:SPS_mixing_kappa_concave,cor:SPS_mixing_CD} also hold for SCPS with any fixed $l_o\in[1,2)$, under the same target and warm-start assumptions, with $R=c_R\sqrt d$ and $h=c_hd^{-1}$. The sufficiently small constant $c_h$ may depend on $l_o$. A proof is given in \Cref{sec:proofs-extension-SCPS}.
\end{remark}

\subsection{SCPS for heavier-tailed targets}
\label{sec:SCPS_results}

When $\kappa\in(-1,0)$ is fixed, an exponential bound on the tail probability is no longer guaranteed. Nevertheless, \Cref{lem:tail_bound_kappa_concave_kappa_const} provides a polynomial tail bound. For SCPS in this regime, we compensate for this weaker tail bound by taking a larger projection scale $R$ together with a smaller proposal step size $h$.

\begin{theorem}[Mixing for heavier-tailed targets]
\label{cor:SCPS_mixing_kappa_concave}
Let $\bar\pi$ be a $\kappa$-concave probability distribution on $\mathbb R^d$, with a positive twice continuously differentiable density and fixed $\kappa\in(-1,0)$. Let $X\sim\bar\pi$, and assume that the $2/3$-quantile $r$ of $|X|$ satisfies $r=\mathcal O(\sqrt d)$. Fix $l_o\in[1,2)$ and $c_R>0$, independently of $d$, and set
\begin{align}
\label{eq:scale-parameter-SCP}
     R=c_Rd^\beta,     \qquad    \beta>\tfrac{1}{2}+\tfrac{3|\kappa|}{2(1-|\kappa|)}.
\end{align}
Suppose that, for every fixed $C>0$,
\begin{equation}
    \label{eq:grad-Hess-bounds-SCPS-results}
    \sup_{|x|\le C\cdot R}|\nabla_x\log\bar\pi(x)|=\mathcal O(d),\qquad    \sup_{|x|\le C\cdot R}\|\nabla_x^2\log\bar\pi(x)\|_{\rm op}=\mathcal O(d),
\end{equation}
where the constants in the $\mathcal{O}(\cdot)$ may depend on $C$. Then, there exists a sufficiently small constant $c_h>0$ such that the SCPS with step size $h=c_h\cdot d^{-(\beta+1)}$ satisfies
\begin{align}
\label{eq:mixing_time_UB_SCPS}
    n_{\rm mix}(\varepsilon,\bar\pi_0 ) =\mathcal O(d^{2\beta+2})
\end{align}
for every fixed $\varepsilon\in(0,1)$ and every $M$-warm initial distribution with $M=\mathcal O(1)$.
\end{theorem}

\begin{proof}
    The proof of \Cref{cor:SCPS_mixing_kappa_concave} is provided in \Cref{sec:proofs-main-results}.
\end{proof}

Unlike \Cref{cor:SPS_mixing_kappa_concave}, this theorem does not require a finite covariance matrix. Its assumption on the concentration of the norm is formulated through a condition on its quantile, thus it also accommodates targets with infinite second moments.

As in the moderately heavy-tailed case, \Cref{cor:SCPS_mixing_kappa_concave} can be sharpened under the conditions of \Cref{prop:sharper_isoperimetry}. The larger Cheeger constant also relaxes the condition on the exponent $\beta$ of the projection scale.

\begin{theorem}[Mixing for heavier-tailed targets under a curvature condition]
\label{cor:SCPS_mixing_CD}
Consider the setting of \Cref{cor:SCPS_mixing_kappa_concave}, with the condition~\eqref{eq:scale-parameter-SCP} replaced by
\begin{align*}
     R=c_Rd^\beta,     \qquad    \beta>\tfrac{1}{2}+\tfrac{|\kappa|}{1-|\kappa|},
\end{align*}
and assume in addition that $\bar\pi$ is radially symmetric and satisfies the additional curvature condition of \Cref{prop:sharper_isoperimetry}. Then, there exists a sufficiently small constant $c_h>0$ such that the SCPS with step size $h=c_h\cdot d^{-(\beta+1)}$ satisfies
\begin{align}
\label{eq:mixing_time_UB_SCPS_CD}
    n_{\rm mix}(\varepsilon,\bar\pi_0)=\mathcal O(d^{2\beta+1})
\end{align}
for every fixed $\varepsilon\in(0,1)$ and every $M$-warm initial distribution with $M=\mathcal O(1)$.
\end{theorem}

\begin{proof}
    The proof of \Cref{cor:SCPS_mixing_CD} is provided in \Cref{sec:proofs-main-results}.
\end{proof}

For the same $\beta$, this improves the bound of \Cref{cor:SCPS_mixing_kappa_concave} by a factor of order $d$. Moreover, the threshold on $\beta$ converges to $1/2$ as $\kappa\to0$, so that the exponent $2\beta+1$ can be chosen close to $2$ when $|\kappa|$ is small.

An overview of the proof of these results is given in \Cref{sec:proof-sketch}. The proofs combine weighted isoperimetric inequalities for the Euclidean target, $(\delta,\rho)$-coupling estimates for the SPS and SCPS transition kernels, and bounds on the tail probability of the Euclidean target. 

\subsection{Applications to Bayesian models}
\label{sec:applications-overview}

The mixing-time results above yield explicit guarantees for several classes of Bayesian posterior distributions. In particular, we consider multivariate Student's $t$ targets, exponential-power regression with an unknown scale, and Gaussian regression with an unknown variance and censored responses. The latter two examples give $\kappa$-concave targets that need not be elliptical or symmetric and for which MCMC sampling is naturally relevant. In \Cref{sec:applications-results}, we introduce these models, establish their $\kappa$-concavity, and apply the preceding theorems to obtain explicit SPS and SCPS mixing-time bounds under conditions on the corresponding sequences of statistical models.
\section{Stereographic and Sub-Cauchy projection samplers}
\label{sec:SPS-and-SCPS}

This section introduces the projection maps and algorithms underlying the Stereographic Projection Sampler (SPS) and the Sub-Cauchy Projection Sampler (SCPS), together with the geometric identities used later in the analysis. Both samplers map the Euclidean state space $\mathbb R^d$ to a subset of the shifted unit sphere
\begin{align*}\mathcal{S}^d\coloneqq\left\{u\in\mathbb{R}^{d+1}:|u-e_{d+1}|=1\right\},\end{align*}
inducing a corresponding target distribution on the transformed state space. The Markov chain is then constructed on this transformed state space, and the resulting samples are mapped back to $\mathbb R^d$.

Observe that the south and north poles of the sphere $\mathcal{S}^d$ are $0$ and $2\,e_{d+1}$, respectively. We identify the Euclidean space $\mathbb{R}^d$ with the hyperplane $\{(x,0):x\in\mathbb{R}^d\}$, tangent to $\mathcal{S}^d$ at its south pole. We write $u=(u_{1:d},u_{d+1})$, with $u_{1:d}\in\mathbb{R}^d$, and we denote the zero in $\mathbb{R}^d$ by $\boldsymbol{0}$. Throughout this section, we let $X\sim\bar\pi$, and when considering a projection $F\in\{{\rm SP},{\rm SCP}\}$, we write $\pi=(F^{-1})_{\#}\bar\pi$ for the corresponding transformed target.

\subsection{Projection maps}
\label{sec:subsec:projection_maps}

\subsubsection{Stereographic projection}
\label{sec:stereographic_projection}

The stereographic projection is obtained by placing an observer at the north pole $2e_{d+1}$ and intersecting the line passing through the observer and $u\in\mathcal{S}^d$ with the tangent hyperplane at the south pole, which we identify with $\mathbb{R}^d$. After scaling by $R$, the map is
\begin{align*}
{\rm SP}:\mathcal{S}^d\setminus\{2e_{d+1}\}\longrightarrow\mathbb{R}^d,\qquad {\rm SP}(u)=\frac{2R}{2-u_{d+1}}u_{1:d}.
\end{align*}
The stereographic projection is a diffeomorphism between $\mathcal{S}^d\setminus\{2\,e_{d+1}\}$ and $\mathbb{R}^d$, with inverse
\begin{equation}
\label{eq:SP_inverse}
u_{1:d}=\frac{4Rx}{|x|^2+4R^2},\qquad u_{d+1}=\frac{2|x|^2}{|x|^2+4R^2}.\end{equation}
Note that $x=\boldsymbol{0}\in\mathbb{R}^d$ corresponds to the south pole $u=(\boldsymbol{0},0)\in\mathbb{R}^{d+1}$, while $|x|\to\infty$ corresponds to points approaching the north pole $u=2 e_{d+1}$. Moreover,
\begin{align}\label{eq:SP_radius_height_relation}|x|^2=4\, R^2\frac{u_{d+1}}{2-u_{d+1}},\end{align}
and therefore, for every $c\in(0,2)$,
\begin{align}
\label{eq:relate_u_d+1_|x|_SP}
u_{d+1}>2-c\quad\Longleftrightarrow\quad|x|^2>4\tfrac{2-c}{c}R^2.
\end{align}

Although the transformed measure $\pi$ is supported on $\mathcal{S}^d\setminus\{2e_{d+1}\}$, we regard $\pi$ as a probability measure on all of $\mathcal{S}^d$ by assigning zero mass to the north pole. In our proofs, we will derive a $(\delta,\rho)$-coupling condition on the whole state space $\mathcal{S}^d$, excluding the spherical cap
\begin{align}
\label{eq:def_cap_SP}
\mathcal{C}_{\rm SP}\coloneqq\left\{u\in\mathcal{S}^d:u_{d+1}>2-c\right\}.
\end{align}
Note that the relation~\eqref{eq:relate_u_d+1_|x|_SP} yields
\begin{equation}
\label{eq:cap_tail_probability_SP}
\pi(\mathcal{C}_{\rm SP})=\mathbb{P}\left(|X|^2>4\tfrac{2-c}{c}R^2\right).
\end{equation}
Thus, controlling the probability of the excluded region $\mathcal{C}_{\rm SP}$ reduces to controlling a Euclidean tail probability of $\bar\pi$.

If $\bar\pi$ admits a density with respect to Lebesgue measure, then the transformed density with respect to Riemannian volume on $\mathcal{S}^d$ satisfies \cite[Example~2]{grazzi2026sub}
\begin{align}\label{eq:SP_transformed_density}\pi(u)=\bar\pi(x)\cdot J_{\rm SP}(x),\qquad x={\rm SP}(u),\qquad J_{\rm SP}(x)=\left(\frac{4R^2+|x|^2}{4R}\right)^d.\end{align}

\subsubsection{Sub-Cauchy projection}
\label{sec:subcauchy_projection}

The Sub-Cauchy projection is obtained by placing an observer at a point $o$ inside the sphere $\mathcal{S}^d$. Here, we restrict the analysis to observers $o=(\boldsymbol{0},l_o)$, with $l_o\in[1,2)$. The portion of the sphere visible from the observer is then
\begin{align}
\label{eq:def_bright_side}
\mathcal{S}_o^d\coloneqq\left\{u\in\mathcal{S}^d:u_{d+1}<l_o\right\},
\end{align}
which we call the \emph{bright side} of the sphere, while its complement $\mathcal{S}^d\setminus\mathcal{S}_o^d$ is the \emph{dark side}.

Following~\cite[Section~2.1]{grazzi2026sub}, after scaling by $R$, the Sub-Cauchy projection map is
\begin{align}\label{def:SCP_projection}{\rm SCP}:\mathcal{S}_o^d\longrightarrow\mathbb{R}^d,\qquad {\rm SCP}(u)=\frac{R\cdot l_o}{l_o-u_{d+1}}u_{1:d}.\end{align}
The Sub-Cauchy projection is a diffeomorphism between $\mathcal{S}_o^d$ and $\mathbb{R}^d$, with inverse
\begin{equation}\label{eq:SCP_inverse}u_{1:d}=\frac{L(x)}{R}x,\qquad u_{d+1}=(1-L(x))l_o,\end{equation}
where
\begin{align}
\label{def:SCP_D-S-L}
    D(x)\coloneqq\frac{|x|^2}{R^2}+l_o^2,\qquad S(x)\coloneqq\sqrt{l_o^2(l_o-1)^2+D(x)l_o(2-l_o)}, \qquad L(x)\coloneqq\frac{l_o(l_o-1)+S(x)}{D(x)}.
\end{align}
As for the stereographic projection, the Euclidean point $x=\boldsymbol{0}$ corresponds to the south pole $u=(\boldsymbol{0},0)\in\mathbb{R}^{d+1}$. Instead, $|x|\to\infty$ corresponds to points approaching the boundary $u_{d+1}=l_o$ of the bright side. Moreover, for $x={\rm SCP}(u)$,
\begin{align}\label{eq:relate_u_d+1_norm_x}\frac{|x|^2}{R^2}=\frac{l_o^2\cdot u_{d+1}\cdot (2-u_{d+1})}{(l_o-u_{d+1})^2}.\end{align}
Consequently, for every $c\in(0,l_o)$,
\begin{align}\label{eq:SCP_cap_tail_relation}u_{d+1}>l_o-c\quad\Longleftrightarrow\quad|x|^2>\bar c(c,l_o)\cdot R^2,\end{align}
where
\begin{align}\label{eq:SCP_bar_c}\bar c(c,l_o)\coloneqq\frac{l_o^2\cdot (l_o-c)\cdot (2-l_o+c)}{c^2}.\end{align}
For the $(\delta,\rho)$-coupling argument in the proof, we will exclude the region
\begin{align}
\label{eq:def_cap_SCP}
\mathcal{C}_{\rm SCP}\coloneqq\left\{u\in\mathcal{S}_o^d:u_{d+1}>l_o-c\right\},
\end{align}
which is a neighborhood of the boundary of the bright side. The relation~\eqref{eq:SCP_cap_tail_relation} then yields
\begin{align}
\label{eq:cap_tail_probability_SCP}
\pi(\mathcal{C}_{\rm SCP})= \mathbb{P}\left(|X|^2>\bar c(c,l_o)\cdot R^2\right).
\end{align}
Thus, controlling the probability of the excluded region $\mathcal{C}_{\rm SCP}$ reduces to controlling a Euclidean tail probability of $\bar\pi$.

If $\bar\pi$ admits a density with respect to the Lebesgue measure, then the transformed density with respect to Riemannian volume on $\mathcal{S}_o^d$ satisfies \cite[Section~2.1]{grazzi2026sub}
\begin{align*}
\pi(u)=\bar\pi(x)\cdot J_{\rm SCP}(x),\qquad x={\rm SCP}(u),\qquad J_{\rm SCP}(x)=\frac{R^d\cdot S(x)}{l_o\cdot L(x)^d}.\end{align*}

\subsection{Projection samplers}
\label{sec:projection_samplers}

For each sampler, we construct a Random Walk Metropolis chain targeting the distribution $\pi$ obtained by transforming the Euclidean target $\bar\pi$ to the spherical state space. The proposal kernel $Q$ is defined as follows.

For $u\in\mathcal{S}^d$ and step size $h>0$, let $Z\sim\mathcal{N}_{d+1}(0,h^2I_{d+1})$ and define its projection onto the tangent space at $u$ by
\begin{align*}
W\coloneqq Z-\langle Z,u-e_{d+1}\rangle(u-e_{d+1}).
\end{align*}
The proposal $V_0$ is obtained by mapping $u-e_{d+1}+W$ back onto the unit sphere centered at $e_{d+1}$,
\begin{align*}
V_0\coloneqq e_{d+1}+\frac{u-e_{d+1}+W}{\sqrt{1+|W|^2}}.
\end{align*}
Its geodesic displacement from $u$ is
\begin{align}
\label{eq:proposal_angle}
\Theta\coloneqq\mathsf{dist}(u,V_0)=\arctan|W|.
\end{align}
The resulting proposal kernel $Q$ is symmetric with respect to Riemannian volume on $\mathcal{S}^d$.

We next describe the corresponding Metropolis-Hastings kernels for SPS and SCPS. 

\subsubsection{Stereographic Projection Sampler}
\label{sec:SPS_algorithm}

The SPS algorithm~\cite[Algorithm~1]{yang2022stereographic} uses the proposal $Q$ directly and applies a Metropolis-Hastings acceptance/rejection mechanism. Given the current state $u$ and a proposal $v$ drawn from $Q(u,\cdot)$, let $x={\rm SP}(u)$ and $y={\rm SP}(v)$. Since $Q$ is symmetric, the proposal is accepted with probability
\begin{align*}
\min\left\{1,\frac{\pi(v)}{\pi(u)}\right\}=\min\left\{1,\frac{\bar\pi(y)\cdot (4R^2+|y|^2)^d}{\bar\pi(x)\cdot (4R^2+|x|^2)^d}\right\}.\end{align*}
If the proposal is rejected, the chain remains at $u$. The resulting Markov kernel on the sphere, denoted by $T_{\rm SP}$, is reversible with respect to $\pi$.

\subsubsection{Sub-Cauchy Projection Sampler}
\label{sec:SCPS_algorithm}

The SCPS uses the same raw spherical proposal. However, a proposal drawn from $Q$ may fall outside the bright side $\mathcal{S}_o^d$. In that case, we implement the stepping-out mechanism of~\cite[Section~2.2]{grazzi2026sub}, which returns the proposal to $\mathcal{S}_o^d$. We denote the resulting proposal kernel by $Q_{\rm SCP}$. The stepping-out construction preserves symmetry with respect to Riemannian volume on $\mathcal{S}_o^d$~\cite[Section~2.2 and Appendix~A.2]{grazzi2026sub}.

Given the current state $u$ and a proposal $v$ drawn from $Q_{\rm SCP}(u,\cdot)$, let $x={\rm SCP}(u)$ and $y={\rm SCP}(v)$. By symmetry of $Q_{\rm SCP}$, the proposal is accepted with probability
\begin{align*}
\min\left\{1,\frac{\pi(v)}{\pi(u)}\right\}=\min\left\{1,\frac{\bar\pi(y)S(y)L(x)^d}{\bar\pi(x)S(x)L(y)^d}\right\}.\end{align*}
If the proposal is rejected, the chain remains at $u$. The resulting Markov kernel on $\mathcal{S}_o^d$, denoted by $T_{\rm SCP}$, is reversible with respect to $\pi$.

\section{$\kappa$-concave distributions}
\label{sec:kappa-concave}

In this section, we first review the definition and basic properties of $\kappa$-concave distributions in \Cref{sec:kappa-concave-def}, for which the mixing-time results of \Cref{sec:main_results} are derived. Examples of $\kappa$-concave target distributions and the corresponding applications of our results are then given in \Cref{sec:applications-results}.

\subsection{Definition and properties of $\kappa$-concave distributions}
\label{sec:kappa-concave-def}

The theory of $\kappa$-concave measures was developed by~\citet{borell1974convex,borell1975convex} in the 1970s, extending the log-concave setting studied by~\citet{prekopa1973logarithmic}. For $\kappa\in[-\infty,1/d]$, a probability measure $\mu$ on $\mathbb{R}^d$ is called $\kappa$-concave if
\begin{equation}
\label{eq:kappa-concavity-definition}
\mu\bigl((1-t)A+tB\bigr)\ge\left((1-t)\mu(A)^\kappa+t\mu(B)^\kappa\right)^{1/\kappa}
\end{equation}
for every $t\in(0,1)$ and all compact sets $A,B$ of positive $\mu$-measure, where 
\[(1-t)A+tB=\{(1-t)x+ty:x\in A,\ y\in B\}.\]
For $\kappa=0$, the right-hand side of~\eqref{eq:kappa-concavity-definition} is interpreted by continuity as $\mu(A)^{1-t}\mu(B)^t$, so that $0$-concavity coincides with the more popular concept of log-concavity. The range of admissible concavity parameters is dimension-dependent: in particular, if $\mu$ is a $\kappa$-concave probability measure whose support has affine hull $\mathbb{R}^d$, then necessarily $\kappa\le 1/d$ (see~\cite[Section~2]{bobkov2007large}). Observe that the inequality~\eqref{eq:kappa-concavity-definition} becomes weaker as $\kappa$ decreases. In particular, a $\kappa$-concave measure is also $\kappa'$-concave for every $\kappa'\le\kappa$. Negative values of $\kappa$ therefore enlarge the class of log-concave measures and include heavy-tailed distributions. The limiting case $\kappa=-\infty$, referred to as the class of convex or hyperbolic measures, corresponds to replacing the right-hand side of~\eqref{eq:kappa-concavity-definition} by $\min\{\mu(A),\mu(B)\}$ and gives the largest class in the hierarchy. Although negative values of $\kappa$ allow for heavy-tailed distributions, they do not impose heavy tails, since every log-concave measure is also $\kappa$-concave for every $\kappa<0$. 

Borell's characterization~\cite[Lemma~2.1]{bobkov2007large} provides an equivalent formulation of $\kappa$-concavity in terms of the density for measures satisfying $\operatorname{aff}(\operatorname{supp}\mu)=\mathbb{R}^d$, where $\operatorname{aff}(\operatorname{supp}\mu)$ denotes the smallest affine subspace containing the support of $\mu$. In particular, for $\kappa<0$, a probability measure $\mu$ satisfying this condition is $\kappa$-concave if and only if it admits a density $f$, positive on an open convex set $D$ and equal to zero on $D^\complement$, such that $f^{\,\kappa/(1-d\kappa)}$ is convex on $D$. Equivalently,
\begin{equation}
\label{eq:kappa-concave-density}
f(x)\propto V(x)^{\frac{1}{\kappa}-d},\qquad x\in D,
\end{equation}
where $V$ is a positive convex function. For $\kappa=0$, the corresponding representation is $f(x)\propto e^{-U(x)}$ with $U$ convex. The representation~\eqref{eq:kappa-concave-density} also explains why negative values of $\kappa$ accommodate heavy tails. For $\kappa<0$, the exponent $\frac{1}{\kappa}-d$ in~\eqref{eq:kappa-concave-density} is negative. Thus, when $V$ has linear growth at infinity, the density has polynomial decay, in contrast with the exponential form arising when $\kappa=0$. More generally, for $\kappa<0$ and under the full-dimensional assumption $\operatorname{aff}(\operatorname{supp}\mu)=\mathbb{R}^d$, the density satisfies the bound
\begin{equation}
f(x)\le\frac{C}{1+|x|^{d-\frac{1}{\kappa}}},
\end{equation}
for some constant $C>0$ (see~\cite[Equation~(2.2)]{bobkov2007large}).
Moreover, since the exponent in~\eqref{eq:kappa-concave-density} is negative for $\kappa<0$, the upper level sets of $f$ are sublevel sets of the convex function $V$ and are therefore convex.

In the literature, $\kappa$-concave measures are also referred to as $s$-concave measures, with $s$ used in place of $\kappa$. The terminology $s$-concavity is also used for densities, but with a different parameterization, so care is needed when comparing results for $s$-concave measures with those for $s$-concave densities. The latter also appear in the statistics literature under the terminology of convex-transformed densities~\citep{seregin2010nonparametric}. By the Borell--Brascamp--Lieb inequality~\citep{borell1975convex,brascamp1976extensions}, an $s$-concavity condition on a density, with $s>-1/d$, induces a $\kappa$-concavity condition on the corresponding probability measure with $\kappa=\frac{s}{1+ds}$, or equivalently, $s=\frac{\kappa}{1-d\kappa}$. The case $s=0$ yields the Pr\'ekopa--Leindler inequality~\citep{prekopa1973logarithmic,leindler1972certain}, while the limiting case $s=+\infty$, applied to indicator functions of sets, gives the Brunn--Minkowski inequality~\citep{brascamp1976extensions}, which has the form of~\eqref{eq:kappa-concavity-definition} for the Lebesgue measure with $\kappa=1/d$.

Here, we mention several structural properties of $\kappa$-concave measures. Affine transformations, and hence marginals, preserve $\kappa$-concavity with the same parameter. Conditioning on a convex set of positive probability also preserves $\kappa$-concavity. These two facts follow directly from~\eqref{eq:kappa-concavity-definition}. Moreover, for $\kappa<0$, the polynomial tail behavior implied by~\eqref{eq:kappa-concave-density} yields moment guarantees: if $X\sim\mu$, then
\begin{equation}
\label{eq:kappa-conc-existence-of-moments}
\mathbb{E}[|X|^p]<\infty\qquad\text{for every }0<p<-\tfrac{1}{\kappa},
\end{equation}
see~\cite[Lemma~2.1 and Equation~(2.2)]{bobkov2007large}. Thus, $\kappa\in(-1,0)$ guarantees a finite first moment, while a finite second moment is guaranteed when $\kappa\in(-\frac{1}{2},0)$. Note that these are uniform guarantees for the entire class, and individual $\kappa$-concave distributions may possess additional finite moments.

\subsection{Application of mixing-time results to Bayesian posteriors}
\label{sec:applications-results}

We now apply the mixing-time results of \Cref{sec:main_results} to three families of $\kappa$-concave target distributions arising in Bayesian inference: multivariate Student's $t$ distributions, exponential-power regression posteriors, and Gaussian regression posteriors with censored responses. For each family, we introduce the underlying Bayesian model, establish the $\kappa$-concavity of the resulting posterior, and verify the assumptions of the corresponding mixing-time results. We obtain bounds for both SPS and SCPS in the moderately heavy-tailed regime, using \Cref{cor:SPS_mixing_kappa_concave} and \Cref{rmk:vanishing-kappa-SCPS}, and for SCPS in the heavier-tailed regime, using \Cref{cor:SCPS_mixing_kappa_concave}.

All mixing-time bounds are stated for fixed $\varepsilon\in(0,1)$, with $c_R>0$ fixed and $c_h>0$ sufficiently small and independent of $d$. For SCPS, $l_o\in[1,2)$ is also fixed. Proofs verifying the remaining assumptions of the main theorems are given in \Cref{sec:proofs-examples}. Throughout this section, Gamma distributions are parameterized by shape and rate.

\subsubsection{Student's $t$ distributions}
\label{sec:example-student}

A canonical example of a $\kappa$-concave distribution is the multivariate Student's $t$ distribution. This distribution is parameterized by its degrees of freedom $\nu>0$, location $m\in\mathbb R^d$, and positive definite scale matrix $S$. Its density satisfies
\begin{equation}
\label{eq:student-t-density-kappa}
f(x)\propto\left(1+\frac{(x-m)^\top S^{-1}(x-m)}{\nu}\right)^{-(d+\nu)/2}.
\end{equation}
To identify its concavity parameter, note that
\[
V(x)=\sqrt{1+\frac{(x-m)^\top S^{-1}(x-m)}{\nu}}
\]
is positive and convex, being the Euclidean norm of an affine map. Thus, $f\propto V^{-(d+\nu)}$, and the characterization in~\eqref{eq:kappa-concave-density} gives
\[
\kappa=-\tfrac{1}{\nu}.
\]
The parameter $\nu$ therefore directly controls the concavity parameter and the heaviness of the tails. A special case of the Student's $t$ distribution is the multivariate Cauchy distribution, which corresponds to $\nu=1$ and $\kappa=-1$. Moreover, for $p>0$, $\mathbb E[|X|^p]$ is finite if and only if $p<\nu$, so the Student's $t$ family attains the moment threshold in~\eqref{eq:kappa-conc-existence-of-moments}.

Student's $t$ distributions arise naturally in Bayesian linear regression after marginalizing an unknown error variance. Consider the Gaussian linear model
\[
Y\mid\theta,\tau\sim\mathcal N_n(D\theta,\tau^{-1}I_n),\qquad \theta\mid\tau\sim\mathcal N_d(m_0,\tau^{-1}V_0),\qquad \tau\sim\operatorname{Gamma}(a_0,b_0),
\]
where $Y\in\mathbb R^n$ is the response vector, $D\in\mathbb R^{n\times d}$ is the design matrix, $\theta\in\mathbb R^d$ is the vector of regression coefficients, and $\tau>0$ is the error precision. The prior parameters satisfy $m_0\in\mathbb R^d$, $V_0\succ0$, and $a_0,b_0>0$. Integrating out $\tau$ yields a multivariate Student's $t$ marginal posterior for $\theta$ with $\nu=2a_0+n$ degrees of freedom~\citep{tiao1964bayesian,tiao1964bayes}.

The relation $\kappa=-1/\nu$ connects the Student's $t$ family directly with the two regimes considered in our main results from \Cref{sec:main_results}. If $\nu_d\asymp d$, then $\kappa_d\asymp-d^{-1}$, corresponding to the moderately heavy-tailed regime. If instead $\nu_d=\nu>1$ is fixed, then $\kappa=-1/\nu\in(-1,0)$ is fixed, corresponding to the heavier-tailed regime. The resulting mixing-time guarantees are summarized in the following corollary, whose proof is given in \Cref{sec:proofs-examples-students-t}.

\begin{corollary}[Student's $t$ targets]
\label{cor:student-applications}
Let $\bar\pi_d$ be a multivariate Student's $t$ distribution with location $m_d$, degrees of freedom $\nu_d$, and scale matrix $S_d$. Assume that the eigenvalues of $S_d$ are bounded above and away from zero uniformly in $d$. After recentering at $m_d$, the following hold.
\begin{enumerate}[label=(\roman*)]
\item \label{cor:students-t-moderately-heavy-tailed} If $\nu_d\asymp d$, then the mixing times of the SPS and SCPS chains with $R=c_R\sqrt d$ and $h=c_hd^{-1}$ satisfy
\[
n_{\rm mix}(\varepsilon,\bar\pi_0)=\mathcal O(d^2\log d)
\]
from $M$-warm starts with $M=\mathcal O(d^q)$ for fixed $q\ge0$.

\item If $\nu_d=\nu>1$ is fixed, then the mixing time of the SCPS chain with $R=c_Rd^\beta$ and $h=c_hd^{-(\beta+1)}$ satisfies
\[
n_{\rm mix}(\varepsilon,\bar\pi_0)=\mathcal O(d^{2\beta+2}),\qquad \beta>\tfrac{\nu+2}{2(\nu-1)},
\]
from $M$-warm starts with $M=\mathcal O(1)$.
\end{enumerate}
\end{corollary}

\begin{remark}[Curvature-improved bound for spherically symmetric Student's $t$ targets]
\label{rmk:student-curvature}
In case~\ref{cor:students-t-moderately-heavy-tailed} of \Cref{cor:student-applications}, suppose additionally that $S_d=I_d$ and that $c_R>0$ is chosen sufficiently small so that $R^2=c_R^2d\le\nu_d$ for all sufficiently large $d$. As discussed in \Cref{sec:shaper_iso}, the assumptions of \Cref{prop:sharper_isoperimetry} are then satisfied. Consequently, under the same warm-start assumptions, the mixing times of the SPS and SCPS chains improve to $n_{\rm mix}(\varepsilon,\bar\pi_0)=\mathcal O(d\log d)$, by \Cref{cor:SPS_mixing_CD} and \Cref{rmk:vanishing-kappa-SCPS}.
\end{remark}

For SCPS with fixed $\nu>1$, taking $\beta=(\nu+2)/(2(\nu-1))+\xi/2$ yields the more explicit bound
\[
n_{\rm mix}(\varepsilon,\bar\pi_0)=\mathcal O\left(d^{3\nu/(\nu-1)+\xi}\right)
\]
for any fixed $\xi>0$. In particular, the result covers $1<\nu\le2$, for which the Student's $t$ distribution has infinite second moments. Although Student's $t$ distributions can be sampled directly and therefore do not require MCMC, they provide a canonical heavy-tailed benchmark for illustrating how the mixing guarantees provided in \Cref{sec:main_results} depend on the tail parameter. The same corollary applies to the Gaussian regression posteriors above whenever the resulting degrees of freedom and scale matrices satisfy its assumptions.

\subsubsection{Exponential-power regression with an unknown scale}
\label{sec:example-quartic}

We next consider a Bayesian linear regression model with exponential-power errors, which allows for tail behavior different from that of Gaussian errors. See~\citep{ferreira2014bayesian} for Bayesian analyses of this class of models. We focus on the smooth quartic case, corresponding to shape parameter $4$. Conditional on the regression coefficients $\theta\in\mathbb R^d$ and an inverse scale parameter $\tau>0$, assume that the observations $Y_1,\ldots,Y_n$ are independent with densities
\[
p(y_i\mid\theta,\tau)=\frac{2\tau^{1/4}}{\Gamma(1/4)}\exp\{-\tau(y_i-d_i^\top\theta)^4\},\qquad i=1,\ldots,n,
\]
where $d_i^\top$ is the $i$th row of a full-column-rank design matrix $D\in\mathbb R^{n\times d}$. We use the flat prior $p_0(\theta)\propto1$ for the regression coefficients $\theta$ and the independent prior $\tau\sim\operatorname{Gamma}(a,b)$, with $a,b>0$.

The joint posterior density of $(\theta,\tau)$, up to a normalizing constant, is
\[
\tau^{a+n/4-1}\exp\{-\tau(b+\|y-D\theta\|_4^4)\},
\]
where $y=(y_1,\ldots,y_n)^\top$, and $\|x\|_4=(\sum_i|x_i|^4)^{1/4}$ for all $x\in\mathbb R^n$. Integrating out the unknown scale parameter $\tau$ therefore yields the marginal posterior
\begin{equation}
\label{eq:quartic-posterior-kappa}
\bar\pi(\theta\mid y)\propto\left[b+\|y-D\theta\|_4^4\right]^{-(a+n/4)}.
\end{equation}
Thus, although the conditional error distribution has lighter tails than a Gaussian distribution, marginalizing over its common scale produces a posterior with polynomial tails.

Observe that the function
\[
V(\theta)=\left[b+\|y-D\theta\|_4^4\right]^{1/4}=\left\|(b^{1/4},y-D\theta)\right\|_4
\]
is positive and convex, and the posterior density can be written as $\bar\pi(\theta\mid y)\propto V(\theta)^{-(4a+n)}$. Hence, the characterization in~\eqref{eq:kappa-concave-density} implies that $\bar\pi(\theta\mid y)$ is $\kappa$-concave with
\[
\kappa=-\frac{1}{4a+n-d}.
\]
Moreover, since $D\in\mathbb R^{n\times d}$ has full column rank, necessarily $n\ge d$, and $V(\theta)\asymp 1+|\theta|$ as $|\theta|\to\infty$. Since $\bar\pi(\theta\mid y)\propto V(\theta)^{-(4a+n)}$, it follows that the posterior density $\bar\pi(\theta\mid y)$ decays as $|\theta|^{-(4a+n)}$ in the tails. Together with $a>0$ and $n\ge d$, this gives $4a+n>d$, so the posterior is integrable on $\mathbb R^d$ and hence proper. The same tail behavior gives, for every $p>0$,
\[
\mathbb E[|\theta|^p]<\infty\qquad\Longleftrightarrow\qquad p<4a+n-d.
\]
Thus, the posterior is positive and smooth, while still allowing genuinely heavy polynomial tails. 

We now specialize to a sequence of models for which the concavity parameter remains fixed as the dimension increases. Specifically, let $n=d+k$, where $k\ge1$ is fixed, and let $a>0$ be fixed. Then
\[
\kappa=-\frac{1}{4a+k}.
\]
Writing
\[
\nu=4a+k,
\]
we have $\kappa=-1/\nu$, with $\nu$ fixed independently of $d$.

Recall that $D\in\mathbb R^{(d+k)\times d}$ has full column rank. Hence, there exist $d$ rows of $D$ that form an invertible $d\times d$ submatrix. After relabeling the observations if necessary, partition the design matrix and response vector as
\[D=\begin{pmatrix}D_0\\D_1\end{pmatrix},\qquad y=\begin{pmatrix}y_0\\y_1\end{pmatrix},
\]
where $D_0\in\mathbb R^{d\times d}$ is invertible and $y_0\in\mathbb R^d$ contains the corresponding responses, while $D_1\in\mathbb R^{k\times d}$ and $y_1\in\mathbb R^k$ correspond to the remaining $k$ observations. Define $\theta_0=D_0^{-1}y_0$. The following corollary gives the resulting mixing-time bound. The corresponding proof is given in \Cref{sec:proofs-examples-exponential-power}.

\begin{corollary}[Quartic regression posteriors]
\label{cor:quartic-applications}
Consider the posterior density~\eqref{eq:quartic-posterior-kappa} with $n=d+k$, where $k\ge1$ is fixed, and let $a,b>0$ be fixed with $\nu=4a+k>1$. Assume that $D_0$ is invertible and that
\begin{align}
    \label{eq:control-norms-quartic-appl}
    \|D_0\|_{\rm op}+\|D_0^{-1}\|_{\rm op}+\|D_1\|_{\rm op}+|y_1-D_1\theta_0|=\mathcal O(1).
\end{align}
After recentering at $\theta_0$, the mixing time of the SCPS chain with $R=c_Rd^\beta$ and $h=c_hd^{-(\beta+1)}$ satisfies
\[
n_{\rm mix}(\varepsilon,\bar\pi_0)=\mathcal O(d^{2\beta+2}),\qquad \beta>\frac{\nu+2}{2(\nu-1)},
\]
from $M$-warm starts with $M=\mathcal O(1)$.
\end{corollary}

The bound~\eqref{eq:control-norms-quartic-appl} ensures the tail and derivative bounds that are needed to apply \Cref{cor:SCPS_mixing_kappa_concave}. As a concrete example, if $k=1$ and $a=1/8$, then $\nu=3/2$ and hence $\kappa=-2/3$. Since moments of order $p$ are finite only for $p<\nu$, the posterior has infinite second moments in this case. Thus, the corollary covers genuinely heavy-tailed targets without a finite covariance matrix, including targets that need not be elliptical.

\subsubsection{Gaussian regression with censored responses}
\label{sec:example-censored}

We finally consider Gaussian linear regression with an unknown variance when some of the responses are censored. Let $y_o\in\mathbb R^{n_o}$ denote observed responses and let $Z\in\mathbb R^{n_c}$ denote censored responses that are known only to lie below specified censoring thresholds $c\in\mathbb R^{n_c}$. Conditional on regression coefficients $\theta\in\mathbb R^d$ and a precision parameter $\tau>0$, assume
\[
Y_o\sim\mathcal N_{n_o}(D_o\theta,\tau^{-1}I_{n_o}),\qquad Z\sim\mathcal N_{n_c}(D_c\theta,\tau^{-1}I_{n_c}),
\]
with $Y_o$ and $Z$ conditionally independent given $(\theta,\tau)$. Here, $D_o\in\mathbb R^{n_o\times d}$ and $D_c\in\mathbb R^{n_c\times d}$ are the corresponding design matrices. The observed censoring event is
\[E=\{Z\le c\},\]
where the inequality is interpreted coordinatewise. The censoring thresholds $c$ are treated as fixed and known.

We adopt the conjugate prior
\[
\theta\mid\tau\sim\mathcal N_d(m_0,\tau^{-1}V_0),\qquad \tau\sim\operatorname{Gamma}(a_0,b_0),
\]
where $m_0\in\mathbb R^d$, $V_0\succ0$, and $a_0,b_0>0$ are fixed prior parameters. 

Conditioning on the observed responses $y_o$ preserves the Normal-Gamma form of the posterior. Define the updated parameters
\[V=(V_0^{-1}+D_o^\top D_o)^{-1},\qquad m=V(V_0^{-1}m_0+D_o^\top y_o),\]
and
\[a=a_0+\frac{n_o}{2},\qquad b=b_0+\frac12\left(|y_o|^2+m_0^\top V_0^{-1}m_0-m^\top V^{-1}m\right).\]
Then,
\[\theta\mid\tau,y_o\sim\mathcal N_d(m,\tau^{-1}V),\qquad \tau\mid y_o\sim\operatorname{Gamma}(a,b).\]

Writing $\nu=2a$, integrating out $\tau$ and then integrating the censored responses over the region ${z\le c}$ yields the marginal posterior of the regression coefficients $\theta$
\begin{equation}
\label{eq:censored-posterior-kappa}
\bar\pi(\theta\mid y_o,E)\propto\int_{{z\le c}}\left[2b+(\theta-m)^\top V^{-1}(\theta-m)+|z-D_c\theta|^2\right]^{-(d+n_c+\nu)/2}\,dz.
\end{equation}

Before showing that $\bar\pi(\theta\mid y_o,E)$ is $\kappa$-concave, we show the concavity of the joint distribution of $(\theta,Z)$. Conditional on $y_o$ and before imposing the censoring event $E=\{Z\le c\}$, the joint density of $(\theta,Z)$ is proportional to
\[\left[2b+(\theta-m)^\top V^{-1}(\theta-m)+|z-D_c\theta|^2\right]^{-(d+n_c+\nu)/2}.\]
Define
\[W(\theta,z)=\left|\left(\sqrt{2b},V^{-1/2}(\theta-m),z-D_c\theta\right)\right|,\]
which is positive and convex. The joint density can then be written as $W(\theta,z)^{-(d+n_c+\nu)}$. The characterization in~\eqref{eq:kappa-concave-density}, applied in dimension $d+n_c$, therefore shows that the distribution of $(\theta,Z)$ is $\kappa$-concave with
\[\kappa=-\tfrac{1}{\nu}.\]

The property of $\kappa$-concavity is preserved, with the same value of $\kappa$, under restriction to a convex set followed by renormalization, as well as under marginalization. Since $\mathbb R^d\times\{z:z\le c\}$ is convex, conditioning on the censoring event and then marginalizing over $z$ shows that~\eqref{eq:censored-posterior-kappa} is $\kappa$-concave with
\[
\kappa=-\frac{1}{\nu}=-\frac{1}{2a_0+n_o}.
\]

Note that the number $n_c$ of censored responses does not enter the concavity parameter. Moreover, the posterior is proper, positive, and smooth on $\mathbb R^d$ and has polynomial tails. Conditioning on the censoring event generally breaks the symmetry of the joint distribution, so the resulting posterior need not be elliptical or symmetric~\citep[Section 3.3]{karling2024conjugacy}.

We now consider a sequence of censored Gaussian regression models indexed by the parameter dimension $d$. We allow the number $n_{o,d}$ of observed responses, as well as the updated posterior parameters and censoring quantities, to vary with $d$. We denote the corresponding quantities by $V_d,m_d,a_d,b_d,D_{c,d}$, and $c_d$. In particular,
\[
\nu_d=2a_d=2a_0+n_{o,d}.
\]
The quantity $m_d$ is the posterior location obtained after conditioning only on the observed responses. After incorporating the censoring event, it need not coincide with the mean of the full coefficient posterior~\eqref{eq:censored-posterior-kappa}. The following corollary gives the corresponding mixing-time bounds. Its proof is available in \Cref{sec:proofs-examples-censored}.

\begin{corollary}[Censored Gaussian regression posteriors]
\label{cor:censored-applications}
Consider the posterior density~\eqref{eq:censored-posterior-kappa} with a fixed number $n_c\ge1$ of censored responses. Let $\nu_d=2a_d$, and assume that there exist constants $v_-,v_+,s_-,s_+>0$, independent of $d$, such that
\[
v_-\le\lambda_{\min}(V_d)\le\lambda_{\max}(V_d)\le v_+,\qquad s_-\le\frac{b_d}{a_d}\le s_+,
\]
and
\[
\|D_{c,d}\|_{\rm op}+|c_d-D_{c,d}m_d|=\mathcal O(1).
\]
After recentering, the following hold.
\begin{enumerate}[label=(\roman*)]
\item If $\nu_d\asymp d$, then the SPS and SCPS chains with $R=c_R\sqrt d$ and $h=c_hd^{-1}$ have mixing time
\[
n_{\rm mix}(\varepsilon,\bar\pi_0)=\mathcal O(d^2\log d)
\]
from $M$-warm starts with $M=\mathcal O(d^q)$ for fixed $q\ge0$.

\item If $\nu_d=\nu>1$ is fixed, then the SCPS chain with $R=c_Rd^\beta$ and $h=c_hd^{-(\beta+1)}$ has mixing time
\[
n_{\rm mix}(\varepsilon,\bar\pi_0)=\mathcal O(d^{2\beta+2}),\qquad \beta>\tfrac{\nu+2}{2(\nu-1)},
\]
from $M$-warm starts with $M=\mathcal O(1)$.
\end{enumerate}
\end{corollary}

The assumptions of \Cref{cor:censored-applications} ensure that the eigenvalues of the updated scale matrix $V_d$ and the scale ratio $b_d/a_d$ remain bounded above and away from zero, while the censored-response design matrix $D_{c,d}$ and the centered censoring thresholds $c_d-D_{c,d}m_d$ remain uniformly bounded. As shown in \Cref{sec:proofs-examples-censored}, these conditions yield the covariance bound in the regime $\nu_d\asymp d$, the required high-probability radius bound when $\nu_d$ is fixed, and the derivative bounds in both regimes.

Finally, since $\nu_d=2a_0+n_{o,d}$, the two mixing regimes can be interpreted directly in terms of the number of observed responses. If $a_0$ is fixed and $n_{o,d}\asymp d$, then $\nu_d\asymp d$, corresponding to the first regime. If instead $n_{o,d}$ remains fixed, then $\nu_d$ is fixed, and the second regime applies whenever $2a_0+n_{o,d}>1$.

\section{Overview of proof techniques}
\label{sec:proof-sketch}

This section provides an overview of the proof techniques used to establish the total variation mixing-time upper bounds for the SPS and SCPS Markov chains stated in \Cref{sec:main_results}. In \Cref{sec:proof-sketch}, we first reduce the problem of obtaining mixing-time upper bounds to that of lower-bounding the $s$-conductance of the corresponding Markov chains on the sphere. In particular, \Cref{prop:lbCs} shows that such a lower bound follows from three ingredients: a three-set isoperimetric inequality, a $(\delta,\rho)$-coupling condition outside an excluded region $\mathcal C$, and a sufficiently small upper bound on $\pi(\mathcal C)$, which converts into a tail bound for the target in Euclidean space.

These ingredients are developed in \Cref{sec:conductance-ingredients}. The isoperimetric inequalities are established in \Cref{sec:isoperimetric_inequalities} and sharpened under additional curvature conditions in \Cref{sec:shaper_iso}. The coupling and tail bounds are developed in \Cref{sec:coupling,sec:tail_bounds}, respectively.

\subsection{Mixing-time upper bound via $s$-conductance lower bound}
\label{sec:proof-sketch-s-cond}

Recall that the stereographic and Sub-Cauchy samplers generate Markov chains on $\mathcal{S}^d$ and $\mathcal{S}^d_o$, respectively, targeting probability distributions $\pi$ obtained as the pushforwards of the Euclidean target $\bar\pi$ under the inverse stereographic and Sub-Cauchy projections. That is,
$$
\pi=({\rm SP}^{-1})_{\#}\bar\pi\in\mathcal{P}(\mathcal{S}^d),
\qquad\text{and}\qquad
\pi=({\rm SCP}^{-1})_{\#}\bar\pi\in\mathcal{P}(\mathcal{S}_o^d),
$$
respectively. We use $\bar T$ to denote the transition kernel of the Euclidean chain and $T$ for the transition kernel of the corresponding transformed chain. If $\bar\pi_0$ is an initial distribution for the Euclidean chain and $\pi_0$ is its pushforward under the corresponding inverse projection, then
$$
\pi_0T^n=F_\#(\bar\pi_0\bar T^n),
$$
where $F\in\{{\rm SP}^{-1},{\rm SCP}^{-1}\}$ denotes the corresponding inverse projection.

Total variation distance is preserved under measurable bijections with measurable inverses: if $F:\mathcal X\to\mathcal Y$ is such a bijection and $\mu,\nu$ are probability measures on $\mathcal X$, then 
\begin{equation}
    \label{eq:TV-preserved-udner-bijection}
    \|F_\#\mu-F_\#\nu\|_{\rm TV}=\|\mu-\nu\|_{\rm TV}.
\end{equation}
Since ${\rm SP}^{-1}$ and ${\rm SCP}^{-1}$ satisfy this property between $\mathbb{R}^d$ and $\mathcal{S}^d\setminus\{2\,e_{d+1}\}$, and between $\mathbb{R}^d$ and $\mathcal{S}_o^d$, respectively, it follows that
$$
\|\bar\pi_0\bar T^n-\bar\pi\|_{\rm TV}=\|\pi_0T^n-\pi\|_{\rm TV}.
$$
Hence, a mixing-time bound established for the transformed chain immediately yields the same bound for its Euclidean counterpart.

Therefore, we first analyse the problem on the sphere in \Cref{sec:subsec-lower-bound-s-conductance}. Later, in \Cref {sec:subsec-transfer-properties}, we convert assumptions required on the target $\pi$ on the sphere into assumptions for the target distribution $\bar\pi$ on $\mathbb{R}^d$.

\subsubsection{Lower-bounding $s$-conductance on the sphere}
\label{sec:subsec-lower-bound-s-conductance}
Throughout this section, we consider a target distribution $\pi$ on a Riemannian manifold $\mathcal{X}$ and a Markov transition kernel $T$ with $\pi$ as its invariant measure. The set $\mathcal{X}$ is to be interpreted as $\mathcal{X}=\mathcal{S}^d$ in the case of SPS, and $\mathcal{X}=\mathcal{S}_o^d$ in the case of SCPS. We assume that $\pi$ admits a density $\pi(u)$ with respect to the Riemannian volume of $\mathcal{X}$.

Here, we establish mixing time upper bounds for the Markov chain associated with $\pi$ and $T$. A common technique consists in relating the mixing time $n_{\rm mix}$ to the $s$-conductance of the Markov chain, defined below.
\begin{definition}[$s$-conductance]
    For $s\in(0,\frac{1}{2})$ and a Markov transition kernel $T(u,\cdot)$ with stationary measure $\pi$, we define the \textit{$s$-conductance} of $T$ as 
    \begin{align}
        \label{eq:conductance}
        C_s(T)\coloneqq\inf_{s<\pi(A)\le\frac{1}{2}}\frac{\int_{A}T(u,A^{\complement})\,\pi(du)}{\pi(A)-s},
    \end{align}
    where the infimum is taken over measurable sets $A$ satisfying $s<\pi(A)\le\frac{1}{2}$.
\end{definition}

The $s$-conductance measures how easily the Markov chain associated with $T$ moves from a set $A$ to its complement $A^{\complement}$. The larger $C_s$ is, the easier it is for the chain to forget its past, and therefore mix faster. Unlike the conductance, which is defined as~\eqref{eq:conductance} by replacing $s$ with $0$, the $s$-conductance allows us to rule out measurable sets of negligible probability (smaller than $s$) where the conductance might be difficult to bound.

We relate the mixing time to the $s$-conductance using \Cref{lem:nonlazy_mixing} in \Cref{sec:technical-results}, which builds on~\cite[Lemma~1]{ascolani2026scalability} and~\citep{lovasz1993random}. Specifically, \Cref{lem:nonlazy_mixing} shows that, for an $M$-warm
initial distribution $\pi_0$ and under the hypotheses of \Cref{prop:lbCs} below,
\begin{align}
\label{eq:lovasz1993-laziness-removed}
\|\pi_0T^n-\pi\|_{\rm TV} \le Ms+M\exp\left(-\frac{C_s(T^2)^2}{2}\left\lfloor\frac n2\right\rfloor\right),\qquad n\ge1.
\end{align}
Here, $C_s(T^2)$ denotes the $s$-conductance of the two-step kernel $T^2$. The same lemma shows that the lower bound in \Cref{prop:lbCs} applies to both $C_s(T)$ and $C_s(T^2)$. Note that, although~\citep{lovasz1993random} and~\cite[Lemma~1]{ascolani2026scalability} present estimates analogous to~\eqref{eq:lovasz1993-laziness-removed}, they assume laziness and positive definiteness of $T$, respectively. \Cref{lem:nonlazy_mixing}, instead, does not depend on such assumptions.

Fix $\varepsilon\in(0,1)$ and choose $s=\frac{\varepsilon^2}{4M^2}\in(0,\frac{1}{2})$. If the hypotheses of \Cref{prop:lbCs} hold for this choice of $s$, then~\eqref{eq:lovasz1993-laziness-removed} gives $\|\pi_0T^n-\pi\|_{\rm TV}\le\frac{\varepsilon}{2}<\varepsilon$
whenever
\[
n\ge \frac{8}{C_s(T^2)^2}\log\left(\frac{4M^2}{\varepsilon^2}\right).
\]
Consequently,
\begin{align}
\label{eq:n_mix-ub_Cs}
n_{\rm mix}(\varepsilon,\pi_0)\le\left\lceil\frac{8}{C_s(T^2)^2}\log\left(\frac{4M^2}{\varepsilon^2}\right)\right\rceil.
\end{align}

Therefore, a lower bound on the $s$-conductance $C_s(T^2)$ yields an upper bound on the mixing time of the chain associated with $T$ and $\pi$.

\Cref{prop:lbCs} below shows how we can determine such a lower bound on the $s$-conductance of $T$ by assuming three main properties of the Markov transition kernel and the target distribution on the sphere: a three-set isoperimetric inequality with Cheeger's constant $\Phi$, a $(\delta,\rho)$-coupling outside a region $\mathcal{C}$, and a bound on the probability of $\mathcal{C}$. These concepts and related results are described in more detail in \Cref{sec:isoperimetric_inequalities,sec:coupling,sec:tail_bounds}, with \Cref{sec:shaper_iso} further improving the isoperimetric inequalities given additional curvature conditions. Note that the same lower bound proved in the proposition below applies to the $s$-conductance $C_s(T^2)$ of the two-step transition kernel, as shown in \Cref{lem:nonlazy_mixing}. This will allow us to upper bound the right-hand side of~\eqref{eq:n_mix-ub_Cs}.

\begin{proposition}[$s$-conductance lower bound]
\label{prop:lbCs}
Let $\mathcal X$ be a $d$-dimensional Riemannian manifold equipped
with its intrinsic Riemannian distance, and let $\pi\in\mathcal P(\mathcal X)$. Let $\mathcal{C}\subset \mathcal{X}$ be a measurable set and $s\in(0,\frac{1}{2})$. Assume that the following hold.
\begin{enumerate}[label=(A\arabic*)]
    \item \label{ass:3set-isoper} $\pi$ satisfies a three-set isoperimetric inequality with Cheeger's constant $\Phi$.
    \item \label{ass:coupling}  The Markov transition kernel $T$ is reversible with respect to $\pi$ and is $(\delta,\rho)$-coupling on $\mathcal X\setminus\mathcal{C}$.
    \item \label{ass:cap-bound} The probability of the set $\mathcal{C}$ is bounded as 
    \begin{align}
\label{bound_pi(C)}
    \pi(\mathcal{C})\le \min\left\{1,\frac{\delta\Phi}{16}\right\}\cdot \frac{s}{4}.
\end{align}
\end{enumerate}

Then, the $s$-conductance of the Markov chain with transition kernel $T$ can be lower bounded as
\begin{align}
\label{eq:LBCs_generic}
    C_s(T)\ge \frac{\rho}{4}\cdot\min\left\{1,\frac{\delta\Phi}{64}\right\}.
\end{align}
\end{proposition}

\begin{proof}
    The proof of \Cref{prop:lbCs} is based on techniques similar to those in~\citep{wu2022minimax} and is available in \Cref{sec:proof-s-conductance-LB} in the Appendix.
\end{proof}

As the set $\mathcal{C}$, we will choose a neighborhood of the north pole, $2e_{d+1}$, in the SPS case, and a neighborhood of the boundary of the bright side, $\{u_{d+1}=l_o\}$, in the SCPS case.

Combining the $s$-conductance lower bound~\eqref{eq:LBCs_generic}, which also holds for $C_s(T^2)$ by \Cref{lem:nonlazy_mixing}, with~\eqref{eq:n_mix-ub_Cs} yields
\begin{align}
\label{eq:lovasz1993+CsLB}
n_{\rm mix}(\varepsilon,\pi_0)
\le
\left\lceil
\frac{128}{\rho^2\cdot
\min\left\{1,\frac{\delta\Phi}{64}\right\}^2}
\log\left(\frac{4M^2}{\varepsilon^2}\right)
\right\rceil.
\end{align}
provided that the hypotheses of \Cref{prop:lbCs} hold with $s=\varepsilon^2/(4M^2)$.

\Cref{sec:isoperimetric_inequalities,sec:coupling,sec:tail_bounds}
establish the isoperimetric, coupling, and tail estimates needed in
the two regimes of \Cref{sec:main_results}. \Cref{sec:shaper_iso} further sharpens the weighted isoperimetric inequalities under radial symmetry and the additional curvature condition used to obtain the improved mixing-time results of \Cref{sec:main_results} (\Cref{cor:SPS_mixing_CD,cor:SCPS_mixing_CD}). The detailed
arguments use SPS in the regime with $-c_1/d\le\kappa\le0$ and SCPS
in the regime with fixed $\kappa\in(-1,0)$.
\Cref{rmk:vanishing-kappa-SCPS} describes how the moderately
heavy-tailed result for SPS extends to SCPS. 

In addition, for radially symmetric targets, we showed that these weighted isoperimetric inequalities can be sharpened under a curvature condition on the radial potential. This is achieved by viewing a weighted Poincar\'e inequality as a standard Poincar\'e inequality in a conformally changed geometry. The curvature condition is then derived from the associated curvature-dimension condition, which allows us to adapt the Buser-type argument used for log-concave targets. The resulting Cheeger constant is of order $\sqrt d$ instead of order one, which may be of independent interest.

\Cref{table:parameter-bounds} summarizes the parameter scalings
and the corresponding bounds for $\Phi$, $\delta$, and $\rho$
derived in \Cref{sec:conductance-ingredients}.

\begin{table}[ht]
\begin{center}

\renewcommand{\arraystretch}{1.5}

\begin{tabular}{@{}c@{\hspace{0cm}}c@{\hspace{0cm}}c@{}}

\begin{tabular}{|c|c|c|c|}
\hhline{~|---|}
\multicolumn{1}{c|}{}
& \cellcolor{tablegray}$R$
& \cellcolor{tablegray}$l_o$
& \cellcolor{tablegray}$h$
\\
\hline

\cellcolor{tablegray} $-c_1/d\le\kappa\le0$
& $\asymp \sqrt{d}$
& \text{const}
& $\asymp d^{-1}$
\\
\hline

\cellcolor{tablegray} $\kappa\in(-1,0)$ fixed
& $\asymp d^{\beta}$
& \text{const}
& $\asymp d^{-(\beta+1)}$
\\
\hline

\cellcolor{tablegray}
\shortstack{
    $-c_1/d\le\kappa\le0$ $+$ curvature
}
& $\asymp\sqrt{d}$
& \text{const}
& $\asymp d^{-1}$
\\
\hline

\cellcolor{tablegray}
\shortstack{
    $\kappa\in(-1,0)$ fixed $+$ curvature
}
& $\asymp d^{\beta}$
& \text{const}
& $\asymp d^{-(\beta+1)}$
\\
\hline

\end{tabular}

&

\raisebox{0.1ex}{
    \scalebox{1.5}{$\Rightarrow$}
}

&

\begin{tabular}{|c|c|c|}
\hline
\cellcolor{tablegray}$\Phi$
& \cellcolor{tablegray}$\delta$
& \cellcolor{tablegray}$\rho$
\\
\hline

$\Omega(1)$
& $\asymp d^{-1}$
& \text{const}
\\
\hline

$\Omega(1)$
& $\asymp d^{-(\beta+1)}$
& \text{const}
\\
\hline

$\Omega(\sqrt{d})$
& $\asymp d^{-1}$
& \text{const}
\\
\hline

$\Omega(\sqrt{d})$
& $\asymp d^{-(\beta+1)}$
& \text{const}
\\
\hline
\end{tabular}

\end{tabular}

\end{center}
\caption{Parameter scalings and corresponding isoperimetric and
coupling bounds under the assumptions of the main results from \Cref{sec:main_results}.
Rows 1 and 3 apply to SPS and SCPS in the moderately heavy-tailed
regime. Rows 2 and 4 apply to SCPS in the heavier-tailed regime.
Rows 3 and 4 additionally assume radial symmetry and the curvature
condition of \Cref{prop:sharper_isoperimetry}.
In rows 2 and 4, $\beta$ satisfies the condition of
\Cref{cor:SCPS_mixing_kappa_concave} and
\Cref{cor:SCPS_mixing_CD}, respectively.
The observer latitude $l_o\in[1,2)$ applies only to SCPS and is
fixed independently of $d$.}
\label{table:parameter-bounds}
\end{table}

\subsubsection{Transferring assumptions from the sphere to the Euclidean space}
\label{sec:subsec-transfer-properties}
\Cref{prop:lbCs} poses its assumptions~\ref{ass:3set-isoper}-\ref{ass:cap-bound} on the transformed target $\pi$ on the Riemannian manifold $\mathcal{X}$, that is, $\mathcal{S}^d$ in the SPS case, and $\mathcal{S}_o^d$ in the SCPS case. We now discuss how these assumptions translate into conditions on the target $\bar \pi$ on the Euclidean space $\mathbb{R}^d$.

First, Assumption~\ref{ass:3set-isoper} is obtained from a weighted isoperimetric inequality for $\kappa$-concave Euclidean targets $\bar\pi$. As shown in \Cref{sec:isoperimetric_inequalities}, the weight functions $w_{\rm SP}$ and $w_{\rm SCP}$ associated with the two projections allow weighted isoperimetric inequalities on $\mathbb{R}^d$ to be transferred to unweighted isoperimetric inequalities for the transformed targets on $\mathcal{S}^d$ and $\mathcal{S}^d_o$, respectively. \Cref{lem:isoperimetryToThreeSet} then yields the three-set inequality required in Assumption~\ref{ass:3set-isoper}.

The $(\delta,\rho)$-coupling Assumption~\ref{ass:coupling} is established directly for the Markov transition kernels on the sphere. As shown in \Cref{sec:coupling}, the coupling bounds are obtained under regularity conditions on the Euclidean target $\bar\pi$, expressed in terms of the gradient and Hessian of its log-density on suitable regions of $\mathbb{R}^d$. These conditions allow us to control the Metropolis-Hastings transition kernels away from the excluded region $\mathcal{C}$ and thereby establish $(\delta,\rho)$-coupling on $\mathcal{X}\setminus\mathcal{C}$.

Finally, Assumption~\ref{ass:cap-bound} is translated into a bound on the tail probability of the target $\bar\pi$ on the Euclidean space, using the relations~\eqref{eq:cap_tail_probability_SP} and~\eqref{eq:cap_tail_probability_SCP}. In fact, under both projections, the region $\mathcal{C}$ corresponds to a tail event of the form $|x|>f(R)$ on the Euclidean space, where $f$ is a function of the scale parameter $R$ of SPS or SCPS.

\subsection{Main ingredients for the $s$-conductance lower bound}
\label{sec:conductance-ingredients}

\subsubsection{Weighted and Unweighted Isoperimetric Inequalities}
\label{sec:isoperimetric_inequalities}

The first main ingredient in our conductance argument is an isoperimetric inequality for the target distribution. Informally, the isoperimetric inequality relates the probability mass of a set to the size of its interface with its complement. A formal definition is provided below.

\begin{definition}[Isoperimetric inequality]
\label{def:isoperimetric_inequality}
We say that a probability measure $\pi$ satisfies an \textit{isoperimetric inequality} with Cheeger constant $\Phi$ if, for every measurable set $A$,
\begin{align*}
    \pi^+(A)\ge \Phi\cdot \pi(A)\cdot(1-\pi(A)).
\end{align*}
Here, $\pi^+(A)$ denotes the Minkowski content (or perimeter) of $A$, defined by
\begin{align}
\label{eq:def:Minkowski_content}
    \pi^+(A)=\liminf_{\epsilon\downarrow0}\frac{\pi(A^\epsilon\setminus A)}{\epsilon},
\end{align}
where $A^\epsilon\coloneqq\{x:\mathsf{dist}(x,A)<\epsilon\}$ denotes the $\epsilon$-fattening of $A$.
\end{definition}

Thus, the Cheeger constant $\Phi$ quantifies how large the interface between a set $A$ and its complement $A^\complement$ must be relative to their probability masses. A larger value of $\Phi$ forces sets of non-zero probability to have a larger interface $\pi^+(A)$ with their complements.

A closely related formulation, which will be used in our conductance argument, is the three-set isoperimetric inequality.

\begin{definition}[Three-set isoperimetric inequality]
\label{def:three_set_isoperimetric_inequality}
We say that a probability measure $\pi$ satisfies a \textit{three-set isoperimetric inequality} with Cheeger constant $\Phi$ if, for every measurable partition $S_1\sqcup S_2\sqcup S_3$ of the state space with $S_1,S_2\neq\varnothing$,
\begin{align}
\label{eq:3set_isoper}
    \pi(S_3)\ge\Phi\cdot\mathsf{dist}(S_1,S_2)\cdot\pi(S_1)\cdot\pi(S_2).
\end{align}
\end{definition}
The isoperimetric inequality in \Cref{def:isoperimetric_inequality} implies the three-set inequality with the same constant $\Phi$, as we prove in \Cref{lem:isoperimetryToThreeSet}.

In our conductance argument, this geometric property is combined with the $(\delta,\rho)$-coupling condition of \Cref{sec:coupling}, which controls the overlap of transition kernels starting from nearby points, and with a bound on the target probability of a small region $\mathcal C$ where the coupling condition is not required to hold. Together, these three ingredients yield the conductance lower bound in \Cref{prop:lbCs} (see \Cref{sec:proof-sketch-s-cond} for details).

We next introduce the weighted version of the isoperimetric inequality.
\begin{definition}[Weighted isoperimetric inequality]
\label{def:weighted_isoperimetric_inequality}
Given a nonnegative measurable function $w:\mathcal{X}\to [0,+\infty)$, define the weighted measure $\pi_w$ by $\pi_w(A)\coloneqq\int_A w\,d\pi$ for every measurable set $A$. We say that $\pi$ satisfies a \textit{weighted isoperimetric inequality} with weight function $w$ and Cheeger constant $\Phi$ if
\begin{align*}
    \pi_w^+(A)\ge \Phi\cdot \pi(A)\cdot(1-\pi(A)),
\end{align*}
for every measurable set $A$, where
$\pi_w^+(A)\coloneqq\liminf_{\epsilon\downarrow0}\pi_w(A^\epsilon\setminus A)/\epsilon$ denotes the Minkowski content of $A$ with respect to $\pi_w$, with $A^\epsilon$ defined as in \Cref{def:isoperimetric_inequality}.
\end{definition}

Using results of \citet{bobkov2009weighted} for
$\kappa$-concave distributions, we derive weighted isoperimetric
inequalities on $\mathbb{R}^d$ adapted to the stereographic and Sub-Cauchy projections.

Let $r$ denote the quantile of order $2/3$ of the Euclidean norm under $\bar\pi$. We first consider the weight function
\begin{align}
\label{eq:def-w_SP}
    w_{\rm SP}(x)\coloneqq\frac{4R^2+|x|^2}{4R},
\end{align}
associated with the stereographic projection. \Cref{lem:weightedIsoperK} below establishes a weighted isoperimetric inequality for $\kappa$-concave distributions with weight function $w_{\rm SP}$.

\begin{lemma}[Weighted isoperimetric inequality for the stereographic projection]
\label{lem:weightedIsoperK}
Let $\bar\pi\in\mathcal P(\mathbb R^d)$ be a $\kappa$-concave distribution with $-1\le\kappa<0$. Then, for every Borel set $A\subseteq\mathbb R^d$,
\begin{align}
\label{eq:isoper_w_finite}
    \bar\pi^+_{w_{\rm SP}}(A)\ge\Phi_{\rm SP}\cdot\bar\pi(A)\cdot(1-\bar\pi(A)),
\end{align}
where $\Phi_{\rm SP}\coloneqq\left(C\left(\frac{r}{R}+1\right)\right)^{-1}$ and $C>0$ is a universal constant.
\end{lemma}
\begin{proof}
    The proof of \Cref{lem:weightedIsoperK} is available in \Cref{sec:proofs-weighted-isoperim}.
\end{proof}
If $r=\mathcal O(\sqrt d)$ and $R\asymp\sqrt d$,
then \Cref{lem:weightedIsoperK} yields $\Phi_{\rm SP}\asymp1$,
providing the isoperimetric estimate used in the proof of
\Cref{cor:SPS_mixing_kappa_concave}.

For the Sub-Cauchy projection, we use the weight function
\begin{align}
\label{eq:def-w_SCP}
    w_{\rm SCP}(x)\coloneqq\frac{R}{L(x)},
\end{align}
with $L(x)$ defined as in~\eqref{def:SCP_D-S-L}, and derive the following analogous lemma.

\begin{lemma}[Weighted isoperimetric inequality for the Sub-Cauchy projection]
\label{lem:weightedIsoperK_SCP}
Let $\bar\pi\in\mathcal P(\mathbb R^d)$ be a $\kappa$-concave distribution with $-1\le\kappa<0$, and let $l_o\in[1,2)$. Then, for every Borel set $A\subseteq\mathbb R^d$,
\begin{align}
\label{eq:isoper_w_SCP}
    \bar\pi^+_{w_{\rm SCP}}(A)\ge\Phi_{\rm SCP}\cdot\bar\pi(A)\cdot(1-\bar\pi(A)),
\end{align}
where $\Phi_{\rm SCP}\coloneqq\left(4C\max\left\{1,\frac{r}{R}\right\}\right)^{-1}$ and $C>0$ is a universal constant.
\end{lemma}
\begin{proof}
    The proof of \Cref{lem:weightedIsoperK_SCP} is available in \Cref{sec:proofs-weighted-isoperim}.
\end{proof}
Note that if $r=\mathcal O(\sqrt d)$ and $R\asymp d^\beta$ with $\beta\ge\frac{1}{2}$, then \Cref{lem:weightedIsoperK_SCP} gives $\Phi_{\rm SCP}\asymp1$.

The preceding results provide weighted isoperimetric inequalities for the Euclidean target $\bar\pi$ with weight functions specifically adapted to the stereographic and Sub-Cauchy projections. These weight functions are chosen so that the corresponding weighted isoperimetric inequalities on $\mathbb{R}^d$ convert into unweighted isoperimetric inequalities for the transformed targets on $\mathcal{S}^d$ under stereographic projection, and on $\mathcal{S}^d_o$ under Sub-Cauchy projection.
 More precisely, suppose that $\Phi>0$, $\int w\,d\bar\pi<\infty$, and that the Euclidean target $\bar\pi$ satisfies the weighted isoperimetric inequality
\begin{align*}
    \bar\pi_w^+(A)
    \ge
    \Phi\cdot\bar\pi(A)\cdot(1-\bar\pi(A))
\end{align*}
for all Borel sets $A\subseteq\mathbb R^d$. Then the corresponding transformed target $\pi$ satisfies
\begin{align*}
    \pi^+(B)\ge\Phi\cdot\pi(B)\cdot(1-\pi(B))
\end{align*}
for every regular set $B$, in the sense of \Cref{def:regular_set}, on $\mathcal{S}^d$ in the case of stereographic projection or on $\mathcal{S}^d_o$ in the case of Sub-Cauchy projection.
This property relies on the geometry of the projection maps and is established for the SPS in \Cref{lem:weightedToUnweighted} and for the SCPS in \Cref{lem:weightedToUnweighted_SCP}.

We have thus established unweighted isoperimetric inequalities on the sphere for SPS and SCPS and, as discussed above, these lead to three set isoperimetric inequalities. These, together with the $(\delta,\rho)$-coupling bounds developed in \Cref{sec:coupling} and the control of the probability of the region $\mathcal{C}$ discussed in \Cref{sec:tail_bounds}, provide the ingredients needed for the conductance lower bound in \Cref{prop:lbCs}.

\subsubsection{Sharper isoperimetry under additional curvature conditions}\label{sec:shaper_iso}

The Cheeger constants obtained in \Cref{lem:weightedIsoperK,lem:weightedIsoperK_SCP} are of order \(1\), which leads to the \(\mathcal{O}(d^2 \log d)\) mixing time bound of \Cref{cor:SPS_mixing_kappa_concave}. A sharper Cheeger constant of order \(\sqrt{d}\) would give \(\mathcal{O}(d \log d)\), matching the dimension dependence of Random Walk Metropolis for well-conditioned log-concave targets \citep{andrieu2022explicit}. In this section, we show that Cheeger constants of order \(\sqrt{d}\) can be obtained for heavy-tailed targets that satisfy an additional curvature condition, analogous to log-concavity. For simplicity, throughout this section, we restrict to the setting where \(\bar{\pi}\) is radially symmetric, with
\begin{equation*}
    \bar{\pi}(dx) \propto \exp(-f(|x|)) dx, \qquad f: [0, \infty) \to \mathbb{R}.
\end{equation*}

Our approach is based on weighted Poincar\'e inequalities. We say that \(\bar{\pi}\) satisfies a weighted Poincar\'e inequality with weight function \(w\) and constant \(C_{\text{PI}}\) if,
\begin{equation}\label{eq:main_wpi}
    \operatorname{Var}_{\bar{\pi}}(\varphi) \leq C_{\text{PI}} \int w(x)^2 |\nabla \varphi(x)|^2 \bar{\pi}(dx),
\end{equation}
for every smooth compactly supported \(\varphi\). It is known that in the unweighted case, if the target is log-concave, a Poincar\'e inequality can be converted into an isoperimetric inequality via a Buser-type inequality \citep{Ledoux2004-ry}. To extend this to the weighted setting, we first reformulate \eqref{eq:main_wpi} as an unweighted Poincar\'{e} inequality on a conformal manifold determined by the weighting. We then utilise the celebrated \emph{curvature-dimension condition} as the natural extension of log-concavity under more general geometry \citep{Bakry2014-ut}. For the weighting \(w(x) = (R^2 + |x|^2)^{1/2}\) and radially symmetric targets, the curvature condition reduces to a condition on the radial potential, producing the following lemma.

\begin{lemma}[From weighted Poincar\'e to weighted isoperimetric inequalities]
\label{lem:wpi_to_cheeger}
Suppose that \(d \geq 2\), \(f\) is twice continuously differentiable and \(\bar{\pi}\) satisfies a weighted Poincar\'e inequality with weight function \(w(x) = (R^2 + |x|^2)^{1/2}\) and constant \(C_{\text{PI}}\). Suppose further that the function
\begin{equation}\label{eq:def_q}
    q(r) = \frac{(R^2 + r^2) f'(r)}{r},
\end{equation}
is non-negative and non-decreasing on \((0, \infty)\). Then, for every measurable set \(A\),
\begin{equation*}
    \bar{\pi}^+_w(A) \geq \frac{1}{6 \sqrt{C_{\text{PI}}}} \bar{\pi}(A) (1 - \bar{\pi}(A)).
\end{equation*}
\end{lemma}
\begin{proof}
    This is \Cref{prop:wp_to_wc2} with \(K = 0\), which is proved in \Cref{sec:cheeger}.
\end{proof}

Since \(w_{\rm SP}\) and \(w_{\rm SCP}\) dominate the weighting in \Cref{lem:wpi_to_cheeger} up to a universal constant factor, this applies to both projections. Thus, a Cheeger constant of order \(\sqrt{d}\) follows from a weighted Poincar\'e inequality with \(C_{\text{PI}} = \mathcal{O}(1/d)\). In \Cref{sec:cheeger}, we give two sufficient conditions for this: a thin-shell condition on \(|X|\) (\Cref{prop:wpi_thin_shell}), and a lower bound on \(q\) (\Cref{prop:wpi_q_lower_bound}). Combining these with \Cref{lem:wpi_to_cheeger}, we obtain the following.

\begin{proposition}[Sharper weighted isoperimetric inequalities]
\label{prop:sharper_isoperimetry}
Suppose that \(f\) is twice continuously differentiable on \([0, \infty)\), and that \(q\) in \eqref{eq:def_q} is non-negative and non-decreasing on \((0, \infty)\). Suppose further that either,
\begin{equation}\label{eq:sharper_isoperimetry_conditions}
    \frac{\operatorname{Var}_{X \sim \bar{\pi}}(|X|)}{R^2} = \mathcal{O}(1/d), \qquad \text{or} \qquad \inf_{r > 0} \big ( q(r) - d \big ) = \Omega(d).
\end{equation}
Then the weighted isoperimetric inequalities \eqref{eq:isoper_w_finite} and \eqref{eq:isoper_w_SCP} hold with \(\Phi_{\rm SP} = \Omega(\sqrt{d})\) and \(\Phi_{\rm SCP} = \Omega(\sqrt{d})\).
\end{proposition}
\begin{proof}
The relevant weighted Poincar\'{e} inequalities are obtained across \Cref{prop:wpi_thin_shell,prop:wpi_q_lower_bound} and are then converted to Cheeger constants via \Cref{prop:wp_to_wc2}.
\end{proof}

When \(R \asymp \sqrt{d}\), the first condition in \eqref{eq:sharper_isoperimetry_conditions} is the thin-shell condition \(\operatorname{Var}(|X|) = \mathcal{O}(1)\).

For example, we can consider the conditions of \eqref{eq:sharper_isoperimetry_conditions} for the Student's \(t\) distribution in \eqref{eq:student-t-density-kappa} with \(m = 0, S = I_d\). Immediately it can be computed that \(q(r) = (d + \nu)(R^2 + r^2)/(\nu + r^2)\) which is non-negative and non-decreasing as soon as \(R^2 \leq \nu\). We also have that \(\inf_{r > 0} q(r) \geq (d + \nu)R^2/\nu\) so the right-hand side of \eqref{eq:sharper_isoperimetry_conditions} is satisfied whenever $\frac{R^2}{\nu}>\frac{d}{d+\nu}$, by a fixed positive margin, as is the case, for example, when $\nu=R^2\asymp d$ in the moderately heavy-tailed regime. 
On the other hand, we can show that the thin shell condition also holds. Writing \(X = (\nu/V)^{1/2} Z\) with \(Z \sim N(0, I_d)\) and \(V \sim \chi^2_\nu\) independent, we have that \(X\) satisfies the relevant student's \(t\) distribution. We can then express the variance as,
\begin{equation*}
    \operatorname{Var}(|X|) = \mathbb{E}[\nu/V] \operatorname{Var}(|Z|) + (\mathbb{E}|Z|)^2 \operatorname{Var}((\nu/V)^{1/2}),
\end{equation*}
where we used that \(V\) and \(|Z|\) are independent. The first term is at most \(\nu/(\nu - 2)\), since \(\mathbb{E}[\nu/V] = \nu/(\nu - 2)\) and \(\operatorname{Var}(|Z|) \leq 1\) by the Gaussian Poincar\'e inequality. For the second term, we have \((\mathbb{E}|Z|)^2 \leq \mathbb{E}|Z|^2 = d\), and since \(|(\nu/V)^{1/2} - 1| \leq |V - \nu|/(\nu V)^{1/2}\), the Cauchy--Schwarz inequality gives,
\begin{equation*}
    \operatorname{Var}((\nu/V)^{1/2}) \leq \mathbb{E}[((\nu/V)^{1/2} - 1)^2] \leq \frac{1}{\nu} \Big ( \mathbb{E}[(V - \nu)^4] \Big )^{1/2} \Big ( \mathbb{E}[V^{-2}] \Big )^{1/2} = \frac{1}{\nu} \bigg ( \frac{12 \nu (\nu + 4)}{(\nu - 2)(\nu - 4)} \bigg )^{1/2},
\end{equation*}
for \(\nu > 4\). Thus, \(\operatorname{Var}(|X|) \leq \nu/(\nu - 2) + \mathcal{O}(d/\nu)\), which is of order one when \(\nu = \Omega(d)\), and so the left-hand side of \eqref{eq:sharper_isoperimetry_conditions} is satisfied when \(R^2 = \Omega(d)\). In either case, with \(R\asymp\sqrt d\) and \(R^2\le\nu\), \Cref{prop:sharper_isoperimetry} gives a Cheeger constant of order \(\sqrt{d}\), and \Cref{cor:SPS_mixing_CD} gives a mixing time of order \(\mathcal{O} (d \log d)\).
\subsubsection{$(\delta,\rho)$-coupling}
\label{sec:coupling}

The second main ingredient in our conductance argument is a local coupling property of the Markov transition kernel.

Informally, we require that transition kernels starting from sufficiently close points are close in total variation. This prevents two nearby points from having one-step transition distributions concentrated almost entirely on opposite sides of a partition of the state space.

We formalize this property through the notion of $(\delta,\rho)$-coupling.

\begin{definition}[$(\delta,\rho)$-coupling]
\label{def:delta-rho-coupling}
Let $(\mathcal X,\mathsf{dist})$ be a metric space and let $T$ be a Markov transition kernel on $\mathcal X$. For $\delta>0$ and $\rho\in(0,1]$, we say that $T$ is \textit{$(\delta,\rho)$-coupling} on a set $\mathcal D\subseteq\mathcal X$ if, for every $u,v\in\mathcal D$ satisfying $\mathsf{dist}(u,v)\le\delta$ we have
\begin{align}
\label{eq:delta-rho-coupling}
\|T(u,\cdot)-T(v,\cdot)\|_{\rm TV}\le 1-\rho.
\end{align}
\end{definition}

Equivalently, by the maximal coupling characterization of the total variation distance~\citep{wang2021maximal}, condition~\eqref{eq:delta-rho-coupling} guarantees that, for every pair $u,v\in\mathcal D$ at distance at most $\delta$, there exists a coupling of one-step transitions $U'\sim T(u,\cdot)$ and $V'\sim T(v,\cdot)$ such that $\mathbb P(U'=V')\ge \rho$. Thus, $\delta$ quantifies the spatial scale over which neighboring transition kernels overlap, while $\rho$ provides a lower bound on the amount of this overlap.

In our setting, the state space is a subset of the sphere $\mathcal{S}^d$, equipped with the geodesic distance $\mathsf{dist}$. To obtain our mixing-time results, it is sufficient to establish $(\delta,\rho)$-coupling for the transition kernels associated with the stereographic and sub-Cauchy samplers on a subset of the state space, excluding regions of sufficiently small probability where this property is difficult to establish. In fact, the geometry of the projection maps degenerates near the north pole, in the case of stereographic projection, and near the boundary of the bright side, in the case of Sub-Cauchy projection. This makes it difficult to obtain uniform bounds on the transition kernels in regions of high latitude. We therefore introduce an excluded region $\mathcal{C}$ containing these areas and establish the coupling property on its complement. That is, we establish that
\begin{align}
\label{eq:coupling-outside-cap}
\|T(u,\cdot)-T(v,\cdot)\|_{\rm TV}\le 1-\rho,\qquad\text{for all }u,v\in\mathcal X\setminus\mathcal{C}\text{ such that }\mathsf{dist}(u,v)\le\delta.
\end{align}
The effect of the excluded region is controlled separately by requiring that its target probability $\pi(\mathcal C)$ is sufficiently small (see assumption~\ref{ass:cap-bound} in \Cref{prop:lbCs}).

The $(\delta,\rho)$-coupling condition is formulated in terms of total variation distance. Since total variation distance is preserved under measurable bijections (see Equation~\eqref{eq:TV-preserved-udner-bijection}), the overlap between the transition kernels, $T(u,\cdot)$ and $T(v,\cdot)$, on the sphere and the corresponding transition kernels on $\mathbb{R}^d$, $\bar T(x,\cdot)$ and $\bar T(y,\cdot)$, remains unchanged under the corresponding transformation. That is,
\begin{align*}
\|T(u,\cdot)-T(v,\cdot)\|_{\rm TV}=\|\bar T(x,\cdot)-\bar T(y,\cdot)\|_{\rm TV},
\end{align*}
where $x={\rm SP}(u), y={\rm SP}(v)$ in the case of stereographic projection, and $x={\rm SCP}(u), y={\rm SCP}(v)$ in the case of Sub-Cauchy projection.
Thus, once the notion of closeness between $u$ and $v$ is translated through the stereographic or sub-Cauchy transformation, the local overlap estimate itself is unchanged. Note, however, that both the distance function and the excluded set $\mathcal C$ change under the bijection. 

We first establish a coupling bound for the random walk proposal kernel $Q$ on the sphere. This result will then be used to derive analogous bounds for the Metropolis-Hastings transition kernels of the SPS and SCPS.

Recall that $Q(u,dv)$ denotes the random walk proposal kernel on the sphere $\mathcal{S}^d$. The SPS proposal kernel coincides with $Q$. For the SCPS, we first sample from $Q$ and, if the proposal falls on the dark side, that is outside the SCPS state space $\mathcal{S}_o^d$, a stepping-out mechanism is applied. We denote the resulting SCPS proposal kernel by $Q_{\rm SCP}$. A first step in the proof of $(\delta,\rho)$-coupling for SPS and SCPS consists in establishing a coupling result for the proposal kernel $Q$. The following proposition gives such a result.

\begin{proposition}[$(\delta,\rho)$-coupling for the proposal kernel $Q(u,\cdot)$]
\label{prop:coupling_uniform}
For $u\in\mathcal{S}^d$, let $Q(u,\cdot)$ denote the random walk proposal kernel on the sphere with step size $h$, and let $F$ denote the cdf of a standard Gaussian random variable. Suppose that $h\le(d+1)^{-\frac{1}{2}}$.

Then, for every $\rho\in(0,1)$, the proposal kernel $Q$ is $(\delta,\rho)$-coupling with $\delta=2\arctan\left(F^{-1}\left(1-\frac{\rho}{2}\right)h\right)$. That is, for all $u,v\in\mathcal{S}^d$,
\begin{align*}
\mathsf{dist}(u,v)\le2\arctan\left(F^{-1}\left(1-\frac{\rho}{2}\right)h\right)\qquad\Rightarrow\qquad\|Q(u,\cdot)-Q(v,\cdot)\|_{\rm TV}\le 1-\rho.
\end{align*}
\end{proposition}

\begin{proof}
The proof is provided in \Cref{sec:proof_coupling}.
\end{proof}

\begin{remark}
\label{rmk:coupling_uniform}
Suppose that $\bar\pi$ is the standard $d$-dimensional Student's $t$ distribution with $\nu=d$ degrees of freedom, and choose $R=\sqrt d/2$. Then the corresponding distribution $\pi$ is the uniform distribution on $\mathcal{S}^d$. In this case, every proposal is accepted, so the SPS transition kernel $T(u,\cdot)$ coincides with the proposal kernel $Q(u,\cdot)$. \Cref{prop:coupling_uniform} therefore implies that, if $h\le(d+1)^{-1/2}$, the SPS kernel is $(\delta,\rho)$-coupling with
\begin{align*}
\delta=2\arctan(0.6744\,h), \qquad \rho=\tfrac{1}{2}.
\end{align*}
\end{remark}

We next extend the above proposal kernel coupling bound to the SPS transition kernel. The additional ingredient is a uniform bound on the total variation distance between the proposal kernel and the Metropolis-Hastings transition kernel away from the spherical cap $\mathcal C$, established in \Cref{lem:couplingTQ_SPS_bound_g'}.

\begin{proposition}[$(\delta,\rho)$-coupling for the SPS transition kernel]
\label{prop:coupling_general}
Assume that $R=c_R\sqrt d$ for a constant $c_R>0$. Suppose there exists a constant $c\in(0,1]$, independent of $d$, such that $\bar\pi$ is twice continuously differentiable and satisfies
\begin{align*}
\sup_{|x|^2\le4\frac{4-c}{c}R^2}|\nabla_x\log\bar\pi(x)|=\mathcal{O}(\sqrt d)\qquad\qquad\text{and}\qquad\qquad\sup_{|x|^2\le4\frac{4-c}{c}R^2}\|\nabla_x^2\log\bar\pi(x)\|_{\rm op}=\mathcal{O}(1).
\end{align*}
Then there exist constants $c_h\in\left(0,\frac{\sqrt2}{2}\right]$ and $d_0\in\mathbb N$ such that, for all $d\ge d_0$, the SPS transition kernel $T_{\rm SP}(u,\cdot)$ with step size $h\le c_h d^{-1}$ is $(\delta,\rho)$-coupling on $\mathcal S^d\setminus\mathcal{C}_{\rm SP}$, where $\mathcal{C}_{\rm SP}$ is defined in~\eqref{eq:def_cap_SP},
\begin{align*}
\delta=2\arctan(0.3184\,h),\qquad \text{and}\qquad \rho=\tfrac{1}{4}.
\end{align*}
That is, for all $u,v\in\mathcal S^d\setminus\mathcal{C}_{\rm SP}$,
\begin{align*}
\mathsf{dist}(u,v)\le2\arctan(0.3184\,h)\qquad\Rightarrow\qquad\|T_{\rm SP}(u,\cdot)-T_{\rm SP}(v,\cdot)\|_{\rm TV}\le\frac{3}{4}.
\end{align*}
\end{proposition}

\begin{proof}
The result is obtained by combining \Cref{prop:coupling_uniform} with \Cref{lem:couplingTQ_SPS_bound_g'}.

Let $\tilde c_h>0$ be the constant in \Cref{lem:couplingTQ_SPS_bound_g'}, and define $c_h
\coloneqq
\min\left\{
\tilde c_h,\frac{\sqrt2}{2}
\right\}$. If $h\le c_h d^{-1}$, then both $h\le(d+1)^{-\frac{1}{2}}$ (required in \Cref{prop:coupling_uniform}) and $h\le\tilde c_h \,d^{-1}$ (required in \Cref{lem:couplingTQ_SPS_bound_g'}) hold.

Applying \Cref{prop:coupling_uniform} with $\rho=\frac{3}{4}$, we obtain
\begin{align}
\label{eq:TVboundQ-Q}
\|Q(u,\cdot)-Q(v,\cdot)\|_{\rm TV}\le\frac{1}{4}
\end{align}
for all $u,v\in\mathcal S^d$ satisfying $\mathsf{dist}(u,v)\le2\arctan(0.3184\,h)$. Moreover, \Cref{lem:couplingTQ_SPS_bound_g'} shows that there exists $\tilde c_h>0$ such that, whenever $h\le\tilde c_h d^{-1}$,
\begin{align}
\label{eq:TVboundT-Q}
\|T_{\rm SP}(u,\cdot)-Q(u,\cdot)\|_{\rm TV}\le\frac{1}{4}
\end{align}
for every $u\in\mathcal S^d\setminus\mathcal{C}_{\rm SP}$.

Following an argument analogous to that in~\citet{dwivedi2019log}, we combine~\eqref{eq:TVboundQ-Q} and~\eqref{eq:TVboundT-Q} using the triangle inequality: for all $u,v\in\mathcal S^d\setminus\mathcal{C}_{\rm SP}$ satisfying $\mathsf{dist}(u,v)\le2\arctan(0.3184\,h)$, we have
\begin{equation*}
\begin{split}
\|T_{\rm SP}(u,\cdot)-T_{\rm SP}(v,\cdot)\|_{\rm TV}&\le\|T_{\rm SP}(u,\cdot)-Q(u,\cdot)\|_{\rm TV}+\|Q(u,\cdot)-Q(v,\cdot)\|_{\rm TV}+\|Q(v,\cdot)-T_{\rm SP}(v,\cdot)\|_{\rm TV}\\
&\le\frac{1}{4}+\frac{1}{4}+\frac{1}{4}=\frac{3}{4}.    
\end{split}
\end{equation*}
This proves the result.
\end{proof}

The same argument, together with the corresponding bound between the SCPS transition kernel and the raw proposal kernel established in \Cref{lem:couplingTQ_SCPS_bound_g'}, yields the coupling result for the SCPS.

\begin{proposition}[$(\delta,\rho)$-coupling for the SCPS transition kernel]
\label{prop:coupling_general_SCP}
Assume that $l_o\in[1,2)$ is independent of $d$, and let $R=c_R\cdot  d^\beta$ for constants $\beta\ge0$ and $c_R>0$. Suppose there exists a constant $c\in(0,l_o)$, independent of $d$, such that $\bar\pi$ is twice continuously differentiable and satisfies
\begin{align*}
\sup_{|x|^2\le c_1d^{2\beta}}|\nabla_x\log\bar\pi(x)|=\mathcal{O}(d)\qquad\qquad\text{and}\qquad\qquad\sup_{|x|^2\le c_1d^{2\beta}}\|\nabla_x^2\log\bar\pi(x)\|_{\rm op}=\mathcal{O}(d),
\end{align*}
where $c_1\coloneqq\frac{4\cdot l_o^2\cdot c_R^2}{c^2}$.

Then there exist constants $c_h\in\left(0,\frac{\sqrt2}{2}\right]$ and $d_0\in\mathbb N$ such that, for all $d\ge d_0$, the SCPS transition kernel $T_{\rm SCP}(u,\cdot)$ with step size $h\le c_h d^{-(\beta+1)}$ is $(\delta,\rho)$-coupling on $\mathcal S_o^d\setminus\mathcal{C}_{\rm SCP}$, where $\mathcal{C}_{\rm SCP}$ is defined in~\eqref{eq:def_cap_SCP},
\begin{align*}
\delta=2\arctan(0.3184\,h),\qquad\text{and}\qquad\rho=\tfrac{1}{4}.
\end{align*}
That is, for all $u,v\in\mathcal S_o^d\setminus\mathcal{C}_{\rm SCP}$,
\begin{align*}
\mathsf{dist}(u,v)\le2\arctan(0.3184\,h)\qquad\Rightarrow\qquad
\|T_{\rm SCP}(u,\cdot)-T_{\rm SCP}(v,\cdot)\|_{\rm TV}\le\frac34.
\end{align*}
\end{proposition}

\begin{proof}

The proof follows the same argument as for \Cref{prop:coupling_general}. Let $\tilde c_h>0$ be the constant in \Cref{lem:couplingTQ_SCPS_bound_g'}, and define $c_h\coloneqq\min\left\{\tilde c_h,\frac{\sqrt2}{2}\right\}$. For all sufficiently large $d$, if $h\le c_h d^{-(\beta+1)}$, then both $h\le(d+1)^{-\frac{1}{2}}$ and $h\le\tilde c_h d^{-(\beta+1)}$ hold. Let $u,v\in\mathcal S_o^d\setminus\mathcal{C}_{\rm SCP}$ satisfy $\mathsf{dist}(u,v)\le2\arctan(0.3184\,h)$. By \Cref{prop:coupling_uniform}, $\|Q(u,\cdot)-Q(v,\cdot)\|_{\rm TV}\le\frac{1}{4}$. Moreover, \Cref{lem:couplingTQ_SCPS_bound_g'} gives $\|T_{\rm SCP}(u,\cdot)-Q(u,\cdot)\|_{\rm TV}\le\frac{1}{4}$ for all $u\in\mathcal S_o^d\setminus\mathcal{C}_{\rm SCP}$. Here, $T_{\rm SCP}(u,\cdot)$ is interpreted as the trivial extension of the SCPS transition kernel to the whole sphere $\mathcal{S}^d$: that is, $T_{\rm SCP}(u,A)=T_{\rm SCP}(u,A\cap\mathcal{S}^d_o)$ for all measurable $A$ and $u\in\mathcal{S}^d_o$, while for $u\in\mathcal{S}^d\setminus\mathcal{S}^d_o$, $T_{\rm SCP}(u,\cdot)$ is arbitrarily defined.
The result follows by applying the triangle inequality as in the end of the proof of \Cref{prop:coupling_general}.
\end{proof}

\subsubsection{Tail bounds}
\label{sec:tail_bounds}

\Cref{prop:lbCs} in \Cref{sec:proof-sketch-s-cond} requires the $(\delta,\rho)$-coupling condition to hold on a subset $\mathcal{X}\setminus\mathcal{C}$ of the state space $\mathcal{X}$ (see Assumption~\ref{ass:coupling}), provided that the probability of the set $\mathcal{C}$ satisfies  $\pi(\mathcal{C})\le\min\left\{1,\frac{\delta\Phi}{16}\right\}\cdot\frac{s}{4}$ (see Assumption~\ref{ass:cap-bound}). For the two samplers, we defined $\mathcal{C}$ to be $\mathcal{C}_{\rm SP}$ as in~\eqref{eq:def_cap_SP} for SPS, and $\mathcal{C}_{\rm SCP}$ as in~\eqref{eq:def_cap_SCP} for SCPS. 

From Equations~\eqref{eq:cap_tail_probability_SP} and~\eqref{eq:cap_tail_probability_SCP}, we see that bounding the probability $\pi(\mathcal{C})$ of these sets on the sphere is equivalent to determining tail bounds of the corresponding probability distribution $\bar\pi$ on the Euclidean space. Establishing these tail bounds guarantees that the region where the $(\delta,\rho)$-coupling is not available carries sufficiently little probability mass.

In this section, we derive tail bounds for the two $\kappa$-concave regimes considered in \Cref{sec:main_results}, that is, $-c_1/d\le\kappa\le0$ and fixed $\kappa\in(-1,0)$.

\begin{lemma}[Tail bound for $\kappa$-concave distributions with moderately heavy tails]
\label{lem:tail_bound_kappa_concave_light_tail}
Let $X\in\mathbb{R}^d$ be a centered $\kappa$-concave random vector with $-\frac{c_1}{d}\le\kappa\le0$ for a constant $c_1>0$. Assume that its covariance matrix $\Sigma_d$ satisfies
\begin{align*}0<\lambda_L \le\lambda_{\min}(\Sigma_d)\le\lambda_{\max}(\Sigma_d)\le\lambda_U <\infty,\end{align*}
for constants $\lambda_L ,\lambda_U $ independent of $d$. Then, there exist constants $\ell>1$ and $c_2>0$, independent of $d$, such that, for all sufficiently large $d$,
\begin{align}\label{eq:exponential_tail_bound}\mathbb{P}\big(|X|>\ell\sqrt d\big)\le e^{-c_2\sqrt d}.\end{align}
\end{lemma}

\begin{proof}
    The proof of \Cref{lem:tail_bound_kappa_concave_light_tail} is provided in \Cref{sec:proofs-tail-bounds}.
\end{proof}

In the regime $-c_1/d\le\kappa\le0$, \Cref{cor:SPS_mixing_kappa_concave} uses $R=c_R\sqrt d$ for SPS, and \Cref{rmk:vanishing-kappa-SCPS} extends this result to SCPS with the same scaling. We now verify the required excluded-region probability bound for both samplers. By~\eqref{eq:cap_tail_probability_SP} and~\eqref{eq:cap_tail_probability_SCP},
\[
\pi(\mathcal C_{\rm SP})=\mathbb{P}\Big(|X|>2R\sqrt{\tfrac{2-c}{c}}\Big),
\qquad
\pi(\mathcal C_{\rm SCP})=\mathbb{P}\Big(|X|>R\sqrt{\bar c(c,l_o)}\Big).
\]
Since $R=c_R\sqrt d$ and both $2\sqrt{(2-c)/c}$ and $\sqrt{\bar c(c,l_o)}$ become arbitrarily large as $c\downarrow0$, we may choose a sufficiently small fixed $c>0$ such that both Euclidean thresholds are at least $\ell\sqrt d$, where $\ell$ is the constant in \Cref{lem:tail_bound_kappa_concave_light_tail}. Consequently,
\[
\pi(\mathcal C_{\rm SP})\le e^{-c_2\sqrt d},
\qquad
\pi(\mathcal C_{\rm SCP})\le e^{-c_2\sqrt d}
\]
for all sufficiently large $d$. Since $e^{-c_2\sqrt d}=o(d^{-p})$ for every fixed $p>0$, these bounds imply the excluded-region condition~\eqref{bound_pi(C)} for all sufficiently large $d$, under the parameter scalings and warm-start assumptions of the moderately heavy-tailed results.

For heavier-tailed targets, where $\kappa\in(-1,0)$ is fixed independently of the dimension, we instead use the polynomial tail estimate of \citet{bobkov2007large}.

\begin{lemma}[Tail bound for $\kappa$-concave distributions with heavier tails]
\label{lem:tail_bound_kappa_concave_kappa_const}
Let $X$ be a $\kappa$-concave random vector in $\mathbb{R}^d$ with fixed $\kappa\in(-1,0)$, and let $m$ denote the median of $|X|$. Then, for every $\ell\ge1$,
\begin{align}
\label{eq:tail_bound_kappa_concave_general}
\mathbb{P}(|X|>m\cdot\ell)\le C_\kappa\cdot\left(\tfrac{\ell}{2}\right)^{-\frac{1}{|\kappa|}},
\end{align}
where $C_\kappa>0$ depends only on $\kappa$.
\end{lemma}

\begin{proof}
    The proof of \Cref{lem:tail_bound_kappa_concave_kappa_const} is provided in \Cref{sec:proofs-tail-bounds}.
\end{proof}

If $m=\mathcal{O}(\sqrt d)$ and $R\asymp d^\beta$ with $\beta>\frac{1}{2}$, then for every fixed cap parameter the corresponding Euclidean threshold is of order $d^\beta$, and \Cref{lem:tail_bound_kappa_concave_kappa_const} yields
\begin{align}\label{eq:tail_bound_kappa_concave_Rbeta}\mathbb{P}\left(|X|>C \cdot R\right)=\mathcal{O}\Big(d^{-\frac{\beta-\frac12}{|\kappa|}}\Big),\end{align}
for every fixed constant $C>0$. Thus, in the heavier-tailed regime, increasing the SCPS projection scale $R$ moves the image of the excluded region $\mathcal C_{\rm SCP}$ under SCP farther into the Euclidean tails and makes its probability polynomially small. This provides the tail estimate used in \Cref{cor:SCPS_mixing_kappa_concave}.

Together with the weighted isoperimetric inequalities of \Cref{sec:isoperimetric_inequalities} and the $(\delta,\rho)$-coupling bounds of \Cref{sec:coupling}, these tail estimates provide the third ingredient required by \Cref{prop:lbCs}. 

\section{Discussion and Future Directions}
\label{sec:discussion}
In this paper, we established mixing-time upper bounds for SPS and SCPS targeting heavy-tailed distributions. More specifically, we studied $\kappa$-concave distributions with negative values of $\kappa$. The results also extend to log-concave distributions. In fact, this light-tailed class satisfies the $\kappa$-concavity condition (Equation~\eqref{eq:kappa-concavity-definition}) for any $\kappa\le0$.

The main results distinguish two regimes of the concavity parameter: $-c_1/d\le\kappa\le0$ and fixed $\kappa\in(-1,0)$. The first yields mixing-time guarantees for both SPS and SCPS from polynomially warm initial distributions, while the second yields guarantees for SCPS from initial distributions with uniformly bounded warmness. The assumptions in the second regime allow heavier-tailed targets, including distributions with infinite second moments.

A main contribution of this paper lies in the proof techniques. Although the $s$-conductance argument that we used to bound mixing times is common to several recent works~\citep{dwivedi2019log,chen2020fast,wu2022minimax,andrieu2022explicit}, a substantial part of our proof consisted of deriving the requires isoperimetric inequalities on the sphere under inverse stereographic and Sub-Cauchy projection. In particular, we established that, in order to obtain (unweighted) isoperimetric inequalities on the sphere, it suffices for the Euclidean target to satisfy weighted isoperimetric inequalities on $\mathbb{R}^d$ with specific weight functions that we determined for the two projections.
For radially symmetric targets satisfying an additional curvature condition, we further obtain the stronger bound $\Phi=\Omega(\sqrt d)$, leading to the improved mixing-time bounds in \Cref{cor:SPS_mixing_CD,cor:SCPS_mixing_CD}. This refinement follows by interpreting a weighted Poincar\'e inequality as an unweighted one in a conformally changed geometry and applying a Buser-type argument under the associated curvature-dimension condition. We believe that the dimension dependence of this stronger isoperimetric bound is optimal under these additional assumptions.

Another important novel contribution concerns the coupling conditions: we controlled the probability of the excluded region on which the coupling condition is not established, and we related it to a control on tail probabilities of the target on Euclidean space. A main difficulty in establishing the coupling condition was accounting for the distortion induced by the projection, particularly near the regions corresponding to the Euclidean tails.

We conjecture that the dimension dependence of the SCPS mixing time can be improved to $\mathcal{O}(d^2)$, at least for multivariate Student’s $t$ distributions with degrees of freedom fixed at $\nu>2$. We believe that the weighted isoperimetric inequality developed here has optimal dimension dependence under the additional curvature condition, yielding a mixing-time bound of $\mathcal{O}(d^{2\beta+1})$. Our current $s$-conductance analysis, however, imposes a restriction on $\beta$ that requires $\beta>\frac{1}{2}$, as in \Cref{cor:SCPS_mixing_kappa_concave}. We believe this restriction is suboptimal and conjecture that the natural projection scale $R\asymp\sqrt{d}$, corresponding to $\beta=\frac{1}{2}$, is optimal. Extending the analysis to this scaling would yield the conjectured $\mathcal{O}(d^2)$ mixing-time bound. Establishing this improvement remains an important direction for future work.

Another important direction is to relax the warm-start assumptions in our main results. Constructing a suitable initialization with a controlled computational cost is itself a challenging problem. Recent advances in algorithmic warm starts for MALA and HMC demonstrate how initialization can be incorporated into quantitative sampling guarantees; see \citep{altschuler2024faster,zhang2026algorithmic}. Developing analogous guarantees for projection samplers targeting heavy-tailed distributions remains an open problem.

More broadly, our analysis provides a basis for studying the non-asymptotic convergence of other stereographic MCMC methods, including stereographic multiple-try Metropolis \citep{wang2025stereographic} and stereographic slice sampler \citep{bell2024adaptive}. The weighted isoperimetric inequalities developed here describe the geometry of the transformed target and can therefore serve as a common ingredient in these analyses. A central challenge is to obtain suitable coupling or local-overlap estimates for each sampler and combine them with an appropriate mixing-time argument. Extending this approach to nonreversible and adaptive stereographic algorithms is a promising direction for future work.

\section*{Acknowledgments}FM gratefully acknowledges travel support from the Danish Data Science Academy (Grant ID 2025-4796). FM's research is supported by the Knut and Alice Wallenberg Foundation under grant KAW 2024.0327. TF's research is supported by the European Research Council Starting Grant DYNASTY (101039676). JY’s research is supported by the Independent Research Fund Denmark (DFF) through the Sapere Aude Starting Grant (5251-00032B). 
The authors acknowledge the use of ChatGPT 5.6 Pro solely as a reference search aid for Section~\ref{sec:applications-results} and for post-drafting grammatical polishing; all selection, synthesis, and mathematical results remain strictly human work.

\bibliographystyle{plainnat}
\bibliography{sample}

@article{yang2022stereographic,
  title={Stereographic {M}arkov Chain {M}onte {C}arlo},
  author={Yang, Jun and {\L}atuszy{\'n}ski, Krzysztof and Roberts, Gareth O},
  journal={The Annals of Statistics},
  volume={52},
  number={6},
  pages={2692--2713},
  year={2024},
  publisher={Institute of Mathematical Statistics}
}

@article{wu2022minimax,
  title={Minimax mixing time of the Metropolis-adjusted Langevin algorithm for log-concave sampling},
  author={Wu, Keru and Schmidler, Scott and Chen, Yuansi},
  journal={The Journal of Machine Learning Research},
  volume={23},
  number={1},
  pages={12348--12410},
  year={2022},
  publisher={JMLRORG}
}

@article{andrieu2022explicit,
  title={Explicit convergence bounds for Metropolis Markov chains: isoperimetry, spectral gaps and profiles},
  author={Andrieu, Christophe and Lee, Anthony and Power, Sam and Wang, Andi Q},
  journal={arXiv preprint arXiv:2211.08959},
  year={2022}
}

@article{bobkov2009weighted,
  title={WEIGHTED POINCAR{\'E}-TYPE INEQUALITIES FOR CAUCHY AND OTHER CONVEX MEASURES1},
  author={Bobkov, Sergey G and Ledoux, Michel},
  journal={The Annals of Probability},
  volume={37},
  number={2},
  pages={403--427},
  year={2009},
  publisher={Citeseer}
}

@inproceedings{bobkov2009weighted2,
  title={On weighted isoperimetric and Poincar{\'e}-type inequalities},
  author={Bobkov, Sergey G and Ledoux, Michel},
  booktitle={High dimensional probability V: the Luminy volume},
  pages={1--29},
  year={2009}
}

@book{bobkov1997some,
  title={Some connections between isoperimetric and Sobolev-type inequalities},
  author={Bobkov, Serguei Germanovich and Houdr{\'e}, Christian},
  volume={129},
  year={1997},
  publisher={American Mathematical Soc.}
}

@article{dwivedi2019log,
  title={Log-concave sampling: Metropolis-Hastings algorithms are fast},
  author={Dwivedi, Raaz and Chen, Yuansi and Wainwright, Martin J and Yu, Bin},
  journal={Journal of Machine Learning Research},
  volume={20},
  pages={1--42},
  year={2019}
}

@article{lovasz1993random,
  title={Random walks in a convex body and an improved volume algorithm},
  author={Lov{\'a}sz, L{\'a}szl{\'o} and Simonovits, Mikl{\'o}s},
  journal={Random structures \& algorithms},
  volume={4},
  number={4},
  pages={359--412},
  year={1993},
  publisher={Wiley Online Library}
}

@book{lee2018introduction,
  title={Introduction to Riemannian manifolds},
  author={Lee, John M},
  volume={2},
  year={2018},
  publisher={Springer}
}

@article{chen2020fast,
  title={Fast mixing of Metropolized Hamiltonian Monte Carlo: Benefits of multi-step gradients},
  author={Chen, Yuansi and Dwivedi, Raaz and Wainwright, Martin J and Yu, Bin},
  journal={Journal of Machine Learning Research},
  volume={21},
  number={92},
  pages={1--72},
  year={2020}
}

@article{wang2025stereographic,
  title={Stereographic Multiple-Try Metropolis},
  author={Wang, Zhihao and Yang, Jun},
  journal={arXiv preprint arXiv:2505.12487},
  year={2025}
}

@article{zhang2026algorithmic,
  title={Algorithmic warm starts for Hamiltonian Monte Carlo},
  author={Zhang, Matthew S and Altschuler, Jason M and Chewi, Sinho},
  journal={arXiv preprint arXiv:2603.22741},
  year={2026}
}

@article{altschuler2024faster,
  title={Faster high-accuracy log-concave sampling via algorithmic warm starts},
  author={Altschuler, Jason M and Chewi, Sinho},
  journal={Journal of the ACM},
  volume={71},
  number={3},
  pages={1--55},
  year={2024},
  publisher={ACM New York, NY}
}

@article{borell1974convex,
  title={Convex measures on locally convex spaces},
  author={Borell, Christer},
  journal={Arkiv f{\"o}r matematik},
  volume={12},
  number={1},
  pages={239--252},
  year={1974},
  publisher={Springer}
}

@article{borell1975convex,
  title={Convex set functions in d-space},
  author={Borell, Christer},
  journal={Periodica Mathematica Hungarica},
  volume={6},
  number={2},
  pages={111--136},
  year={1975},
  publisher={Springer}
}

@article{prekopa1973logarithmic,
  title={On logarithmic concave measures and functions},
  author={Pr{\'e}kopa, Andr{\'a}s},
  journal={Acta Sci. Math.},
  volume={34},
  pages={335},
  year={1973}
}

@book{robert2004monte,
  title={Monte Carlo statistical methods},
  author={Robert, Christian P and Casella, George and Casella, George},
  volume={2},
  year={2004},
  publisher={Springer}
}

@book{brooks2011handbook,
  title={Handbook of markov chain monte carlo},
  author={Brooks, Steve and Gelman, Andrew and Jones, Galin and Meng, Xiao-Li},
  year={2011},
  publisher={CRC press}
}

@article{geweke1993bayesian,
  title={Bayesian treatment of the independent Student-t linear model},
  author={Geweke, John},
  journal={Journal of applied econometrics},
  volume={8},
  number={S1},
  pages={S19--S40},
  year={1993},
  publisher={Wiley Online Library}
}

@article{stuart2010inverse,
  title={Inverse problems: a Bayesian perspective},
  author={Stuart, Andrew M},
  journal={Acta numerica},
  volume={19},
  pages={451--559},
  year={2010},
  publisher={Cambridge University Press}
}

@article{bobkov2007large,
  title={Large deviations and isoperimetry over convex probability measures with heavy tails},
  author={Bobkov, Sergey},
  year={2007}
}

@article{grazzi2026sub,
  title={Sub-Cauchy Sampling: Escaping the Dark Side of the Moon},
  author={Grazzi, Sebastiano and Liu, Sifan and Roberts, Gareth O and Yang, Jun},
  journal={arXiv preprint arXiv:2601.11066},
  year={2026}
}

@article{adamczak2012moment,
  title={Moment estimates for convex measures},
  author={Adamczak, Rados{\l}aw and Gu{\'e}don, Olivier and Lata{\l}a, Rafa{\l} and Litvak, Alexander and Oleszkiewicz, Krzysztof and Pajor, Alain and Tomczak-Jaegermann, Nicole},
  year={2012}
}

@ARTICLE{Ledoux2004-ry,
  title     = "Spectral gap, logarithmic Sobolev constant, and geometric bounds",
  author    = "Ledoux, Michel",
  journal   = "Surveys in differential geometry",
  publisher = "International Press of Boston",
  volume    =  9,
  number    =  1,
  pages     = "219--240",
  year      =  2004
}

@article{jarner2000geometric,
  title={Geometric ergodicity of Metropolis algorithms},
  author={Jarner, S{\o}ren Fiig and Hansen, Ernst},
  journal={Stochastic processes and their applications},
  volume={85},
  number={2},
  pages={341--361},
  year={2000},
  publisher={Elsevier}
}

@article{bell2024adaptive,
  title={Adaptive stereographic {MCMC}},
  author={Bell, Cameron and {\L}atuszy{\'n}ski, Krzysztof and Roberts, Gareth},
  journal={arXiv preprint arXiv:2408.11780},
  year={2024}
}

@article{seregin2010nonparametric,
  title={Nonparametric estimation of multivariate convex-transformed densities},
  author={Seregin, Arseni and Wellner, Jon A},
  journal={Annals of statistics},
  volume={38},
  number={6},
  pages={3751},
  year={2010}
}

@article{christen2024bayesian,
  title={Bayesian Linear Models: A compact general set of results},
  author={Christen, J Andres},
  journal={arXiv preprint arXiv:2406.01819},
  year={2024}
}

@inproceedings{wang2021maximal,
  title={Maximal couplings of the Metropolis-Hastings algorithm},
  author={Wang, Guanyang and O’Leary, John and Jacob, Pierre},
  booktitle={International Conference on Artificial Intelligence and Statistics},
  pages={1225--1233},
  year={2021},
  organization={PMLR}
}

@article{brascamp1976extensions,
  title={On extensions of the Brunn-Minkowski and Pr{\'e}kopa-Leindler theorems, including inequalities for log concave functions, and with an application to the diffusion equation},
  author={Brascamp, Herm Jan and Lieb, Elliott H},
  journal={Journal of functional analysis},
  volume={22},
  number={4},
  pages={366--389},
  year={1976},
  publisher={Elsevier}
}

@inproceedings{leindler1972certain,
  title={On a certain converse of H{\"o}lder’s inequality},
  author={Leindler, L{\'a}szl{\'o}},
  booktitle={Linear Operators and Approximation/Lineare Operatoren und Approximation: Proceedings of the Conference held at the Oberwolfach Mathematical Research Institute, Black Forest, August 14--22, 1971/Abhandlungen zur Tagung im Mathematischen Forschungsinstitut Oberwolfach, Schwarzwald, vom 14. bis 22. August 1971},
  pages={182--184},
  year={1972},
  organization={Springer}
}

@book{coles2001introduction,
  title={An introduction to statistical modeling of extreme values},
  author={Coles, Stuart and Bawa, Joanna and Trenner, Lesley and Dorazio, Pat},
  volume={208},
  year={2001},
  publisher={Springer}
}

@article{gelman2008weakly,
  title={A weakly informative default prior distribution for logistic and other regression models},
  author={Gelman, Andrew and Jakulin, Aleks and Pittau, Maria Grazia and Su, Yu-Sung},
  journal = {The Annals of Applied Statistics},
  volume  = {2},
  number  = {4},
  pages   = {1360--1383},
  year    = {2008},
  doi     = {10.1214/08-AOAS191}
}

@book{asmussen2007stochastic,
  title={Stochastic simulation: algorithms and analysis},
  author={Asmussen, S{\o}ren and Glynn, Peter W},
  volume={57},
  year={2007},
  publisher={Springer}
}

@article{andrieu2003introduction,
  title={An introduction to MCMC for machine learning},
  author={Andrieu, Christophe and De Freitas, Nando and Doucet, Arnaud and Jordan, Michael I},
  journal={Machine learning},
  volume={50},
  number={1},
  pages={5--43},
  year={2003},
  publisher={Springer}
}

@article{metropolis1953equation,
  title={Equation of state calculations by fast computing machines},
  author={Metropolis, Nicholas and Rosenbluth, Arianna W and Rosenbluth, Marshall N and Teller, Augusta H and Teller, Edward},
  journal={The journal of chemical physics},
  volume={21},
  number={6},
  pages={1087--1092},
  year={1953},
  publisher={American Institute of Physics}
}

@article{hastings1970monte,
  title={Monte Carlo sampling methods using Markov chains and their applications},
  author={Hastings, W Keith},
  year={1970},
  publisher={Oxford University Press}
}

@article{ascolani2026scalability,
  title={Scalability of Metropolis-within-Gibbs schemes for high-dimensional Bayesian models},
  author={Ascolani, Filippo and Roberts, Gareth O and Zanella, Giacomo},
  journal={Journal of the Royal Statistical Society Series B: Statistical Methodology},
  pages={qkaf084},
  year={2026},
  publisher={Oxford University Press UK}
}

@article{tiao1964bayesian,
  author = {Tiao, George C. and Zellner, Arnold},
  title = {On the Bayesian Estimation of Multivariate Regression},
  journal = {Journal of the Royal Statistical Society: Series B (Methodological)},
  volume = {26},
  number = {2},
  pages = {277--285},
  year = {1964}
}

@article{tiao1964bayes,
  title={Bayes's theorem and the use of prior knowledge in regression analysis},
  author={Tiao, George C and Zellner, Arnold},
  journal={Biometrika},
  volume={51},
  number={1/2},
  pages={219--230},
  year={1964},
  publisher={JSTOR}
}

@article{ferreira2014bayesian,
  title={Bayesian reference analysis for exponential power regression models},
  author={Ferreira, Marco AR and Salazar, Esther},
  journal={Journal of Statistical Distributions and Applications},
  volume={1},
  number={1},
  pages={12},
  year={2014},
  publisher={Springer}
}

@article{karling2024conjugacy,
  title={Conjugacy properties of multivariate unified skew-elliptical distributions},
  author={Karling, Maicon J and Durante, Daniele and Genton, Marc G},
  journal={Journal of Multivariate Analysis},
  volume={204},
  pages={105357},
  year={2024},
  publisher={Elsevier}
}

@article{kramer2023uqsa,
  title={UQSA--An R-Package for Uncertainty Quantification and Sensitivity Analysis for Biochemical Reaction Network Models},
  author={Kramer, Andrei and Milinanni, Federica and Kotaleski, Jeanette Hellgren and Nyquist, Pierre and Jauhiainen, Alexandra and Eriksson, Olivia},
  journal={arXiv preprint arXiv:2308.05527},
  year={2023}
}

@ARTICLE{Bonnefont2016-if,
  title    = "Spectral gap for spherically symmetric log-concave probability
              measures, and beyond",
  author   = "Bonnefont, Michel and Joulin, Aldéric and Ma, Yutao",
  journal  = "Journal of Functional Analysis",
  volume   =  270,
  number   =  7,
  pages    = "2456--2482",
  year     =  2016
}

@ARTICLE{Chen1997-it,
  title     = "Estimation of Spectral Gap for Elliptic Operators",
  author    = "Chen, Mu-Fa and Wang, Feng-Yu",
  journal   = "Transactions of the American Mathematical Society",
  publisher = "American Mathematical Society",
  volume    =  349,
  number    =  3,
  pages     = "1239--1267",
  year      =  1997
}

@BOOK{Bakry2014-ut,
  title     = "Analysis and Geometry of Markov Diffusion Operators",
  author    = "Bakry, Dominique and Gentil, Ivan and Ledoux, Michel",
  publisher = "Springer International Publishing",
  year      =  2014
}

@ARTICLE{Bobkov1999-vu,
  title     = "Isoperimetric and analytic inequalities for log-concave
               probability measures",
  author    = "Bobkov, S G",
  journal   = "Ann. Probab.",
  publisher = "Institute of Mathematical Statistics",
  volume    =  27,
  number    =  4,
  pages     = "1903--1921",
  month     =  oct,
  year      =  1999
}

@BOOK{Chewi2023-hc,
  title  = "Log-concave sampling",
  author = "Chewi, Sinho",
  year   =  2023
}

\clearpage
\appendix
\section{Proofs of Main Results}
\label{sec:appendix-proofs}
\subsection{Proofs of the main mixing-time results}
\label{sec:proofs-main-results}

This section proves \Cref{cor:SPS_mixing_kappa_concave} for SPS in the regime $-c_1/d\le\kappa\le0$ and \Cref{cor:SCPS_mixing_kappa_concave} for SCPS in the regime of fixed $\kappa\in(-1,0)$, together with the corresponding results, \Cref{cor:SPS_mixing_CD,cor:SCPS_mixing_CD}, under the additional curvature condition of \Cref{prop:sharper_isoperimetry}. The adaptation extending the moderately heavy-tailed result to SCPS is described in \Cref{rmk:vanishing-kappa-SCPS} and proved in \Cref{sec:proofs-extension-SCPS}.

\begin{proof}[Proof of \Cref{cor:SPS_mixing_kappa_concave}]
Let $X\sim\bar\pi$, and let $\pi_0 =({\rm SP}^{-1})_{\#}\bar\pi_0 $ and $\pi=({\rm SP}^{-1})_{\#}\bar\pi$ denote the corresponding initial and target distributions on the sphere. By assumption, the distribution $\pi_0$ is $M$-warm with respect to $\pi$.

We now establish a mixing-time upper bound for the corresponding chain on $\mathcal{S}^d$. Since total variation distance is preserved under the projection, the same mixing-time upper bound applies to the SPS Markov chain on $\mathbb{R}^d$. As described in \Cref{sec:proof-sketch-s-cond}, we obtain a mixing-time upper bound by finding a lower bound for the $s$-conductance of the Markov chain via \Cref{prop:lbCs}. Thus, we first prove that all three assumptions of \Cref{prop:lbCs} hold.

\underline{\textbf{Verify Assumptions of  \Cref{prop:lbCs}.}}
Since $X$ is centered, the covariance assumption gives
\[
\mathbb E[|X|^2]=\operatorname{tr}(\Sigma_d)\le\lambda_U d.
\]
Markov's inequality therefore implies that the $2/3$-quantile $r$ of $|X|$ satisfies $r\le\sqrt{3\lambda_U d}$. For all sufficiently large $d$, we have $\kappa\ge-1$. We recall that a $\kappa$-concave measure is also $\kappa'$-concave for every $\kappa'\le\kappa$ (see \Cref{sec:kappa-concave-def} for details). Then, since $\bar\pi$ is $\kappa$-concave with $\kappa\ge-1$, it is in particular $(-1)$-concave, including the case with $\kappa=0$ (i.e., $\bar\pi$ is log-concave). \Cref{lem:weightedIsoperK} consequently gives a weighted isoperimetric inequality with weight function $w_{\rm SP}$ as in~\eqref{eq:def-w_SP} and Cheeger constant
\[
\Phi=\left(C\left(\frac rR+1\right)\right)^{-1} \asymp1,
\]
where we used $R=c_R\cdot\sqrt d$ and $r=\mathcal O(\sqrt d)$ to get the asymptotics. 
Moreover,
\[
\int w_{\rm SP}(x)\,\bar\pi(dx)=R+\frac{\mathbb E[|X|^2]}{4R}<\infty.
\]
Thus, \Cref{lem:weightedToUnweighted,lem:isoperimetryToThreeSet} imply that $\pi$ satisfies the three-set isoperimetric inequality with Cheeger constant $\Phi$. Hence, Assumption~\ref{ass:3set-isoper} of \Cref{prop:lbCs} is satisfied.

By \Cref{lem:tail_bound_kappa_concave_light_tail}, there exist constants $\ell>1$ and $c_2>0$, independent of $d$, such that
\[
\mathbb P(|X|>\ell\sqrt d)\le e^{-c_2\sqrt d}
\]
for all sufficiently large $d$.

Choose a fixed $c\in(0,1]$ sufficiently small that $4\frac{2-c}{c}c_R^2\ge\ell^2$, and set $\mathcal{C}_{\rm SP}=\{u\in\mathcal S^d:u_{d+1}>2-c\}$.
Using~\eqref{eq:cap_tail_probability_SP} and $R=c_R\cdot \sqrt{d}$, we then obtain 
\begin{align}
    \label{eq:exponential-cap-bound}
    \pi(\mathcal{C}_{\rm SP}) =\mathbb P\left(|X|^2>4\frac{2-c}{c}R^2\right) \le e^{-c_2\sqrt d}.
\end{align}

The theorem assumes that the gradient and Hessian bounds~\eqref{eq:grad-Hess-bounds-SPS-results} hold on every ball whose radius is a fixed multiple of $\sqrt d$. Since $c$ and $c_R$ are independent of $d$, and $R=c_R\cdot\sqrt d$, these bounds hold in particular on the region $\{|x|^2\le4\frac{4-c}{c}R^2\}$, as required in \Cref{prop:coupling_general} for $(\delta,\rho)$-coupling. The proposition then implies that there exists a sufficiently small constant $c_h>0$ such that, for $h=c_h\cdot d^{-1}$ and all sufficiently large $d$, the SPS kernel $T_{\rm SP}$ is $(\delta,\rho)$-coupling on $\mathcal S^d\setminus\mathcal{C}_{\rm SP}$,
where
\[
\delta=2\arctan(0.3184\,h)\asymp d^{-1},\qquad\text{and}\qquad \rho=\frac14.
\]
Since $T_{\rm SP}$ is a Random Walk Metropolis kernel with target distribution $\pi$, it is reversible with respect to $\pi$. Hence, Assumption~\ref{ass:coupling} in \Cref{prop:lbCs} holds.

Fix $\varepsilon\in(0,1)$ and set $s=\frac{\varepsilon^2}{4M^2}\in(0,1/2)$. Since $M=\mathcal O(d^q)$, $\Phi\asymp1$, and
$\delta\asymp d^{-1}$, we have
\[
\frac{s}{4}\min\left\{1,\frac{\delta\Phi}{16}\right\} =\Omega\left(d^{-(2q+1)}\right).
\]
The exponential cap bound~\eqref{eq:exponential-cap-bound} therefore gives
\[
\pi(\mathcal{C}_{\rm SP}) \le\frac{s}{4}\min\left\{1,\frac{\delta\Phi}{16}\right\}
\]
for all sufficiently large $d$. Thus, also Assumption~\ref{ass:cap-bound} of \Cref{prop:lbCs} is satisfied.

\underline{\textbf{Apply \Cref{prop:lbCs}.}}
Applying the argument of \Cref{sec:proof-sketch-s-cond}, based on \Cref{prop:lbCs} and \Cref{lem:nonlazy_mixing} (see in particular~\eqref{eq:n_mix-ub_Cs} and~\eqref{eq:lovasz1993+CsLB}), with $s=\varepsilon^2/(4M^2)$ gives
\[
C_s(T_{\rm SP}^2)\ge\frac{\rho}{4} \min\left\{1,\frac{\delta\Phi}{64}\right\}=\Omega(d^{-1}),
\]
and
\[
n_{\rm mix}(\varepsilon,\bar\pi_0)\le \left\lceil \frac{8}{C_s(T_{\rm SP}^2)^2} \log\left(\frac{4M^2}{\varepsilon^2}\right)\right\rceil=\mathcal O\left(\frac{\log(4M^2/\varepsilon^2)}{\rho^2\min\{1,\delta\Phi/64\}^2}\right),
\]
where we used preservation of total variation under stereographic projection.

Since $\rho=1/4$, $\Phi\asymp1$, $\delta\asymp d^{-1}$, and $\log(4M^2/\varepsilon^2)=\mathcal O(\log d)$, we conclude that $n_{\rm mix}(\varepsilon,\bar\pi_0)=\mathcal O(d^2\log d)$.
\end{proof}

\begin{proof}[Proof of \Cref{cor:SPS_mixing_CD}]
    Applying \Cref{prop:sharper_isoperimetry}, we obtain that $\Phi=\Omega(\sqrt{d})$.
    The proof is then analogous to that of \Cref{cor:SPS_mixing_kappa_concave}, replacing the asymptotics $\Phi\asymp 1$ with $\Phi=\Omega(\sqrt{d})$.
\end{proof}

\begin{proof}[Proof of \Cref{cor:SCPS_mixing_kappa_concave}]
As in the proof of \Cref{cor:SPS_mixing_kappa_concave}, we work with the transformed target $\pi=({\rm SCP}^{-1})_{\#}\bar\pi$ and initial distribution
$\pi_0=({\rm SCP}^{-1})_{\#}\bar\pi_0$. Warmness and total variation distance are preserved under the projection. Thus, as discussed in \Cref{sec:proof-sketch-s-cond}, it suffices to establish a mixing-time upper bound on $\mathcal S_o^d$ using \Cref{prop:lbCs} and \Cref{lem:nonlazy_mixing}. Then, the result directly transfers to the chain on $\mathbb{R}^d$.

We now verify that the three assumptions of \Cref{prop:lbCs} hold.

\underline{\textbf{Verify Assumptions of  \Cref{prop:lbCs}.}} Since we assume that $\beta>\frac{1}{2}$, $r=\mathcal O(\sqrt d)$, and $R=c_R\cdot d^\beta$, \Cref{lem:weightedIsoperK_SCP} gives a weighted isoperimetric inequality with weight function $w_{\rm SCP}$ as in~\eqref{eq:def-w_SCP} and Cheeger constant
\[
\Phi=\left(4C\max\left\{1,\frac rR\right\}\right)^{-1}\asymp1.
\]
Moreover, \Cref{lem:tail_bound_kappa_concave_kappa_const}
implies $\mathbb E[|X|]<\infty$, since its tail exponent $1/|\kappa|$ is greater than one.
The definitions of $S(x)$ and $L(x)$ in~\eqref{def:SCP_D-S-L} give
\[
w_{\rm SCP}(x) =\frac{R\bigl(S(x)-l_o(l_o-1)\bigr)}{l_o(2-l_o)} \le \frac{R}{2-l_o}+\frac{|x|}{\sqrt{l_o(2-l_o)}}.
\]
Thus, $\int w_{\rm SCP}\,d\bar\pi<\infty$.

As in the proof of \Cref{cor:SPS_mixing_kappa_concave}, applying \Cref{lem:weightedToUnweighted_SCP}, followed by \Cref{lem:isoperimetryToThreeSet}, establishes Assumption~\ref{ass:3set-isoper} of \Cref{prop:lbCs} with Cheeger constant $\Phi$.

Fix any $c\in(0,l_o)$ independently of $d$, and set
\[
\mathcal{C}_{\rm SCP}=\{u\in\mathcal S_o^d:u_{d+1}>l_o-c\}.
\]
The theorem assumes that the gradient and Hessian bounds~\eqref{eq:grad-Hess-bounds-SCPS-results} hold on every ball whose radius is a fixed multiple of $R$. Since $R=c_R\cdot d^{\beta}$, the assumption of \Cref{prop:coupling_general_SCP} is satisfied and we obtain that, for a sufficiently small constant $c_h>0$ and $h=c_hd^{-(\beta+1)}$, the Markov kernel $T_{\rm SCP}$ is $(\delta,\rho)$-coupling on $\mathcal{S}^d_o\setminus\mathcal{C}_{\rm SCP}$, with
\[
\delta=2\arctan(0.3184\,h)\asymp d^{-(\beta+1)}, \qquad\text{and}\qquad \rho=\frac14.
\]
Together with reversibility of $T_{\rm SCP}$ with respect to $\pi$, established in~\cite[Section~2.2 and Appendix~A.2]{grazzi2026sub}, this verifies Assumption~\ref{ass:coupling} of \Cref{prop:lbCs}.

It remains to verify the bound on the probability of $\mathcal{C}_{\rm SCP}$. Let $m$ denote the median of $|X|$, and let $\bar c=\bar c(c,l_o)>0$ be defined as in~\eqref{eq:SCP_bar_c}. Since $m\le r=\mathcal O(\sqrt d)$, we have that $\ell\coloneqq \tfrac{\sqrt{\bar c}\cdot R}{m}=\Omega(d^{\beta-1/2})$, which satisfies $\ell\ge1$ for all sufficiently large $d$. Thus, \Cref{lem:tail_bound_kappa_concave_kappa_const} and~\eqref{eq:cap_tail_probability_SCP} give
\begin{align}
    \label{eq:bound-pi-C_SCP}
\pi(\mathcal{C}_{\rm SCP}) =\mathbb P\left(|X|>\sqrt{\bar c}\,R\right) =\mathcal O\left(d^{-\frac{\beta-\frac{1}{2}}{|\kappa|}}\right).
\end{align}
Because the theorem's assumption~\eqref{eq:scale-parameter-SCP} on $\beta$ is equivalent to $\tfrac{\beta-\frac{1}{2}}{|\kappa|}>\beta+1$, we obtain that $\pi(\mathcal{C}_{\rm SCP})=o(d^{-(\beta+1)})$.

For fixed $\varepsilon\in(0,1)$, choose $s=\varepsilon^2/(4M^2)$ as in the proof of
\Cref{cor:SPS_mixing_kappa_concave}. Since $M=\mathcal O(1)$ and $\Phi\asymp1$,
\[
\frac{s}{4}\min\left\{1,\frac{\delta\Phi}{16}\right\} \asymp d^{-(\beta+1)}.
\]
Consequently, Assumption~\ref{ass:cap-bound} of \Cref{prop:lbCs} holds for all sufficiently large $d$.

\underline{\textbf{Apply \Cref{prop:lbCs}.}} Applying \Cref{prop:lbCs} and \Cref{lem:nonlazy_mixing}, exactly as at the end of the proof of \Cref{cor:SPS_mixing_kappa_concave}, here yields
\[
n_{\rm mix}(\varepsilon,\bar\pi_0) = \mathcal O\left( \frac{\log(4M^2/\varepsilon^2)}{\rho^2\min\{1,\delta\Phi/64\}^2}\right)=\mathcal O(d^{2\beta+2}),
\]
where the last equality uses $\rho=1/4$, $\Phi\asymp1$, $\delta\asymp d^{-(\beta+1)}$, and $\log(4M^2/\varepsilon^2)=\mathcal O(1)$.
\end{proof}

\begin{proof}[Proof of \Cref{cor:SCPS_mixing_CD}]
Applying \Cref{prop:sharper_isoperimetry}, we obtain that $\Phi=\Omega(\sqrt d)$. The proof is then analogous to that of \Cref{cor:SCPS_mixing_kappa_concave}, replacing the asymptotics $\Phi\asymp1$ with $\Phi=\Omega(\sqrt d)$. In particular, since $\delta\asymp d^{-(\beta+1)}$, we have $\delta\,\Phi=\Omega(d^{-(\beta+1/2)})$. Moreover, note that the assumption $\beta>\frac{1}{2}+\frac{|\kappa|}{1-|\kappa|}$ of \Cref{cor:SCPS_mixing_CD} is equivalent to $\frac{\beta-\frac{1}{2}}{|\kappa|}>\beta+\frac{1}{2}$. Consequently, the bound~\eqref{eq:bound-pi-C_SCP} on $\pi(\mathcal C_{\rm SCP})$ obtained in the preceding proof is sufficient to verify Assumption~\ref{ass:cap-bound}. 
Substituting the improved bound on $\Phi$ into the mixing-time estimate
yields $n_{\rm mix}(\varepsilon,\bar\pi_0)=\mathcal O(d^{2\beta+1})$.
\end{proof}
\subsection{Proof of $s$-conductance lower bound}
\label{sec:proof-s-conductance-LB}
In this section, we determine lower bounds on the $s$-conductance of a reversible Markov transition kernel $T$ on a Riemannian state space $\mathcal X$, assuming a three-set isoperimetric inequality, a $(\delta,\rho)$-coupling condition on the state space $\mathcal{X}$ excluding a region $\mathcal{C}$, and a control on the probability of the excluded region $\mathcal{C}$. The proof is based on an approach similar to that in~\citet{wu2022minimax}.

\begin{proof}[Proof of Proposition~\ref{prop:lbCs}]
Recall the definition~\eqref{eq:conductance} of $s$-conductance. To prove the lower bound~\eqref{eq:LBCs_generic}, it therefore suffices to show that for all measurable sets $A$ with $s< \pi(A)\le\tfrac{1}{2}$,
\begin{equation}
    \label{eq:boundFixA}
    \frac{\int_AT(u,A^\complement)\pi(du)}{\pi(A)-s} \ge \frac{\rho}{4}\cdot \min\left\{1,\frac{\delta\Phi}{64}\right\}.
\end{equation}

Define the partition $\mathcal{X}=S_1\sqcup S_2\sqcup S_3$ of the state space $\mathcal{X}$ with
\begin{equation}
\label{eq:S1S2S3}
\begin{split}
      S_1\coloneqq\left\{u\in A\,:\,T(u,A^\complement)<\frac{\rho}{2}\right\}, \qquad
    S_2\coloneqq\left\{u\in A^\complement\,:\,T(u,A)<\frac{\rho}{2}\right\},\qquad
    S_3\coloneqq(S_1\cup S_2)^\complement.
\end{split}
\end{equation}

Using an approach similar to that in~\cite{wu2022minimax}, we distinguish two complementary cases: the first case corresponds to $\pi(S_1)\le \frac{\pi(A)}{2}$ \textbf{or} $\pi(S_2)\le \frac{\pi(A^\complement)}{2}$, whereas the second case is $\pi(S_1)> \frac{\pi(A)}{2}$ \textbf{and} $\pi(S_2)> \frac{\pi(A^\complement)}{2}$.

\textbf{\underline{Case $\pi(S_1)\le \frac{\pi(A)}{2}$ \textbf{or} $\pi(S_2)\le \frac{\pi(A^\complement)}{2}$.}} We first show that in this case 
\begin{align}
\label{eq:numerator_UB_case1}
    \int_AT(u,A^\complement)\pi(du)\ge \frac{\rho}{4}\pi(A).
\end{align}
In fact, in the sub-case $\pi(S_1)\le \frac{\pi(A)}{2}$, we have by definition of $S_1$ that 
\begin{align*}
   \int_AT(u,A^\complement)\pi(du)\ge  \int_{A\setminus S_1}T(u,A^\complement)\pi(du)\ge \int_{A\setminus S_1}\frac{\rho}{2}\pi(du)\ge \frac{\rho}{2}(\pi(A)-\pi(S_1))\ge \frac{\rho}{2}\frac{\pi(A)}{2}.
\end{align*}
On the other hand, if $\pi(S_2)\le \frac{\pi(A^\complement)}{2}$, we can write
\begin{align*}
    \int_AT(u,A^\complement)\pi(du)&\stackrel{(i)}{=}\int_{A^\complement}T(u,A)\pi(du)\ge \int_{A^\complement\setminus S_2}T(u,A)\pi(du)\\
    &\stackrel{(ii)}{\ge }\int_{A^\complement\setminus S_2}\frac{\rho}{2}\pi(du)=\frac{\rho}{2}(\pi(A^\complement)-\pi(S_2))\ge \frac{\rho}{2}\frac{\pi(A^\complement)}{2}\stackrel{(iii)}{\ge}\frac{\rho}{2}\frac{\pi(A)}{2},
\end{align*}
where in $(i)$ we used the reversibility of the Markov transition kernel $T$, in $(ii)$ we applied the definition of $S_2$, while $(iii)$ follows from the assumption $\pi(A)\le \frac{1}{2}$.

We thus have proved that in the case $\pi(S_1)\le \frac{\pi(A)}{2}$ \textbf{or} $\pi(S_2)\le \frac{\pi(A^\complement)}{2}$, the inequality~\eqref{eq:numerator_UB_case1} holds, which implies that 
\begin{align*}
    \frac{\int_AT(u,A^\complement)\pi(du)}{\pi(A)-s}\ge  \frac{\int_AT(u,A^\complement)\pi(du)}{\pi(A)}\ge \frac{\rho}{4}.
\end{align*}
This shows that~\eqref{eq:boundFixA} holds in this first case.

\textbf{\underline{Case $\pi(S_1)> \frac{\pi(A)}{2}$ \textbf{and} $\pi(S_2)> \frac{\pi(A^\complement)}{2}$.}}
By reversibility of the Markov kernel $T$,
    \begin{equation}
    \label{eq:bound_int_T}
    \begin{split}
        \int_AT(u,A^\complement)\pi(du)&=\frac{1}{2}\left[\int_AT(u,A^\complement)\pi(du)+\int_{A^\complement}T(u,A)\pi(du)\right]\\
        &\ge \frac{1}{2}\left[\int_{A\setminus S_1}T(u,A^\complement)\pi(du)+\int_{A^\complement\setminus S_2}T(u,A)\pi(du)\right]\\
        &\ge \frac{1}{2}\left[\frac{\rho}{2}\pi(A\setminus S_1)+\frac{\rho}{2}\pi(A^\complement\setminus S_2)\right]\\
        &=\frac{\rho}{4}\pi((S_1\cup S_2)^\complement)\\
        &=\frac{\rho}{4}\pi(S_3)
    \end{split}
    \end{equation}
    where in the last equality we used the definition of $S_3$.

    We now define the following new partition of $\mathcal X$:
    \begin{align*}
        \tilde S_1 \coloneqq S_1\setminus \mathcal{C},\qquad  \tilde S_2 \coloneqq S_2\setminus \mathcal{C},\qquad \tilde S_3 \coloneqq S_3 \cup \mathcal{C}.
    \end{align*}
    First, we show that $\mathsf{dist}(\tilde S_1,\tilde S_2)\ge \delta$. Let $u\in \tilde S_1$ and $v\in \tilde S_2$. Noting that $\tilde S_i\subseteq S_i$ for $i=1,2$, the definition \eqref{eq:S1S2S3} of $S_1$ and $S_2$ implies that $T(u,A)> 1-\frac{\rho}{2}$ and $T(v,A)<\frac{\rho}{2}$, therefore,
    \begin{align*}
        T(u,A)-T(v,A)>\left(1-\frac{\rho}{2}\right)-\frac{\rho}{2}=1-\rho.
    \end{align*}
    By the definition of total variation distance,
    \begin{align*}
        \|T(u,\cdot)-T(v,\cdot)\|_{\text{TV}}\ge |T(u,A)-T(v,A)|>1-\rho.
    \end{align*}
    The assumption that $T$ is $(\delta,\rho)$-coupling on $\mathcal X\setminus \mathcal{C}$ implies that $\mathsf{dist}(u,v)\ge \delta$. Since this holds for any $u\in \tilde S_1$ and any $v\in \tilde S_2$, our claim that $\mathsf{dist}(\tilde S_1,\tilde S_2)\ge \delta$ follows.

    Since we are considering the case $\pi(S_1)> \frac{\pi(A)}{2}$ \textbf{and} $\pi(S_2)> \frac{\pi(A^\complement)}{2}$, recalling $s<\pi(A)\le \frac{1}{2}$ and the bound~\eqref{bound_pi(C)} on $\pi(\mathcal{C})$, we can write
    \begin{equation}
    \label{eq:tilde_S_1_tilde_S_2-lower-bounds}
    \begin{split}
        \pi(\tilde S_1) &\ge \pi(S_1) - \pi(\mathcal{C}) \ge \frac{\pi(A)}{2}-\frac{s}{4}\ge \frac{\pi(A)}{2}-\frac{\pi(A)}{4}=\frac{\pi(A)}{4},\qquad\text{and}\\
        \pi(\tilde S_2)&\ge \pi(S_2) -  \pi(\mathcal{C}) \ge \frac{\pi(A^\complement)}{2}-\frac{s}{4}\ge \frac{1}{4}-\frac{\pi(A)}{4}\ge\frac{1}{4}-\frac{1}{8}=\frac{1}{8}.
        \end{split}
    \end{equation}
    
    By the three-set isoperimetric inequality~\eqref{eq:3set_isoper}, we then obtain that
    \begin{align*}
        \pi(\tilde S_3)\ge \Phi\cdot \mathsf{dist}(\tilde S_1,\tilde S_2)\cdot \pi(\tilde S_1)\cdot\pi(\tilde S_2)\ge \delta\cdot \Phi \cdot  \pi(\tilde S_1)\cdot\pi(\tilde S_2).
    \end{align*}

    Together with~\eqref{eq:tilde_S_1_tilde_S_2-lower-bounds}, we obtain $\pi(\tilde S_3)\ge \frac{\delta\cdot\Phi}{32}\pi(A)$. Moreover, since $s<\pi(A)$ and $\min\{1,\delta\Phi/16\}\le\delta\Phi/16$, condition~\eqref{bound_pi(C)} gives $\pi(\mathcal C)\le\frac{\delta\Phi}{64}\pi(A)$. Since $\tilde S_3 = S_3 \cup \mathcal{C}$, we bound $\pi(S_3)$ as 
    \begin{align*}
         \pi(S_3) \ge \pi(\tilde S_3) - \pi(\mathcal{C})\ge \frac{\delta\cdot\Phi}{32}\pi(A) - \frac{\delta\cdot\Phi}{64}\pi(A) = \frac{\delta\cdot\Phi}{64}\pi(A).
    \end{align*}

    Consequently, from~\eqref{eq:bound_int_T} we can write
    \begin{align*}
        \int_AT(u,A^\complement)\pi(du)\ge \frac{\rho}{4}\cdot \frac{\delta\cdot\Phi}{64}\pi(A),
    \end{align*}
    hence,
    \begin{align*}
        \frac{\int_AT(u,A^\complement)\pi(du)}{\pi(A)-s}\ge  \frac{\int_AT(u,A^\complement)\pi(du)}{\pi(A)}\ge \frac{\rho}{4}\cdot \frac{\delta\cdot\Phi}{64}.
    \end{align*}
    This proves~\eqref{eq:boundFixA} in this second case ($\pi(S_1)> \frac{\pi(A)}{2}$ \textbf{and} $\pi(S_2)> \frac{\pi(A^\complement)}{2}$), completing the proof.
\end{proof}

\subsection{Proofs for isoperimetric inequalities}
\label{sec:proofs-isoperim}

The purpose of this section is to establish the isoperimetric results used in our conductance argument described in \Cref{sec:proof-sketch-s-cond}. We first prove the weighted isoperimetric inequalities on $\mathbb{R}^d$ for $\kappa$-concave distributions stated in \Cref{lem:weightedIsoperK,lem:weightedIsoperK_SCP}, for which we use results of \citet{bobkov2009weighted2}.
We then show, in \Cref{lem:weightedToUnweighted,lem:weightedToUnweighted_SCP}, that for specific choices of weight functions, tailored to the stereographic and Sub-Cauchy projections, these weighted isoperimetric inequalities on $\mathbb{R}^d$ are transferred to unweighted isoperimetric inequalities on the sphere. Finally, \Cref{lem:isoperimetryToThreeSet} converts the resulting boundary isoperimetric inequalities into the three-set form used in the conductance argument in \Cref{sec:proof-sketch-s-cond}.

For functions defined on a Riemannian manifold with variable $u$, we write $\nabla_u$ for the Riemannian gradient with respect to $u$, while $\nabla_x$ denotes the Euclidean gradient with respect to $x\in\mathbb R^d$.

\subsubsection{Weighted isoperimetric inequalities on $\mathbb{R}^d$}
\label{sec:proofs-weighted-isoperim}

We first recall two results of \citet{bobkov2009weighted2} that will be used to prove \Cref{lem:weightedIsoperK,lem:weightedIsoperK_SCP}.

\begin{lemma}[Bobkov-Ledoux inequality for $\kappa$-concave measures \citep{bobkov2009weighted2}]
\label{lem:BL09thm1.1}
If $\bar\pi$ is a $\kappa$-concave probability measure on $\mathbb{R}^d$, $-\infty<\kappa< 0$, then for any locally Lipschitz function $f$ on $\mathbb{R}^d$,
\begin{align*}
    \iint|f(x)-f(y)|\bar\pi(dx)\bar\pi(dy)\le C\cdot(1-\kappa)\int|\nabla_x f(x)|(r+|x|)\bar\pi(dx),
\end{align*}
where $C$ is a universal constant and $r$ is the quantile of order $\tfrac{2}{3}$ of the Euclidean norm under $\bar\pi$.
\end{lemma}

\begin{lemma}[Functional characterization of the weighted isoperimetric inequality {\citep[Proposition 6.1]{bobkov2009weighted2}}]
\label{lem:BL09prop6.1}
    Given a Borel probability measure $\bar\pi$ on $\mathbb{R}^d$, the following properties are equivalent:
    \begin{enumerate}
        \item[(a)] For any locally Lipschitz $\bar\pi_w$-finite function $f$ on $\mathbb{R}^d$,
        \begin{align*}
            \int|\nabla_x f(x)|\,w(x)\,\bar\pi(dx)\ge \frac{1}{2}\iint|f(x)-f(y)|\,\bar\pi(dx)\,\bar\pi(dy);
        \end{align*}
        \item[(b)] For any closed set $A$ in $\mathbb{R}^d$ of finite measure $\bar\pi_w(A)$, we have
        \begin{align}
        \label{eq:isoperBL09}
            \bar\pi^+_w(A)\ge \bar\pi(A)\cdot (1-\bar\pi(A)).
        \end{align}
    \end{enumerate}
\end{lemma}

\begin{remark}
\label{rmk:extendLemmaBK09}
    As pointed out in \cite[Remark 6.3]{bobkov2009weighted2}, if $w(x)>0$ on the support of $\bar\pi$, the isoperimetric inequality~\eqref{eq:isoperBL09} can be extended from the class of all closed subsets to the class of all Borel sets $A\subset \mathbb{R}
    ^d$ satisfying $\bar\pi_w(A)<\infty$.
\end{remark}

We now prove the weighted isoperimetric inequalities stated in \Cref{lem:weightedIsoperK,lem:weightedIsoperK_SCP}.

\begin{proof}[Proof of \Cref{lem:weightedIsoperK}]
Define
\[
w'(x)\coloneqq C\left(\frac{r}{R}+1\right)w_{\rm SP}(x).
\]
Since
\[
\frac{r}{R}\,w_{\rm SP}(x)=r\,\frac{4R^2+|x|^2}{4R^2}\ge r, \qquad w_{\rm SP}(x)=\frac{4R^2+|x|^2}{4R}\ge |x|,
\]
and $1\ge\frac{1-\kappa}{2}$ for $-1\le\kappa<0$, we obtain
\begin{align*}
    w'(x)\ge C(r+|x|)\ge C\frac{1-\kappa}{2}(r+|x|).
\end{align*}    
By \Cref{lem:BL09thm1.1}, it follows that
    \begin{align*}
        \iint|f(x)-f(y)|\bar\pi(dx)\bar\pi(dy)\le C(1-\kappa)\int|\nabla_x f(x)|(r+|x|)\bar\pi(dx)\le 2\int|\nabla_x f(x)|w'(x)\bar\pi(dx),
    \end{align*}
    for all locally Lipschitz functions $f$. Then, applying \Cref{lem:BL09prop6.1}, we obtain the isoperimetric inequality
    \begin{align}
    \label{eq:isoper_w'_SP}
        \bar\pi_{w'}^+(A)&\ge \bar\pi(A)\cdot (1-\bar\pi(A)),
    \end{align}
    for any closed subset $A\subset \mathbb{R}^d$ of finite measure $\bar\pi_{w'}(A)$.
    
    Because $w_{\rm SP}(x)=\frac{1}{C\left(\frac{r}{R} +1\right)}w'(x)$, the weighted measures and their corresponding Minkowski boundaries both scale by the constant factor $\Phi_{\rm SP} = \left(C\left(\frac{r}{R} +1\right)\right)^{-1}$. Inequality \eqref{eq:isoper_w'_SP} thus implies that for any closed set $A$ of finite $\bar\pi_{w_{\rm SP}}$ measure, ~\eqref{eq:isoper_w_finite} holds.

    To remove the restriction of finite $\bar\pi_{w_{\rm SP}}$ mass and extend the result to all Borel sets, we apply a truncation argument. For $n\in\mathbb{N}$, define the truncated probability measures
    \[
    \bar\pi_n(\cdot)=\frac{\bar\pi(\,\cdot\,\cap \{|x|<n\})}{p_n},\qquad \text{with }\quad     p_n=\bar\pi(\{|x|<n\}).
    \]
    Note that $p_n>0$ for sufficiently large $n$. As mentioned in \Cref{sec:kappa-concave-def}, conditioning a $\kappa$-concave distribution on a convex ball preserves its $\kappa$-concavity. Let $r_n$ denote the $\frac{2}{3}$-quantile of $|X|$ under $\bar\pi_n$. Because the truncation removes tail mass, the quantiles satisfy $r_n \le r$. 
    
    Consequently, the preceding argument applies to $\bar\pi_n$ using the exact same weight $w_{\rm SP}(x)$ and Cheeger's constant $\Phi_n=\left(C\left(\frac{r_n}{R}+1\right)\right)^{-1}\ge\Phi_{\rm SP}$. 

    Because $\bar\pi_n$ has bounded support, its total $w_{\rm SP}$-weighted mass is finite. Moreover, $w_{\rm SP}>0$, so \Cref{rmk:extendLemmaBK09} extends the preceding inequality to every Borel set $A\subseteq\mathbb R^d$. Thus, for every Borel set $A$,
    \[
    \bar\pi_{w_{\rm SP}}^+(A)
    \ge p_n\cdot (\bar\pi_n)_{w_{\rm SP}}^+(A)
    \ge p_n\cdot \Phi_n\cdot \bar\pi_n(A)\cdot (1-\bar\pi_n(A))\ge p_n\cdot \Phi_{\rm SP}\cdot \bar\pi_n(A)\cdot (1-\bar\pi_n(A)).
    \]
    Taking the limit as $n\to\infty$, we have $p_n \to 1$ and $\bar\pi_n(A) \to \bar\pi(A)$, which yields the desired inequality~\eqref{eq:isoper_w_finite} for all Borel sets $A\subset \mathbb{R}^d$.
\end{proof}

\begin{proof}[Proof of \Cref{lem:weightedIsoperK_SCP}]
The function $L(x)$ defined in \Cref{def:SCP_D-S-L} can be rewritten as
\begin{align*}
    L(x)=\frac{l_o}{\frac{|x|^2}{R^2}+l_o^2}\left(l_o-1+\sqrt{\left(\frac{2}{l_o}-1\right)\frac{|x|^2}{R^2}+1}\right).
\end{align*}
Therefore, the weight function becomes
\begin{align*}
    w_{\rm SCP}(x)=R\frac{\frac{|x|^2}{R^2}+l_o^2}{l_o}\left(l_o-1+\sqrt{\left(\frac{2}{l_o}-1\right)\frac{|x|^2}{R^2}+1}\right)^{-1}.
\end{align*}

Since $l_o \in [1,2)$, the quantity $\frac{2}{l_o}-1$ satisfies $\frac{2}{l_o}-1\le1$. Moreover, $a^2+b^2\le (a+b)^2$ for $a,b\ge0$. We can then bound the square root term in $w_{\rm SCP}$ as
    \begin{align*}
        \sqrt{\left(\frac{2}{l_o}-1\right)\frac{|x|^2}{R^2}+1}\le \sqrt{\frac{|x|^2}{R^2}+1}\le \frac{|x|}{R}+1.
    \end{align*}
    This implies the following lower bound for the weight function:
    \begin{align*}
        w_{\rm SCP}(x)&\ge \frac{R}{l_o}\frac{\frac{|x|^2}{R^2}+l_o^2}{l_o-1+\frac{|x|}{R}+1}=\frac{R}{l_o}\frac{\frac{|x|^2}{R^2}+l_o^2}{\frac{|x|}{R}+l_o}\stackrel{(i)}{\ge} \frac{R}{2l_o}\left(\frac{|x|}{R}+l_o\right)\stackrel{(ii)}{\ge} \frac{R}{4}\left(\frac{|x|}{R}+1\right)=\frac{R+|x|}{4}\\
        &\ge\frac{1}{4}\min\left\{1,\frac{R}{r}\right\}(r+|x|),
    \end{align*}
    where $(i)$ follows from $a^2+b^2\ge \frac{(a+b)^2}{2}$, and in $(ii)$ we used the assumption $l_o\in[1,2)$.

    Let $C$ be the universal constant from \Cref{lem:BL09thm1.1}. We now define $\Phi_{\rm SCP}\coloneqq \frac{1}{4C}\min\left\{1,\frac{R}{r}\right\}$, and $w'(x)\coloneqq\Phi_{\rm SCP}^{-1}\cdot w_{\rm SCP}(x)$. Since $\kappa\ge-1$, we have that $\frac{1-\kappa}{2} \le 1$, and therefore,
    \begin{align*}
        w'(x) \ge C \frac{1-\kappa}{2}(r+|x|).
    \end{align*}
   
    By \Cref{lem:BL09thm1.1}, for all locally Lipschitz functions $f$, we have
    \begin{align*}
        \iint|f(x)-f(y)|\bar\pi(dx)\bar\pi(dy)\le C(1-\kappa)\int|\nabla_x f(x)|(r+|x|)\bar\pi(dx)\le 2\int|\nabla_x f(x)|w'(x)\bar\pi(dx).
    \end{align*}

Then, applying \Cref{lem:BL09prop6.1}, we obtain that
    \begin{align}
    \label{eq:isoper_w'}
        \bar\pi_{w'}^+(A)&\ge \bar\pi(A)\cdot (1-\bar\pi(A)),
    \end{align}
    for any closed subset $A\subset \mathbb{R}^d$ of finite measure $\bar\pi_{w'}(A)$.

    Because $w_{\rm SCP}(x)=\Phi_{\rm SCP}\cdot w'(x)$, the weighted measures and their corresponding Minkowski boundaries scale by the constant factor $\Phi_{\rm SCP} =\left(4 \, C \max\left\{1, \frac{r}{R}\right\}\right)^{-1}$. Inequality \eqref{eq:isoper_w'} thus implies that for any closed set $A$ of finite $\bar\pi_{w_{\rm SCP}}$ measure,~\eqref{eq:isoper_w_SCP} holds.
    
By applying the same truncation argument as in the proof of \Cref{lem:weightedIsoperK}, we can remove the restriction of finite $\bar\pi_{w_{\rm SCP}}$ mass and obtain that \textit{for all} Borel sets $A\subset \mathbb{R}^d$,~\eqref{eq:isoper_w_SCP} holds.

\end{proof}

\subsubsection{From weighted to unweighted isoperimetric inequalities under projection maps}
\label{sec:proofs-transfer-isoper}

Having established the required weighted isoperimetric inequalities on $\mathbb{R}^d$, we now show how the weight functions $w_{\rm SP}$ and $w_{\rm SCP}$ defined above allow them to be transferred to unweighted isoperimetric inequalities for the transformed targets on $\mathcal S^d$ and $\mathcal S_o^d$.

To formulate the transfer results, we introduce the class of regular sets for which the boundary Minkowski content can be represented by a surface integral.

 \begin{definition}[Regular sets]
 \label{def:regular_set}

Let $\mathcal X$ be a connected $d$-dimensional Riemannian manifold, equipped with its intrinsic Riemannian metric, and let $\pi$ be a probability measure admitting a density with respect to the Riemannian volume. 

We say that a measurable set
$A\subseteq\mathcal X$ is \emph{regular} if its relative boundary
$\partial_{\mathcal X}A$ is countably $(d-1)$-rectifiable and
\begin{align*}
    \pi^+(A)
    =
    \int_{\partial_{\mathcal X}A}
    \pi(u)\,\mathcal H_{d-1}(du).
\end{align*}
We denote the collection of such sets by
$\mathcal R_{d-1}(\mathcal X)$.

Similarly, a measurable set $B\subseteq\mathbb R^d$ is
\emph{regular with respect to $w$} if $\partial B$ is countably
$(d-1)$-rectifiable and
\begin{align*}
    \bar\pi_w^+(B)
    =
    \int_{\partial B}
    w(x)\bar\pi(x)\,
    \bar{\mathcal H}_{d-1}(dx).
\end{align*}
The collection of such sets is denoted by
$\mathcal R^w_{d-1}(\mathbb R^d)$.
 \end{definition}

We next establish the transfer result for stereographic projection. The following lemma shows that the weighted Euclidean isoperimetric inequality with weight
$w_{\rm SP}(x)=\frac{4R^2+|x|^2}{4R}$
yields the corresponding unweighted isoperimetric inequality on $\mathcal{S}^d$. 
The relevant property of this weight function is that
\[
w_{\rm SP}(x)\cdot 
\|D({\rm SP}^{-1})_x\|_{\rm op}=1,
\]
as shown in \Cref{lem:singular_values_SP}.

\begin{lemma}[From weighted to unweighted isoperimetric inequality under stereographic projection]
\label{lem:weightedToUnweighted}
Let
\[
    w_{\rm SP}(x)\coloneqq\frac{4R^2+|x|^2}{4R}.
\]

Assume that $\Phi>0$, $\int w_{\rm SP}\,d\bar\pi<\infty$ and that, for every Borel set
$B\subseteq\mathbb R^d$,
\begin{equation}
\label{eq:weightedEuclid}
    \bar\pi_{w_{\rm SP}}^+(B)
    \ge
    \Phi\cdot\bar\pi(B)\cdot (1-\bar\pi(B)).
\end{equation}
Then, for every $A\in\mathcal R_{d-1}(\mathcal S^d)$,
\begin{equation}
    \label{eq:isoper_no_weight}
    \pi^+(A)\ge\Phi\cdot\pi(A)\cdot (1-\pi(A)).
\end{equation}
\end{lemma}

\begin{proof}
We first pass from the weighted isoperimetric inequality to its equivalent functional
form.
Since $\int w_{\rm SP}\,d\bar\pi<\infty$, the measure associated with the rescaled weight $w_{\rm SP}/\Phi$ is finite, thus every locally Lipschitz function is $\bar\pi_{w_{\rm SP}/\Phi}$-finite. Applying \Cref{lem:BL09prop6.1} to $w_{\rm SP}/\Phi$ therefore gives, for every locally Lipschitz function $f$,
\begin{align}
\label{eq:weighted_functional_SP}
    \int_{\mathbb R^d}
    |\nabla_x f(x)|\,w_{\rm SP}(x)\,\bar\pi(dx)
    \ge
    \frac{\Phi}{2}
    \iint_{\mathbb R^d\times\mathbb R^d}
    |f(x)-f(y)|\,\bar\pi(dx)\,\bar\pi(dy).
\end{align}

Let $g$ be any locally Lipschitz function $\mathcal{S}^d\to \mathbb{R}$, and define $f(x)=g({\rm SP}^{-1}(x))$.
By \Cref{lem:singular_values_SP}, the inverse stereographic
projection satisfies
\[
    \|D({\rm SP}^{-1})_x\|_{\rm op}=\frac{4R}{|x|^2+4R^2}=\frac{1}{w_{\rm SP}(x)}.
\]
Hence, by the chain rule,
\[
    w_{\rm SP}(x)\cdot|\nabla_x f(x)|\le w_{\rm SP}(x)\cdot|\nabla_{u}g({\rm SP}^{-1}(x))|\cdot  \|D({\rm SP}^{-1})_x\|_{\rm op}=|\nabla_{u}g({\rm SP}^{-1}(x))|.
\]
Using the relation between $\bar\pi$ and $\pi$ into~\eqref{eq:weighted_functional_SP} therefore yields
\begin{align}
\label{eq:functional_sphere_SP}
    \int_{\mathcal S^d}|\nabla_{u}g(u)|\,\pi(du)\ge\frac{\Phi}{2}\iint_{\mathcal S^d\times\mathcal S^d}|g(u)-g(v)|\,\pi(du)\pi(dv).
\end{align}

We now use~\eqref{eq:functional_sphere_SP} to recover the unweighted isoperimetric inequality~\eqref{eq:isoper_no_weight}. Let $A\in\mathcal R_{d-1}(\mathcal S^d)$. Since $\partial A$ is $(d-1)$-rectifiable and $\pi$ has a density with respect to the Riemannian volume, $\pi(\partial A)=0$. Replacing $A$ by its closure changes neither $\pi(A)$ nor $\pi^+(A)$. For $\varepsilon>0$, define the following approximation of the indicator function $\mathbf 1_A(u)$:
\[
    g_\varepsilon(u)=\left(1-\frac{\mathsf{dist}(u,A)}{\varepsilon}\right)_+,
\]
where $(a)_+ \coloneqq \max\{a,0\}$ denotes the positive part of $a\in\mathbb{R}$. Observe that $g_\varepsilon\to\mathbf 1_A$ $\pi$-almost everywhere as $\varepsilon\downarrow0$,
$|\nabla_u g_\varepsilon|\le\varepsilon^{-1}$, and the gradient is supported on
$A^\varepsilon\setminus A$. Consequently,
\[
    \int|\nabla_u g_\varepsilon|\,d\pi\le\frac{\pi(A^\varepsilon\setminus A)}{\varepsilon}.
\]
Moreover, by dominated convergence,
\[
    \iint|g_\varepsilon(u)-g_\varepsilon(v)|\,\pi(du)\,\pi(dv)\longrightarrow2\,\pi(A)\cdot(1-\pi(A)),
\]
since $|\mathbf 1_A(u)-\mathbf 1_A(v)|=1$ precisely when exactly one of $u$ and $v$ belongs to $A$. Applying~\eqref{eq:functional_sphere_SP} and taking
$\varepsilon\downarrow0$ gives~\eqref{eq:isoper_no_weight}, as claimed.
\end{proof}

We now establish the corresponding transfer result for the Sub-Cauchy projection. In this case, the state space is the bright side $\mathcal S_o^d$ of the sphere $\mathcal{S}^d$, and the weight function is
$w_{\rm SCP}(x)=\frac{R}{L(x)}$,
which, by \Cref{lem:singular_values_inverse_SCP}, satisfies
\[
w_{\rm SCP}(x)\cdot
\|D({\rm SCP}^{-1})_x\|_{\rm op}\le1,
\]

\begin{lemma}[From weighted to unweighted isoperimetric inequality under Sub-Cauchy projection]
\label{lem:weightedToUnweighted_SCP}
Let
\[
    w_{\rm SCP}(x)\coloneqq\frac{R}{L(x)}.
\]
Assume that $\Phi>0$, $\int w_{\rm SCP}\,d\bar\pi<\infty$ and that, for every Borel set
$B\subseteq\mathbb R^d$,
\begin{equation*}
    \bar\pi_{w_{\rm SCP}}^+(B)
    \ge
    \Phi\cdot\bar\pi(B)\cdot (1-\bar\pi(B)).
\end{equation*}
Then, for every $A\in\mathcal R_{d-1}(\mathcal S_o^d)$,
\begin{equation}
    \label{eq:isoper_ineq_sphere}
    \pi^+(A)\ge\Phi\cdot\pi(A)\cdot (1-\pi(A)),
\end{equation}
where the Minkowski content and distance on $\mathcal S_o^d$ are defined with respect to the induced Riemannian metric.
\end{lemma}

\begin{proof}
Proceeding as in the proof of \Cref{lem:weightedToUnweighted}, the assumption $\int w_{\rm SCP}\,d\bar\pi<\infty$ and \Cref{lem:BL09prop6.1}, applied to the rescaled weight $w_{\rm SCP}/\Phi$, give, for every locally Lipschitz function $f$,
\begin{align}
\label{eq:weighted_functional_SCP}
\int_{\mathbb R^d}|\nabla_x f(x)|\,w_{\rm SCP}(x)\,\bar\pi(dx)\ge\frac{\Phi}{2}\iint_{\mathbb R^d\times\mathbb R^d}|f(x)-f(y)|\,\bar\pi(dx)\,\bar\pi(dy).
\end{align}

Let $g$ be a locally Lipschitz function on $\mathcal S_o^d$ and set $f(x)=g({\rm SCP}^{-1}(x))$. By \Cref{lem:singular_values_inverse_SCP},
\[\|D({\rm SCP}^{-1})_x\|_{\rm op}\le \frac{L(x)}{R}=\frac{1}{w_{\rm SCP}(x)}.\]
Hence, by the chain rule,
\[w_{\rm SCP}(x)\cdot|\nabla_x f(x)|\le w_{\rm SCP}\cdot |\nabla_u g({\rm SCP}^{-1}(x))|\cdot \|D({\rm SCP}^{-1}_x)\|_{\rm op}\le |\nabla_u g({\rm SCP}^{-1}(x))|.\]

After applying a change of variable in
\eqref{eq:weighted_functional_SCP}, we obtain
\begin{align}
\label{eq:functional_sphere_SCP}
\int_{\mathcal S_o^d}|\nabla_u g(u)|\,\pi(du)
\ge
\frac{\Phi}{2}
\iint_{\mathcal S_o^d\times\mathcal S_o^d}
|g(u)-g(v)|\,\pi(du)\pi(dv).
\end{align}

The remainder of the proof follows as in the proof of \Cref{lem:weightedToUnweighted}, with distances, closures, and Minkowski content understood relative to $\mathcal{S}_o^d$. Namely, for $A\in\mathcal R_{d-1}(\mathcal S_o^d)$, approximate $\mathbf 1_A$ by
\[g_\varepsilon(u)=\left(1-\frac{\mathsf{dist}(u,A)}{\varepsilon}\right)_+.\]
Applying~\eqref{eq:functional_sphere_SCP} to $g_\varepsilon$ and letting
$\varepsilon\downarrow0$ yields~\eqref{eq:isoper_ineq_sphere}, as claimed.
\end{proof}

Having obtained the required unweighted isoperimetric inequalities for the transformed targets, we finally convert them into the three-set form used in \Cref{prop:lbCs} to derive an $s$-conductance lower bound.

\begin{lemma}[Three-set isoperimetric inequality]
\label{lem:isoperimetryToThreeSet}
Let $\mathcal X$ be a connected $d$-dimensional Riemannian manifold, equipped with its intrinsic Riemannian distance, and let
$\pi$ have a density with respect to the Riemannian volume measure.
Suppose that, for some $\Phi>0$,
\begin{align}
\label{eq:isoper_Minkowski}
    \pi^+(A)\ge
    \Phi\cdot\pi(A)\cdot(1-\pi(A))
\end{align}
for every $A\in\mathcal R_{d-1}(\mathcal X)$.
Then, for every measurable partition
$S_1\sqcup S_2\sqcup S_3=\mathcal X$, with $S_1,S_2\neq\varnothing$,
\begin{align*}
    \pi(S_3)\ge
    \Phi\cdot\mathsf{dist}(S_1,S_2)\cdot
    \pi(S_1)\cdot\pi(S_2),
\end{align*}
where $\mathsf{dist}(S_1,S_2)\coloneqq\inf\{\mathsf{dist}(u,v):u\in S_1,\ v\in S_2\}$.
\end{lemma}

\begin{proof}
Set $\delta\coloneqq\mathsf{dist}(S_1,S_2)$. The claim is immediate
if $\delta=0$. Assume $\delta>0$ and define
\[
f(u)\coloneqq\mathsf{dist}(u,S_1),\qquad\text{and}\qquad S_1^t\coloneqq\{u\in\mathcal X:f(u)<t\},\quad 0<t<\delta.
\]
Then, since $\delta\coloneqq\mathsf{dist}(S_1,S_2)$,
\[
S_1\subseteq S_1^t\subseteq S_1\cup S_3=S_2^\complement,
\]
hence
\begin{align}
\label{eq:three_set_mass_product}
    \pi(S_1^t)\cdot(1-\pi(S_1^t))
    \ge
    \pi(S_1)\cdot\pi(S_2).
\end{align}

The distance function $f$ is $1$-Lipschitz and satisfies
$|\nabla_u f|=1$ almost everywhere on $\{f>0\}$. By the coarea formula,
for $0< a<b<\delta$,
\begin{align*}
    \pi(S_1^b)-\pi(S_1^a)
    =
    \int_a^b
    \int_{f^{-1}(t)}
    \pi(u)\,\mathcal H_{d-1}(du)\,dt,
\end{align*}
where $\mathcal H_{d-1}$ denotes the $(d-1)$-dimensional Hausdorff measure induced by the Riemannian metric on $\mathcal X$. Note that on the rectifiable level sets $f^{-1}(t)$, $\mathcal H_{d-1}$ coincides with the corresponding Riemannian surface measure. Consequently, the function $V(t)\coloneqq\pi(S_1^t)$ is absolutely
continuous on $(0,\delta)$. Moreover, $(S_1^t)^\varepsilon=S_1^{t+\varepsilon}$. For almost every $t\in(0,\delta)$, the
level set $f^{-1}(t)=\partial S_1^t$ is countably
$(d-1)$-rectifiable and
\begin{align*}
    \pi^+(S_1^t)
    =
    V'(t)
    =
    \int_{\partial S_1^t}
    \pi(u)\,\mathcal H_{d-1}(du).
\end{align*}
Thus $S_1^t\in\mathcal R_{d-1}(\mathcal X)$ for almost every
$t\in(0,\delta)$. Applying~\eqref{eq:isoper_Minkowski} and
\eqref{eq:three_set_mass_product} gives
\begin{align}
\label{eq:bound_pi+S1t}
    \pi^+(S_1^t)
    \ge
    \Phi\cdot\pi(S_1)\cdot\pi(S_2)
\end{align}
for almost every $t\in(0,\delta)$. Finally, note that $\{u:0<f(u)<\delta\}\subseteq S_3$, and the coarea formula yields
\begin{align*}
    \pi(S_3)\ge
    \int_{\{0<f<\delta\}}\pi(du)=
    \int_0^\delta
    \int_{f^{-1}(t)}
    \pi(u)\,\mathcal H_{d-1}(du)\,dt=
    \int_0^\delta\pi^+(S_1^t)\,dt.
\end{align*}
Combining this with~\eqref{eq:bound_pi+S1t} proves
\[
\pi(S_3)\ge\delta\cdot\Phi\cdot\pi(S_1)\cdot\pi(S_2),
\]
completing the proof.
\end{proof}

\subsection{Proofs for coupling results}
\label{sec:proof_coupling}

In this section, we establish the $(\delta,\rho)$-coupling estimates used in \Cref{prop:coupling_general,prop:coupling_general_SCP}.

Recall that the SPS proposal $Q(u,dv)$ coincides with the proposal used by the SCPS before the stepping-out mechanism is applied. We use $Q_{\rm SCP}$ to denote the SCPS proposal kernel incorporating the stepping-out mechanism. We indicate the corresponding SPS and SCPS Metropolis-Hastings transition kernels by $T_{\rm SP}$ and $T_{\rm SCP}$, respectively.

We first study the random walk proposal kernel $Q$ on the sphere. \Cref{lem:monotonicity_Q} shows that its density decreases with the geodesic distance from the starting point. This property is used in the proof of \Cref{prop:coupling_uniform} to determine a $(\delta,\rho)$-coupling for the proposal kernel. 

In \Cref{lem:TV_T_Q_acceptance}, we relate the total variation distance between a Metropolis-Hastings transition kernel $T$ and its proposal kernel $Q$ to the Metropolis-Hastings acceptance probability. This is then used in \Cref{lem:couplingTQ_assumptions_g'} to bound the total variation distance between $T$ and $Q$ under suitable bounds of the derivative of the log-density along geodesics, when $Q$ is a random walk kernel on $\mathcal{S}^d$. We verify this derivative bound for the SPS in \Cref{lem:couplingTQ_SPS_bound_g'} and establish the corresponding result for the SCPS in \Cref{lem:couplingTQ_SCPS_bound_g'}.

We begin with the monotonicity property of the random walk proposal density $Q(u,v)$.

\begin{lemma}[Monotonicity of the proposal density]
\label{lem:monotonicity_Q}
    Let $Q(u,dw)$ be the random-walk proposal on the sphere at $u\in\mathcal{S}^d$ with step size $h$. With an abuse of notation, let $Q(u,w)$ denote its density with respect to the Riemannian volume measure on $\mathcal{S}^d$. If $h\le(d+1)^{-\frac{1}{2}}$, then for every $u,v,s,t\in\mathcal{S}^d$ such that $\max\{\mathsf{dist}(u,v), \mathsf{dist}(s,t)\}<\frac{\pi}{2}$,
    \begin{align*}
        Q(u,v)\ge Q(s,t)\iff\mathsf{dist}(u,v)\le\mathsf{dist}(s,t).
    \end{align*}
\end{lemma}

\begin{proof}
Recall that the stochastic kernel $Q(u,\cdot)$ is defined on the sphere of unit radius centered at $e_{d+1}=(0,\dots,0,1)\in\mathbb{R}^{d+1}$. For simplicity, we redefine $Q(u,\cdot)$ on the unit sphere centered at the origin, $\mathbb{S}^d$, applying a shift by $-e_{d+1}$. The monotonicity result automatically translates to the original definition of $Q(u,\cdot)$.

Let $H$ denote the hyperplane tangent to the sphere $\mathbb{S}^d$ at $u$, which we can identify with $\mathbb{R}^d$.

For all $v$ such that $\mathsf{dist}(u,v)\ge\frac{\pi}{2}$, the raw proposal density satisfies $Q(u,v)=0$. Conversely, $Q(u,v)=0$ implies that $\mathsf{dist}(u,v)\ge\frac{\pi}{2}$.

If, instead, $v$ satisfies $\mathsf{dist}(u,v)<\frac{\pi}{2}$, proposing $v$ from $u$ with proposal probability $Q(u,\cdot)$ corresponds to 
    \begin{enumerate}
        \item proposing $w:=\frac{v}{v^Tu}-u\in H$ from a centered multivariate normal distribution on $H$ with covariance matrix $h^2\, I_d$,
        
        \item and then normalizing $u+w$, i.e., letting $v=\frac{u+w}{|u+w|}$.
    \end{enumerate}
    The probability density of proposing $w$ on $H$ is proportional to $\exp\left(-\frac{|w|^2}{2h^2}\right)$. Besides, by \Cref{lem:det_J_normalization_from_tangent_space}, the Jacobian determinant of the function $f_u:H\to\mathbb{S}^d$, defined as $f_u(w)\coloneqq\frac{u+w}{|u+w|}$ is given by $(1+|w|^2)^{-\frac{d+1}{2}}$. As a consequence, the probability density $Q(u,v)$ for $\mathsf{dist}(u,v)<\frac{\pi}{2}$ satisfies
    \begin{align}
    \label{SPS_density}
        Q(u,v)\propto& (1+|w|^2)^{\frac{d+1}{2}} e^{-\frac{|w|^2}{2h^2}},
    \end{align}
    where $w=\frac{v}{v^Tu}-u$.

    Note that the function $g:[0,+\infty)\to\mathbb{R}$ defined as $g(\omega)=(1+\omega)^{\frac{d+1}{2}}e^{-\frac{\omega}{2h^2}}$ for $\omega\in[0,+\infty)$ is strictly decreasing in $\omega$ if $h\le (d+1)^{-\frac{1}{2}}$.

    Since $|w|=\tan(\mathsf{dist}(u,v))$ and $\tan$ is strictly increasing on $[0,\pi/2)$, it follows that $|w|$ is strictly increasing in $\mathsf{dist}(u,v)$. Therefore, if $h\le(d+1)^{-1/2}$, the density $Q(u,v)$ is strictly decreasing in $\mathsf{dist}(u,v)$ on the open hemisphere $\{v:\mathsf{dist}(u,v)<\frac{\pi}{2}\}$. This proves the stated equivalence whenever both distances are smaller than $\frac{\pi}{2}$.
\end{proof}

\begin{proof}[Proof of \Cref{prop:coupling_uniform}]
As in the proof of \Cref{lem:monotonicity_Q}, it suffices to prove the result on the unit sphere $\mathbb{S}^d$ centered at the origin. After translating the sphere $\mathbb{S}^d$ by $+e_{d+1}$, we obtain the desired result on $\mathcal{S}^d$.

Let $u,v\in\mathbb{S}^d$ and define $\theta\coloneqq\mathsf{dist}(u,v)$. Let $H_u=T_u\mathbb{S}^d$ be the tangent space to $\mathbb{S}^d$ at $u$. A proposal $W\sim Q(u,\cdot)$ can be generated as $W=\frac{u+Z}{|u+Z|}$, where $Z\sim\mathcal N(0,h^2I_d)$ on $H_u$.

If $\theta=0$, the result is immediate. Assume $\theta\in(0,\pi)$ and define
\begin{align*}
    \tilde e_1 \coloneqq \frac{v-\langle u,v\rangle u}{|v-\langle u,v\rangle u|}\in H_u.
\end{align*}
Then
\[
v=\cos(\theta)\,u+\sin(\theta)\,\tilde e_1.
\]
Writing $Z_1=\langle Z,\tilde e_1\rangle$, a direct calculation gives
\begin{align*}
    \mathsf{dist}(u,W)>\mathsf{dist}(v,W)\qquad\iff\qquad Z_1>\tan\left(\frac{\theta}{2}\right).
\end{align*}

By \Cref{lem:monotonicity_Q}, under the assumption $h\le(d+1)^{-1/2}$, the proposal density is decreasing as a function of the geodesic distance from its starting point. Therefore, on the set
\[
\{w\in\mathbb{S}^d:\mathsf{dist}(u,w)>\mathsf{dist}(v,w)\},
\]
we have $Q(u,w)\le Q(v,w)$. Indeed, if $\mathsf{dist}(u,w)<\frac{\pi}{2}$, then also $\mathsf{dist}(v,w)<\frac{\pi}{2}$ and the claim follows from \Cref{lem:monotonicity_Q}. If, instead, $\mathsf{dist}(u,w)\ge\frac{\pi}{2}$, then $Q(u,w)=0$.

By symmetry,
\begin{align*}
    \|Q(u,\cdot)-Q(v,\cdot)\|_{\rm TV}&=1-\int_{\mathbb{S}^d}\min\{Q(u,w),Q(v,w)\}\,dw\\
    &\le1-2\,\mathbb P\left(Z_1>\tan\left(\frac{\theta}{2}\right)\right)=2F\left(\frac{1}{h}\tan\left(\frac{\theta}{2}\right)\right)-1.
\end{align*}
Consequently, $\|Q(u,\cdot)-Q(v,\cdot)\|_{\rm TV}\le1-\rho$ whenever $\theta\le2\arctan\left(F^{-1}\left(1-\frac{\rho}{2}\right)h\right)$, which proves the result.
\end{proof}

The following identity turns a high acceptance probability into a total variation bound between a Metropolis kernel $T$ and its proposal $Q$.

\begin{lemma}[Distance between a Metropolis kernel and its proposal kernel]
\label{lem:TV_T_Q_acceptance}
Let $Q(u,dv)$ be a symmetric proposal kernel satisfying $Q(u,\{u\})=0$, and let $T(u,dv)$ be the corresponding Metropolis--Hastings transition kernel with invariant density $\pi$. Let $V\sim Q(u,\cdot)$. Then
\begin{align}
\label{eq:TV_T_Q_rejection}
    \|T(u,\cdot)-Q(u,\cdot)\|_{\rm TV}=1-\mathbb E\left[\min\left\{1,\frac{\pi(V)}{\pi(u)}\right\}\right].
\end{align}

In particular, if there exist constants $a>0$ and $\eta\in(0,1)$ such that
\begin{align}
\label{eq:event_log_ratio}
    \mathbb P\left(|\log\pi(V)-\log\pi(u)|\le a\right)\ge1-\eta,
\end{align}
then
\begin{align}
\label{eq:TV_T_Q_event_bound}
    \|T(u,\cdot)-Q(u,\cdot)\|_{\rm TV}\le1-e^{-a}(1-\eta).
\end{align}
\end{lemma}

\begin{proof}
The Metropolis--Hastings kernel satisfies
\begin{align*}
    T(u,dv)&=\alpha(u,v)Q(u,dv)+\left(1-\int\alpha(u,z)Q(u,dz)\right)\delta_u(dv),
\end{align*}
where $\alpha(u,v)=\min\left\{1,\frac{\pi(v)}{\pi(u)}\right\}$. Since we assume $Q(u,\{u\})=0$, the total variation distance between $T(u,\cdot)$ and $Q(u,\cdot)$ coincides with the Metropolis-Hastings rejection probability, i.e.,
\begin{align*}
    \|T(u,\cdot)-Q(u,\cdot)\|_{\rm TV}=1-\int\alpha(u,v)Q(u,dv),
\end{align*}
which proves~\eqref{eq:TV_T_Q_rejection}.

On the event $\{|\log\pi(V)-\log\pi(u)|\le a\}$ in~\eqref{eq:event_log_ratio}, we have that $\frac{\pi(V)}{\pi(u)}\ge e^{-a}$. Because this event has probability larger than $(1-\eta)$ by~\eqref{eq:event_log_ratio}, it follows that
\begin{align*}
    \mathbb E\left[\min\left\{1,\frac{\pi(V)}{\pi(u)}\right\}\right]\ge e^{-a}(1-\eta).
\end{align*}
The bound~\eqref{eq:TV_T_Q_event_bound} then follows.
\end{proof}

The next lemma shows that, on a subset $\mathcal A$ of the sphere, if the log-density varies slowly enough along the geodesic $\gamma$ connecting a point $u\in\mathcal A$ to the proposed point $V\sim Q(u,\cdot)$, then the Metropolis-Hastings kernel $T(u,\cdot)$ remains close to the proposal kernel $Q(u,\cdot)$ in total variation.

\begin{lemma}[Closeness of kernels from control of the log-density derivative]
\label{lem:couplingTQ_assumptions_g'}
For $h>0$, let $Q(u,dv)$ be the random walk proposal kernel on the sphere with step size $h$, and let $T(u,dv)$ be the corresponding Metropolis-Hastings kernel with invariant density $\pi$. Let $\mathcal A$ be a measurable subset of the state space. For $u\in\mathcal A$, let $V\sim Q(u,\cdot)$ and define $\Theta\coloneqq\mathsf{dist}(u,V)$, where $\mathsf{dist}$ denotes the geodesic distance. Moreover, let $\gamma:[0,\Theta]\to\mathcal S^d$ be the unit-speed geodesic joining $u$ and $V$, and define $g(t)\coloneqq\log\pi(\gamma(t))$.

Assume that there exist constants $\alpha>0$ and $c_0>0$ such that
\begin{align}
\label{ass:bound_g'}
\sup_{t\in[0,\Theta]}|g'(t)|=\mathcal O_{\mathbb P}\left(d^{\alpha-\frac12}\right)
\end{align}
uniformly over $u\in\mathcal A$ and $0<h\le c_0\cdot d^{-\alpha}$. That is, for every $\eta>0$, there exist $K_g<\infty$ and $d_g\in\mathbb N$ such that
\begin{align}
\label{eq:uniform-prob-bound-g-prime}
\sup_{0<h\le c_0\cdot d^{-\alpha}}\sup_{u\in\mathcal A}\mathbb P\left(\sup_{t\in[0,\Theta]}|g'(t)|>K_g\cdot d^{\alpha-\frac{1}{2}}\right)\le\eta
\end{align}
for all $d\ge d_g$.

Then, there exist constants $\tilde c_h\in(0,c_0]$ and $d_1\in\mathbb N$ such that, for all $d\ge d_1$ and $0<h\le\tilde c_h\cdot d^{-\alpha}$,
\begin{align*}
\|T(u,\cdot)-Q(u,\cdot)\|_{\rm TV}\le\frac{1}{4}
\end{align*}
for every $u\in\mathcal A$.
\end{lemma}

\begin{proof}
For the random walk proposal on the sphere, we have $\Theta=\arctan(h\cdot|Z|)$, where $Z\sim\mathcal N(0,I_d)$, and therefore $\Theta\le h\cdot|Z|$.

Fix $a=\log\frac87$ and $\eta=\frac1{14}$. Choose $K_Z<\infty$ and $d_Z\in\mathbb N$ such that
\begin{align*}
\mathbb P\left(|Z|>K_Z\sqrt d\right)\le\eta
\end{align*}
for all $d\ge d_Z$. By assumption, there exist $K_g<\infty$ and $d_g\in\mathbb N$ such that~\eqref{eq:uniform-prob-bound-g-prime} holds for all $d\ge d_g$.

Define $d_1\coloneqq\max\{d_Z,d_g\}$. Recall the definition of $\gamma$, $\Theta$, and $g$. On the intersection of the events
\begin{align*}
\left\{|Z|\le K_Z\sqrt d\right\}\qquad\text{and}\qquad\left\{\sup_{t\in[0,\Theta]}|g'(t)|\le K_gd^{\alpha-\frac12}\right\},
\end{align*}
we have
\begin{align*}
|\log\pi(V)-\log\pi(u)|\le\Theta\cdot \sup_{t\in[0,\Theta]}|g'(t)|\le h\cdot K_Z\cdot \sqrt d\cdot K_g\cdot d^{\alpha-\frac{1}{2}}=K_Z\cdot K_g\cdot h\cdot d^\alpha.
\end{align*}

Choose $\tilde c_h\coloneqq\min\left\{c_0,\frac{a}{K_ZK_g}\right\}$. Then, for all $d\ge d_1$ and $0<h\le\tilde c_h\cdot d^{-\alpha}$, we have
\begin{align*}
\inf_{u\in\mathcal A}\mathbb P\left(|\log\pi(V)-\log\pi(u)|\le a\right)\ge1-2\eta=\frac67.
\end{align*}
Applying \Cref{lem:TV_T_Q_acceptance}, we obtain, for every $u\in\mathcal A$,
\begin{align*}
\|T(u,\cdot)-Q(u,\cdot)\|_{\rm TV}\le1-e^{-a}\frac67=1-\frac{7}{8}\cdot\frac{6}{7}=\frac{1}{4},
\end{align*}
which proves the result.
\end{proof}

We now verify Assumption~\eqref{ass:bound_g'} on the derivative of the log-density for the SPS on a set away from the north-pole cap $\mathcal{C}$. This will allow us to control the closeness between the SPS transition kernel $T$ and the proposal kernel $Q$.

Recall that $\bar\pi\in\mathcal{P}(\mathbb R^d)$ denotes the pushforward measure $\bar\pi=({\rm SP})_{\#}\pi$ of $\pi\in\mathcal{P}(\mathcal{S}^d)$ under stereographic projection. With an abuse of notation, we denote by $\bar\pi(x)$ and $\pi(u)$ the corresponding densities with respect to the Lebesgue measure on $\mathbb{R}^d$ and the Riemannian volume measure on $\mathcal S^d$, respectively.

\begin{lemma}[Closeness of SPS transition kernel and proposal]
\label{lem:couplingTQ_SPS_bound_g'}
Fix $c\in(0,1]$ and define $\mathcal{C}_{\rm SP}$ as in~\eqref{eq:def_cap_SP}. Assume $R=c_R\cdot \sqrt d$ for a constant $c_R>0$. Suppose that
$\bar\pi$ is twice continuously differentiable and satisfies
\begin{align}
\label{eq:SPS_gradient_assumption}
    \sup_{|x|^2\le
    4\frac{4-c}{c}R^2}
    |\nabla_x\log\bar\pi(x)|
    &=
    \mathcal O(\sqrt d),\\
\label{eq:SPS_Hessian_assumption}
    \sup_{|x|^2\le
    4\frac{4-c}{c}R^2}
    \|\nabla_x^2\log\bar\pi(x)\|_{\rm op}
    &=
    \mathcal O(1).
\end{align}
Then, there exist a constant $\tilde c_h>0$ and an integer $d_0\in\mathbb N$ such that for all $d\ge d_0$, if
$h\le\tilde c_h\cdot d^{-1}$, then
\begin{align}
\label{eq:boundTQ_SPS}
    \|T_{\rm SP}(u,\cdot)-Q(u,\cdot)\|_{\rm TV}
    \le
    \frac14
\end{align}
for every $u\in\mathcal S^d\setminus\mathcal C_{\rm SP}$.
\end{lemma}

\begin{proof}
Fix $u\in\mathcal S^d\setminus\mathcal C$. Let
$V\sim Q(u,\cdot)$ and define $\Theta\coloneqq\mathsf{dist}(u,V)$.
Let $U$ denote the random unit vector in the tangent space $T_u\mathcal S^d$ that determines the direction of the random walk proposal $Q$. More precisely, if $Z\sim\mathcal N(0,I_d)$ is the Gaussian vector used to generate the proposal, then $U$ is the corresponding normalized tangent direction. Since the sphere is centered at
$e_{d+1}$, the unit-speed geodesic from $u$ in direction $U$ is
\begin{align}
\label{eq:unit_speed_geodesic}
    \gamma(t)=e_{d+1}+\cos(t)(u-e_{d+1})+\sin(t)\,U,\qquad t\in[0,\Theta].
\end{align}
Its derivative is
\begin{align}
\label{eq:unit_speed_geodesic_derivative}
    \gamma'(t)=-\sin(t)(u-e_{d+1})+\cos(t)U,
\end{align}
hence $|\gamma'(t)|=1$.

Fix an arbitrary constant $c_0>0$ and consider step sizes satisfying $0<h\le c_0\cdot d^{-1}$. Observe that we can write
\begin{align}
\label{eq:Theta_SPS}
    \Theta=\arctan(h\cdot |Z|),
\end{align}
where $Z\sim\mathcal N(0,I_d)$ is independent of $u$. Therefore, uniformly in the starting point $u\in\mathcal{S}^d\setminus\mathcal{C}$ and $0<h\le c_0\cdot d^{-1}$,
\begin{align}
    \label{eq:bound_Theta}
    \Theta=\mathcal{O}_{\mathbb P}(h\sqrt d)=\mathcal{O}_{\mathbb P}(d^{-\frac{1}{2}}).
\end{align}

This implies that the probability of the event $\{\Theta\le \frac{c}{2}\}$ tends to one as $d\to\infty$, uniformly over $u\in\mathcal S^d\setminus\mathcal C$ and $0<h\le c_0\cdot d^{-1}$. In what follows, we derive the required bounds on the event $\{\Theta\le\frac{c}{2}\}$. Since the probability of its complement tends to zero uniformly over $u\in\mathcal{S}^d\setminus\mathcal{C}$ and $0<h\le c_0\cdot d^{-1}$, the resulting bounds imply the corresponding uniform $\mathcal{O}_{\mathbb{P}}$ estimates. 

We first show that, on the event $\{\Theta\le \frac{c}{2}\}$, the whole geodesic segment remains a fixed distance away from the north pole. Since
$|\gamma'(t)_{d+1}|\le1$,
\[
    \gamma(t)_{d+1}
    \le
    u_{d+1}+t
    \le
    u_{d+1}+\Theta.
\]
Because $u_{d+1}\le2-c$, on the event $\{\Theta\le \frac{c}{2}\}$, we have $\sup_{t\in[0,\Theta]}\gamma(t)_{d+1}\le2-\frac c2$. Using~\eqref{eq:relate_u_d+1_|x|_SP}, for
$x_t\coloneqq{\rm SP}(\gamma(t))$ this implies
\begin{align}
\label{eq:SPS_buffer_ball}
    \sup_{t\in[0,\Theta]}|x_t|^2
    \le
    4\frac{4-c}{c}R^2.
\end{align}

We next control the displacement $|x_t-x_0|$ in $\mathbb{R}^d$.
By \Cref{lem:singular_values_SP}, the differential
$D({\rm SP})_z:T_z\mathcal S^d\to\mathbb R^d$ of the stereographic projection satisfies $\|D({\rm SP})_z\|_{\rm op}=\frac{4R^2+|{\rm SP}(z)|^2}{4R}$.
Hence, using $|\gamma'(t)|=1$,
\begin{align*}
    |x_t-x_0|
    =
    \left|
    \int_0^t
    D({\rm SP})_{\gamma(s)}[\gamma'(s)]\,ds
    \right|\le
    \int_0^t
    \frac{4R^2+|x_s|^2}{4R}\,ds.
\end{align*}
On the event $\{\Theta\le \frac{c}{2}\}$,~\eqref{eq:SPS_buffer_ball} gives $\frac{4R^2+|x_s|^2}{4R}\le\frac{4}{c}R$,
and therefore
\begin{align}
\label{eq:SPS_displacement_x}
    \sup_{t\in[0,\Theta]}|x_t-x_0|
    \le
    \frac4cR\cdot \Theta
    =
    \mathcal O_{\mathbb P}(1),
\end{align}
where in the last equality we used the scaling $R=c_R\sqrt{d}$ and~\eqref{eq:bound_Theta}.

We now bound the directional derivative of the Euclidean log-density along the random geodesic. Let $\gamma'(t)_{1:d}\in\mathbb R^d$ denote the vector formed by the first $d$ components of $\gamma'(t)\in\mathbb R^{d+1}$, and decompose
\begin{align}
\label{eq:SPS_gradient_decomposition}
\left\langle\nabla\log\bar\pi(x_t),\gamma'(t)_{1:d}\right\rangle=\left\langle\nabla\log\bar\pi(x_0),\gamma'(t)_{1:d}\right\rangle+\left\langle\nabla\log\bar\pi(x_t)-\nabla\log\bar\pi(x_0),\gamma'(t)_{1:d}\right\rangle.
\end{align}

We first bound the first term. By~\eqref{eq:unit_speed_geodesic_derivative},
\begin{align}
\label{eq:bound-grad-gamma-prime}
\left|\left\langle\nabla\log\bar\pi(x_0),\gamma'(t)_{1:d}\right\rangle\right|&\le|\sin t|\,|\nabla\log\bar\pi(x_0)|+\left|\left\langle(\nabla\log\bar\pi(x_0),0),U\right\rangle\right|.
\end{align}
Since $\sup_{t\le\Theta}|\sin t|\le\Theta$, \eqref{eq:bound_Theta} and the gradient bound \eqref{eq:SPS_gradient_assumption} imply
\[
\sup_{t\in[0,\Theta]}|\sin t|\,|\nabla\log\bar\pi(x_0)|=\mathcal O_{\mathbb P}(1).
\]

Given $u$, the vector $\nabla\log\bar\pi(x_0)$ is
deterministic and independent of $U$. Define
\[
W\coloneqq\begin{cases}
\dfrac{(\nabla\log\bar\pi(x_0),0)}{|\nabla\log\bar\pi(x_0)|},& \nabla\log\bar\pi(x_0)\neq0,\\
e_1,& \nabla\log\bar\pi(x_0)=0.
\end{cases}
\]
Then, $W$ is a unit vector independent of $U$ conditional on $u$, and
\[
\left|\left\langle(\nabla\log\bar\pi(x_0),0),U\right\rangle\right|=|\nabla\log\bar\pi(x_0)|\,|\langle W,U\rangle|.
\]
By \Cref{lem:U_scalar_ei},
\[
|\langle W,U\rangle|=\mathcal O_{\mathbb P}(d^{-1/2}).
\]
Together with \eqref{eq:SPS_gradient_assumption} and~\eqref{eq:bound-grad-gamma-prime}, this yields
\begin{align*}
\sup_{t\in[0,\Theta]}\left|\left\langle\nabla\log\bar\pi(x_0),\gamma'(t)_{1:d}\right\rangle\right|=\mathcal O_{\mathbb P}(1).
\end{align*}

For the second term in~\eqref{eq:SPS_gradient_decomposition}, since the ball in~\eqref{eq:SPS_buffer_ball} is convex, the mean-value theorem, the assumption~\eqref{eq:SPS_Hessian_assumption}, and the bound~\eqref{eq:SPS_displacement_x} yield
\begin{align*}
\sup_{t\in[0,\Theta]}|\nabla\log\bar\pi(x_t)-\nabla\log\bar\pi(x_0)|&\le\sup_{|x|^2\le4\frac{4-c}{c}R^2}\|\nabla^2\log\bar\pi(x)\|_{\rm op}\cdot\sup_{t\in[0,\Theta]}|x_t-x_0|\\
&=\mathcal O_{\mathbb P}(1).
\end{align*}

Since $|\gamma'(t)_{1:d}|\le1$,~\eqref{eq:SPS_gradient_decomposition} therefore implies
\begin{align}
\label{eq:SPS_directional_gradient}
\sup_{t\in[0,\Theta]}\left|\left\langle\nabla\log\bar\pi(x_t),\gamma'(t)_{1:d}\right\rangle\right|=\mathcal O_{\mathbb P}(1).
\end{align}

Similarly, from~\eqref{eq:unit_speed_geodesic_derivative},
\Cref{lem:U_scalar_ei} with $W=e_{d+1}$, and
\eqref{eq:bound_Theta},
\begin{align}
\label{eq:SPS_vertical_velocity}
\sup_{t\in[0,\Theta]}|\gamma'(t)_{d+1}|=\mathcal O_{\mathbb P}(d^{-1/2}).
\end{align}

Using the expressions for
$\partial_{z_i}\log\pi(z)$ derived in
\cite[Section S8.4]{yang2022stereographic}, with the scale parameter
in that reference replaced by $2R$ to match our definition of stereographic projection\footnote{In \citep{yang2022stereographic}, the unit sphere is centered at the origin, whereas in our definition it is centered in $e_{d+1}$.}, we obtain
\begin{align*}
    g'(t)
    =
    \frac{4R^2+|x_t|^2}{4R}
    \left[
    \left\langle
    \nabla\log\bar\pi(x_t),
    \gamma'(t)_{1:d}
    \right\rangle
    +
    \frac{
    \langle\nabla\log\bar\pi(x_t),x_t\rangle+d
    }{2R}
    \gamma'(t)_{d+1}
    \right],
\end{align*}
where $g(t)=\log\pi(\gamma(t))$.

On the event $\{\Theta\le \frac{c}{2}\}$, since~\eqref{eq:SPS_buffer_ball} holds and $R=c_R\cdot\sqrt{d}$, we have that $|x_t|=\mathcal O(\sqrt d)$ and
\[
    \frac{4R^2+|x_t|^2}{4R}
    =
    \mathcal O(\sqrt d).
\]
Moreover, using the assumption~\eqref{eq:SPS_gradient_assumption},
\[
    |\langle\nabla\log\bar\pi(x_t),x_t\rangle|
    \le
    |\nabla\log\bar\pi(x_t)|\cdot|x_t|
    =
    \mathcal O(d).
\]
Combining these bounds with
\eqref{eq:SPS_directional_gradient} and
\eqref{eq:SPS_vertical_velocity} gives
\begin{align}
\label{eq:SPS_gprime}
    \sup_{t\in[0,\Theta]}|g'(t)|
    =
    \mathcal O_{\mathbb P}(\sqrt d)
\end{align}
uniformly over $u\in\mathcal S^d\setminus\mathcal C$ and $0<h\le c_0\cdot d^{-1}$. Thus,~\eqref{eq:SPS_gprime} verifies Assumption~\eqref{ass:bound_g'} of \Cref{lem:couplingTQ_assumptions_g'} with $\alpha=1$ and $\mathcal{A}=\mathcal{S}^d\setminus\mathcal{C}$.

Applying \Cref{lem:couplingTQ_assumptions_g'} therefore shows that there exist $\tilde c_h\in(0,c_0]$ and $d_0\in\mathbb{N}$ such that, for all $d\ge d_0$ and $h\le\tilde c_h\cdot d^{-1}$,
\begin{align*}
\|T_{\rm SP}(u,\cdot)-Q(u,\cdot)\|_{\rm TV}\le\frac14
\end{align*}
for every $u\in\mathcal{S}^d\setminus\mathcal{C}$, which proves the result.
\end{proof}

For the SCPS, we first derive an expression for the derivative of the transformed log-density along a spherical geodesic.

\begin{lemma}[Derivative of the SCPS transformed log-density]
\label{lem:gprime_SCPS_expression}
Let $\gamma(t)$ and $g(t)$ be defined as in \Cref{lem:couplingTQ_assumptions_g'}, and assume that $\gamma(t)\in\mathcal S_o^d$. Define $x_t\coloneqq {\rm SCP}(\gamma(t))$ and 
\begin{align}
\label{def:B}
    B(x)\coloneqq\nabla_x\log\bar\pi(x)+\frac{1}{S(x)R^2}\left(d\cdot L(x)+\frac{l_o(2-l_o)}{S(x)}\right)x,
\end{align}
where the functions $S(x)$ and $L(x)$ are as in~\eqref{def:SCP_D-S-L}. 

Then,
    \begin{equation}
  \label{eq:g'_SCPS}
  \begin{split}
    g'(t)&=\frac{R}{L(x_t)}\langle B(x_t),\gamma'(t)_{1:d}\rangle + \frac{1}{l_o\cdot L(x_t)}\langle B(x_t), x_t\rangle \gamma'(t)_{d+1}.
    \end{split}
\end{equation}
\end{lemma}

\begin{proof}
Recall that the densities on $\mathcal{S}_o^d$ and $\mathbb{R}^d$ are related by
$\pi(u)=J_{\rm SCP}(x)\cdot\bar\pi(x)$, where $x={\rm SCP}(u)$. Then,
\begin{align}
\label{eq:log_pi_change_variables}
\log\pi(u)=\log\bar\pi(x)+\log J_{\rm SCP}(x).
\end{align}
Recall that $J_{\rm SCP}(x) =\frac{R^d}{l_o}\frac{S(x)}{L(x)^d}$.
We first compute the gradient of $\log J_{\rm SCP}$. Using the definition of $S(x)$ and $L(x)$ in~\eqref{def:SCP_D-S-L}, and differentiating $L(x)$, we obtain
\begin{align}
\label{eq:SCP_grad_S_L}
    \nabla_x S(x)
    &=
    \frac{l_o\cdot (2-l_o)}{S(x)\cdot R^2}x,
    &
    \nabla_x L(x)
    &=
    -\frac{L(x)^2}{S(x)\cdot R^2}x.
\end{align}
Therefore,
\begin{align*}
\nabla_x\log J_{\rm SCP}(x)=-d\frac{\nabla_x L(x)}{L(x)}+\frac{\nabla_x S(x)}{S(x)}=\frac{1}{S(x)R^2}\left(d\cdot L(x)+\frac{l_o(2-l_o)}{S(x)}\right)x.
\end{align*}

Combining this with \eqref{eq:log_pi_change_variables} yields
\begin{align*}
\nabla_x\bigl[\log\pi({\rm SCP}^{-1}(x))\bigr]=\nabla_x\log\bar\pi(x)+\frac{1}{S(x)R^2}\left(d\cdot L(x)+\frac{l_o(2-l_o)}{S(x)}\right)x=B(x).
\end{align*}

It remains to differentiate $x_t={\rm SCP}(\gamma(t))$. From $x_j=R\cdot l_o\frac{u_j}{l_o-u_{d+1}}$, and using $l_o-u_{d+1}=l_o\cdot L(x)$, we obtain
\begin{align*}
\frac{\partial x_j}{\partial u_i}&=\frac{R}{L(x)}\delta_{ij},\qquad i=1,\ldots,d,\\
\frac{\partial x_j}{\partial u_{d+1}}&=\frac{R \cdot l_o\cdot u_j}{(l_o-u_{d+1})^2}=\frac{x_j}{l_o\cdot L(x)}.
\end{align*}

Consequently,
\begin{align*}
x_t'=
D\,{\rm SCP}_{\gamma(t)}[\gamma'(t)]=
\frac{R}{L(x_t)}\gamma'(t)_{1:d}
+
\frac{x_t}{l_o\cdot L(x_t)}\gamma'(t)_{d+1}.
\end{align*}

Finally, since
\[
g(t)
=
\log\pi(\gamma(t))
=
\log\bar\pi(x_t)+\log J_{\rm SCP}(x_t),
\]
the chain rule and the definition of $B$ give
\begin{align*}
g'(t)=
\left\langle B(x_t),x_t'\right\rangle=
\frac{R}{L(x_t)}
\left\langle B(x_t),\gamma'(t)_{1:d}\right\rangle
+
\frac{1}{l_o\cdot L(x_t)}
\left\langle B(x_t),x_t\right\rangle
\gamma'(t)_{d+1},
\end{align*}
which proves~\eqref{eq:g'_SCPS}.
\end{proof}

We can now combine the preceding derivative formula with the stepping-out construction to obtain the SCPS analogue of \Cref{lem:couplingTQ_SPS_bound_g'}.

Recall that the SCPS proposal $Q_{\rm SCP}$ is constructed from the same random walk proposal $Q$ used in the SPS. If the proposal generated by $Q$ falls outside of the sphere's bright side $\mathcal{S}_o^d$, that is, on the dark side $\mathcal{S}^d\setminus\mathcal{S}_o^d$, a stepping-out mechanism is applied. As shown in~\cite[Section~2.2 and Appendix~A.2]{grazzi2026sub}, the stepping-out construction preserves the symmetry of the proposal, so that $Q_{\rm SCP}$ is symmetric. Note that our following result compares the SCP transition kernel $T_{\rm SCP}$ with the raw proposal $Q$, rather then with $Q_{\rm SCP}$, where $T_{\rm SCP}$ is considered a kernel on the whole sphere $\mathcal{S}^d$: with an abuse of notation, for points $u\in\mathcal{S}^d_o$, $T_{\rm SCP}(u,A)=T_{\rm SCP}(u,A\cap \mathcal{S}^d_o)$, whereas for points $u$ on the dark side $\mathcal{S}^d\setminus \mathcal{S}$, the definition of $T(u,\cdot)$ is arbitrary).

\begin{lemma}[Closeness of SCPS transition kernel and proposal]
\label{lem:couplingTQ_SCPS_bound_g'}
Assume that $l_o\in[1,2)$ is constant in $d$. Fix
$c\in(0,l_o)$ and define $\mathcal{C}_{\rm SCP}$ as in~\eqref{eq:def_cap_SCP}.
Assume $R=c_R\cdot d^\beta$ for constants $\beta\ge0$ and $c_R>0$, and set
\begin{align}
\label{eq:def_c1}
    c_1
    \coloneqq
    \frac{4\cdot l_o^2\cdot c_R^2}{c^2}.
\end{align}
Suppose that $\bar\pi$ is twice continuously differentiable and
satisfies
\begin{align}
\label{eq:SCPS_gradient}
    \sup_{|x|^2\le c_1\cdot d^{2\beta}}
    |\nabla_x\log\bar\pi(x)|
    &=
    \mathcal O(d),\\
\label{eq:SCPS_Hessian}
    \sup_{|x|^2\le c_1\cdot d^{2\beta}}
    \|\nabla_x^2\log\bar\pi(x)\|_{\rm op}
    &=
    \mathcal O(d).
\end{align}

Let $Q$ denote the raw proposal kernel on the sphere $\mathcal{S}^d$, $Q_{\rm SCP}$ the proposal kernel implementing the SCPS stepping-out mechanism, and $T_{\rm SCP}$ the corresponding SCPS Metropolis transition kernel.

Then, there exist a constant
$\tilde c_h>0$ and an integer $d_0\in\mathbb N$ such that, for all $d\ge d_0$,
if $h\le\tilde c_h\cdot d^{-(\beta+1)}$, then
\begin{align}
\label{eq:boundTQ_SCPS}
    \|T_{\rm SCP}(u,\cdot)-Q(u,\cdot)\|_{\rm TV}
    \le
    \frac{1}{4}
\end{align}
for every $u\in\mathcal S_o^d\setminus\mathcal C_{\rm SCP}$.
\end{lemma}

\begin{proof}
Fix $u\in\mathcal S_o^d\setminus\mathcal C_{\rm SCP}$, generate one raw proposal $V_0\sim Q(u,\cdot)$, and define $\Theta
    \coloneqq
    \mathsf{dist}(u,V_0)$. Fix a constant $c_{h,0}>0$ and throughout the proof consider step sizes that satisfy $0<h\le c_{h,0}\cdot d^{-(\beta+1)}$. All the $\mathcal O_{\mathbb P}$ bounds in the proof are uniform over this range of $h$.
        
    As in~\eqref{eq:Theta_SPS},
\begin{align}
\label{eq:Theta_SCPS}
    \Theta=\arctan(h\cdot |Z|)=\mathcal O_{\mathbb P}(h\cdot \sqrt d)=\mathcal O_{\mathbb P}\left(d^{-(\beta+\frac12)}\right),
\end{align}
uniformly in the starting point $u$.

We first compare the raw and stepped-out proposal kernels, $Q$ and $Q_{\rm SCP}$. Given a raw proposal $V_0\sim Q(u,\cdot)$, the corresponding SCP proposal $V_{\rm SCP}\sim Q_{\rm SCP}(u,\cdot)$ is defined as
\[V_{\rm SCP}=\begin{cases}
V_0,& V_0\in\mathcal S_o^d,\\
{\rm StepOut}_u(V_0),& V_0\notin\mathcal S_o^d,
\end{cases}
\]
where ${\rm StepOut}_u(\cdot)$ indicates the stepping-out mechanism. Since the two proposals coincide whenever $V_0\in \mathcal S_o^d$, then, by the
coupling inequality,
\begin{align*}
    \|Q_{\rm SCP}(u,\cdot)-Q(u,\cdot)\|_{\rm TV}
    \le
    \mathbb{P}(V_0\notin\mathcal S_o^d).
\end{align*}
Recalling that $u_{d+1}\le l_o-c$, the event
$V_0\notin\mathcal S_o^d$ implies
$(V_0)_{d+1}\ge l_o$, and therefore $|V_0-u|\ge c$. The spherical
geodesic distance dominates the Euclidean distance, therefore $|V_0-u|\ge c$ implies $\Theta\ge c$. Consequently,

\begin{align}
\label{eq:stepout_probability}
\sup_{0<h\le c_{h,0}\cdot d^{-(\beta+1)}}\sup_{u\in\mathcal{S}_o^d\setminus\mathcal{C}_{\rm SCP}}\mathbb{P}(V_0\notin\mathcal{S}_o^d)\le\sup_{0<h\le c_{h,0}\cdot d^{-(\beta+1)}}\mathbb{P}(\Theta\ge c)=o(1).
\end{align}

We next control the Metropolis acceptance probability. Let
$\gamma$ be the geodesic connecting $u$ and $V_0$, as defined in~\eqref{eq:unit_speed_geodesic}, and set
$x_t={\rm SCP}(\gamma(t))$. Since $|\gamma'(t)_{d+1}|\le1$, we have that $ \gamma(t)_{d+1}
    \le
    u_{d+1}+t
    \le
    u_{d+1}+\Theta$. Thus, on the event $\{\Theta\le c/2\}$,
\begin{align*}
    \sup_{t\in[0,\Theta]}
    \gamma(t)_{d+1}
    \le
    l_o-\frac c2.
\end{align*}
In particular, the complete raw geodesic remains on the bright side $\mathcal{S}^d_o$,
meaning that $V_{\rm SCP}=V_0$ on this event ($\{\Theta\le \frac{c}{2}\}$).

Using~\eqref{eq:relate_u_d+1_norm_x}, together with
$\gamma(t)_{d+1}\cdot (2-\gamma(t)_{d+1})\le1$ and
$l_o-\gamma(t)_{d+1}\ge c/2$, we obtain
\begin{align*}
    \sup_{t\in[0,\Theta]}|x_t|^2
    \le
    \frac{4\cdot l_o^2}{c^2}R^2
    =
    c_1\cdot d^{2\beta}.
\end{align*}

By \Cref{lem:singular_values_inverse_SCP}, $\|D({\rm SCP})_{\gamma(t)}\|_{\rm op}
 =\frac{R S(x_t)}{l_o\cdot L(x_t)}$.
The definitions of $S(x)$ and $L(x)$ in~\eqref{def:SCP_D-S-L} give $\frac{S(x)}{l_o}
 =l_o-1+\frac{2-l_o}{L(x)}$.
Moreover, on the set
$\{\gamma(t)_{d+1}\le l_o-c/2\}$,
 $L(x_t)=\frac{l_o-\gamma(t)_{d+1}}{l_o}
 \ge\frac{c}{2l_o}$.
Since $1\le l_o<2$, it follows that
\begin{align*}
 \|D{\rm SCP}_{\gamma(t)}\|_{\rm op}
 =R\left(
       \frac{l_o-1}{L(x_t)}
       +\frac{2-l_o}{L(x_t)^2}
     \right)\le R\left(
       \frac{2l_o(l_o-1)}{c}
       +\frac{4l_o^2(2-l_o)}{c^2}
     \right)\le l_o\left(\frac{2}{c}+\frac{4}{c^2}\right)R.
\end{align*}
Thus, recalling that
$|\gamma'(t)|=1$, we obtain
\begin{align}
\label{eq:SCP_x_displacement}
    \sup_{t\in[0,\Theta]}|x_t-x_0|
    \le
   l_o\cdot \left(\frac{2}{c}+\frac{4}{c^2}\right)\cdot R\cdot \Theta
    =
    \mathcal O_{\mathbb P}(d^{-1/2}).
\end{align}

Recall that
\[
B(x)=\nabla_x\log\bar\pi(x)+\frac{1}{S(x)R^2}\left(d\cdot L(x)+\frac{l_o(2-l_o)}{S(x)}\right)x.
\]
On the region $|x|^2\le c_1\cdot d^{2\beta}$, the definition of $c_1$ and the identity $R=c_R\cdot d^\beta$ give $\frac{|x|}{R}\le    \frac{2\cdot l_o}{c}$. Hence, by~\eqref{def:SCP_D-S-L},
$L(x)$, $L(x)^{-1}$, $S(x)$, and $S(x)^{-1}$ are uniformly bounded on this region. Moreover, Equation~\eqref{eq:SCP_grad_S_L} gives
\[
    |\nabla_x L(x)|+|\nabla_x S(x)|
    =
    \mathcal O(R^{-1}).
\]

Consider the second term in the definition of $B$, that is,
\[
B_2(x)
\coloneqq
\frac{1}{S(x)R^2}
\left(
d\cdot L(x)+\frac{l_o(2-l_o)}{S(x)}
\right)x.
\]

Using the previous bounds and $|x|=\mathcal O(R)$, we obtain
\[
    |B_2(x)|
    =
    \mathcal O\left(\frac{d}{R}\right),
    \qquad
    \|D(B_2)(x)\|_{\rm op}
    =
    \mathcal O\left(\frac{d}{R^2}\right),
\]

where the second bound follows directly from the product rule and
Equation~\eqref{eq:SCP_grad_S_L}. Since $R=c_R\cdot d^\beta$ with $\beta\ge0$,
both quantities are $\mathcal O(d)$. Therefore,
Assumptions~\eqref{eq:SCPS_gradient} and
\eqref{eq:SCPS_Hessian} imply
\begin{align}
\label{eq:B_DB_norm_SCPS}
\sup_{|x|^2\le c_1\cdot d^{2\beta}}|B(x)|=
\mathcal O(d),\qquad 
\sup_{|x|^2\le c_1\cdot d^{2\beta}}\|DB(x)\|_{\rm op}=
\mathcal O(d).
\end{align}

We now decompose
\[
\langle B(x_t),\gamma'(t)_{1:d}\rangle
=
\langle B(x_0),\gamma'(t)_{1:d}\rangle
+
\langle B(x_t)-B(x_0),\gamma'(t)_{1:d}\rangle.
\]
Conditional on $u$, $B(x_0)$ is fixed and independent of the random
direction $U$. Using~\eqref{eq:unit_speed_geodesic_derivative},
\Cref{lem:U_scalar_ei}, and the bounds~\eqref{eq:Theta_SCPS} and
\eqref{eq:B_DB_norm_SCPS}, we obtain
\begin{align*}
    \sup_{t\in[0,\Theta]}
    |\langle B(x_0),\gamma'(t)_{1:d}\rangle|
    =
    \mathcal O_{\mathbb P}(\sqrt d).
\end{align*}
On the other hand, the bounds~\eqref{eq:B_DB_norm_SCPS} and
\eqref{eq:SCP_x_displacement} yield
\[
    \sup_{t\in[0,\Theta]}|B(x_t)-B(x_0)|
    =
    \mathcal O_{\mathbb P}(\sqrt d).
\]
Therefore,
\begin{align}
\label{eq:B_direction_SCPS_corrected}
    \sup_{t\in[0,\Theta]}
    |\langle B(x_t),\gamma'(t)_{1:d}\rangle|
    =
    \mathcal O_{\mathbb P}(\sqrt d).
\end{align}
Similarly, by \Cref{lem:U_scalar_ei} with $W=e_{d+1}$, almost surely,
\begin{align}
\label{eq:SCPS_vertical_velocity_corrected}
    \sup_{t\in[0,\Theta]}|\gamma'(t)_{d+1}|=\mathcal O_{\mathbb P}(d^{-\frac{1}{2}}).
\end{align}

Equation~\eqref{eq:g'_SCPS}, together with the uniform bounds 
$\frac{R}{L(x_t)}=\mathcal O(d^\beta)$,
$|B(x_t)|=\mathcal O(d)$, and
$|x_t|=\mathcal O(d^\beta)$ on the set $\{\Theta\le \frac{c}{2}\}$, now gives
\begin{align}
\label{eq:gprime_SCPS_corrected}
    \sup_{t\in[0,\Theta]}|g'(t)|
    =
    \mathcal O_{\mathbb P}
    \left(d^{\beta+\frac12}\right)
\end{align}
uniformly over $u\in\mathcal S_o^d\setminus\mathcal{C}_{\rm SCP}$, on the event $\{\Theta\le \frac{c}{2}\}$.

Fix $a=\log\frac{15}{14}$. Choose constants $K_Z,K_g<\infty$ so that, for all sufficiently large
$d$,
\begin{align*}
    \mathbb P(|Z|>K_Z\cdot \sqrt d)\le\frac{1}{32},\quad \text{and}\quad 
    \sup_{0<h\le \frac{c_{h,0}}{d^{\beta+1}}}\sup_{u\in\mathcal S_o^d\setminus\mathcal{C}_{\rm SCP}}
    \mathbb P\left(
    \Theta\le\frac c2,\quad 
    \sup_{t\in[0,\Theta]}|g'(t)|
    >
    K_g\cdot d^{\beta+\frac{1}{2}}
    \right)&\le\frac{1}{32}.
\end{align*}

By~\eqref{eq:Theta_SCPS}, on the event $\{|Z|\le K_Z\sqrt d\}$, $\Theta\le h\cdot K_Z\sqrt d$. Hence, for all sufficiently large $d$ and every $0<h\le\tilde c_h\cdot d^{-(\beta+1)}$, this event implies
$\Theta\le \frac{c}{2}$. Moreover, on the intersection with the event
\begin{align*}
\left\{\sup_{t\in[0,\Theta]}|g'(t)|\le K_g\cdot d^{\beta+\frac{1}{2}}\right\},
\end{align*}
we have
\begin{align*}
|\log\pi(V_0)-\log\pi(u)|\le\Theta\cdot \sup_{t\in[0,\Theta]}|g'(t)|\le K_Z\cdot K_g\cdot h\cdot d^{\beta+1}.
\end{align*}
By choosing $\tilde c_h\le\min\left\{c_{h,0},\frac{a}{K_Z\cdot K_g}\right\}$, we guarantee that, for every $0<h\le\tilde c_h\cdot d^{-(\beta+1)}$,
\begin{align*}
K_Z\cdot K_g\cdot h\cdot d^{\beta+1}\le K_Z\cdot K_g\cdot\tilde c_h\le a.
\end{align*}
Therefore, by the union bound, for all sufficiently large $d$,
\begin{align}
\label{eq:SCPS_good_acceptance_event}
\inf_{u\in\mathcal S_o^d\setminus\mathcal{C}_{\rm SCP}}
\mathbb P\left(\Theta\le\frac c2,\quad|\log\pi(V_0)-\log\pi(u)|\le a\right)
\ge1-\frac{1}{32}-\frac{1}{32}=\frac{15}{16}.
\end{align}

On the event in~\eqref{eq:SCPS_good_acceptance_event}, we have $\Theta\le \frac{c}{2}$, hence $V_{\rm SCP}=V_0$. Therefore,
\begin{align*}
\inf_{u\in\mathcal S_o^d\setminus\mathcal{C}_{\rm SCP}}\mathbb P\left(|\log\pi(V_{\rm SCP})-\log\pi(u)|\le a\right)\ge\frac{15}{16}.
\end{align*}

Since $Q_{\rm SCP}$ is symmetric and satisfies $Q_{\rm SCP}(u,\{u\})=0$~\cite[Section~2.2 and Appendix~A.2]{grazzi2026sub}, and recalling that $a=\log\frac{15}{14}$, \Cref{lem:TV_T_Q_acceptance} with $\eta=\frac{1}{16}$ gives
\begin{align}
\label{eq:T_QSCP_bound}
\|T_{\rm SCP}(u,\cdot)-Q_{\rm SCP}(u,\cdot)\|_{\rm TV}\le
1-e^{-a}\frac{15}{16}=1-\frac{14}{15}\cdot \frac{15}{16}=\frac{1}{8}
\end{align}

Furthermore, by~\eqref{eq:stepout_probability}, for all sufficiently large $d$,
\begin{align}
\label{eq:QSCP_Q0_small}
\|Q_{\rm SCP}(u,\cdot)-Q(u,\cdot)\|_{\rm TV}\le\frac18
\end{align}
uniformly over $u\in\mathcal{S}_o^d\setminus\mathcal{C}_{\rm SCP}$. Combining~\eqref{eq:T_QSCP_bound} and~\eqref{eq:QSCP_Q0_small}, the triangle inequality gives
\begin{align*}
\|T_{\rm SCP}(u,\cdot)-Q(u,\cdot)\|_{\rm TV}\le\|T_{\rm SCP}(u,\cdot)-Q_{\rm SCP}(u,\cdot)\|_{\rm TV}+\|Q_{\rm SCP}(u,\cdot)-Q(u,\cdot)\|_{\rm TV}\le\frac{1}{8}+\frac{1}{8}=\frac{1}{4},
\end{align*}
which proves~\eqref{eq:boundTQ_SCPS}.

\end{proof}
\subsection{Proofs for tail bounds}
\label{sec:proofs-tail-bounds}

Here, we prove the tail bounds for $\kappa$-concave distributions stated in \Cref{lem:tail_bound_kappa_concave_light_tail,lem:tail_bound_kappa_concave_kappa_const}.

\begin{proof}[Proof of \Cref{lem:tail_bound_kappa_concave_light_tail}]
Set $Y=\Sigma_d^{-1/2}X$. Then $Y$ is centered, isotropic, and $\kappa$-concave. If $\kappa<0$, define $r\coloneqq-\frac{1}{\kappa}$. Then $r\ge\frac{d}{c_1}$. 
Hence, for all sufficiently large $d$, $r\ge2\sqrt d$. Moreover, since $\frac{r}{\sqrt d}\ge\frac{\sqrt d}{c_1}\to\infty$ as $d\to\infty$, the upper endpoint in the range of~\cite[Corollary~5.4]{adamczak2012moment} also diverges as $d\to\infty$. Therefore, we may fix any $t_0$ larger than the universal lower threshold in that corollary, and for all sufficiently large $d$,
\begin{align}\label{eq:5.6Adamczak}\mathbb{P}(|Y|>t_0\sqrt d)\le e^{-c_0t_0\sqrt d},\end{align}
where $c_0>0$ is a universal constant. 

The case $\kappa=0$ follows from the remark following~\cite[Corollary~5.4]{adamczak2012moment}.

Since $|X|\le\sqrt{\lambda_U }\,|Y|$, choosing $\ell=\max\{2,\sqrt{\lambda_U }\,t_0\}$ gives
\begin{align*}\mathbb{P}(|X|>\ell\sqrt d)\le\mathbb{P}(|Y|>t_0\sqrt d)\le e^{-c_0t_0\sqrt d}.\end{align*}
Thus,~\eqref{eq:exponential_tail_bound} holds with $c_2=c_0t_0$.
\end{proof}

\begin{proof}[Proof of \Cref{lem:tail_bound_kappa_concave_kappa_const}]
The bound~\eqref{eq:tail_bound_kappa_concave_general} follows directly from  \cite[Theorem 1.1]{bobkov2007large} with $f(x)=|x|$, using the fact that $\delta_f(\varepsilon)\le2\varepsilon$ for $f(x)=|x|$.

For any fixed $b>0$, choosing $\ell=2d^{b|\kappa|}$ gives
\begin{align*}C_\kappa\left(\frac{\ell}{2}\right)^{-1/|\kappa|}=C_\kappa d^{-b},\end{align*}
and therefore
\begin{align*}
\mathbb{P}\left(|X|^2>4m^2d^{2b|\kappa|}\right)\le C_\kappa d^{-b}.\end{align*}
\end{proof}
\subsection{Extension of the results to SCPS}
\label{sec:proofs-extension-SCPS}

\begin{proof}[Proof of \Cref{rmk:vanishing-kappa-SCPS}]
Assume the hypotheses of
\Cref{cor:SPS_mixing_kappa_concave}, fix
$l_o\in[1,2)$ independently of $d$, and set
$R=c_R\sqrt d$.
We follow the proof of that theorem, using the
Sub-Cauchy projection in place of stereographic projection.

First, the covariance assumption gives
$\mathbb E[|X|^2]=\mathcal O(d)$, so
$r=\mathcal O(\sqrt d)$ and $w_{\rm SCP}$ is integrable.
For all sufficiently large $d$, the target is
$(-1)$-concave by monotonicity of the concavity parameter,
including when $\kappa=0$.
Since $R\asymp\sqrt d$, \Cref{lem:weightedIsoperK_SCP,lem:weightedToUnweighted_SCP,lem:isoperimetryToThreeSet} give the required
three-set isoperimetric inequality with $\Phi\asymp1$.

Next, choose a sufficiently small fixed $c\in(0,l_o)$
and define $\mathcal C_{\rm SCP}$ as in~\eqref{eq:def_cap_SCP}. By the cap--tail relation~\eqref{eq:cap_tail_probability_SCP}
and \Cref{lem:tail_bound_kappa_concave_light_tail},
this choice ensures that
\[
\pi(\mathcal C_{\rm SCP})\le e^{-c_2\sqrt d}
\]
for some $c_2>0$ independent of $d$ and all
sufficiently large $d$.

It remains to establish coupling at step size
$h=c_hd^{-1}$.
Repeat the proof of \Cref{lem:couplingTQ_SCPS_bound_g'}
using~\eqref{eq:grad-Hess-bounds-SPS-results}, which applies
on the ball $\{|x|\le C\sqrt d\}$ for a fixed
$C>0$, since $R\asymp\sqrt d$.
For the vector field $B(x)$ defined in~\eqref{def:B},
these bounds give
\[
\sup_{|x|\le C\sqrt d}|B(x)|=\mathcal O(\sqrt d),
\qquad
\sup_{|x|\le C\sqrt d}\|DB(x)\|_{\rm op}=\mathcal O(1).
\]
The same argument then bounds the derivative of
the transformed log-density along the proposal
geodesic by $\mathcal O_{\mathbb P}(\sqrt d)$,
uniformly over starting points outside
$\mathcal C_{\rm SCP}$.
The proposal angle is
$\mathcal O_{\mathbb P}(h\sqrt d)$, and the
stepping-out probability tends to zero.
Consequently, the acceptance argument and
\Cref{prop:coupling_uniform} give,
for sufficiently small $c_h>0$,
$(\delta,\rho)$-coupling with
\[
\delta=2\arctan(0.3184\,h)\asymp d^{-1},
\qquad
\rho=\frac14.
\]
The constant $c_h$ may depend on $l_o$.

Finally, choose $s=\varepsilon^2/(4M^2)$.
Since $M=\mathcal O(d^q)$, the exponential bound
on $\pi(\mathcal C_{\rm SCP})$ verifies
Assumption~\ref{ass:cap-bound} exactly as in the
proof of \Cref{cor:SPS_mixing_kappa_concave}.
\Cref{prop:lbCs} and
\Cref{lem:nonlazy_mixing}, together with
preservation of total variation under projection,
therefore yield
\[
n_{\rm mix}(\varepsilon,\bar\pi_0)
=\mathcal O(d^2\log d).
\]

To prove that \Cref{cor:SPS_mixing_CD} also extends to the SCPS sampler, we assume that the additional radial symmetry and curvature condition hold. By \Cref{prop:sharper_isoperimetry}, the weighted isoperimetric inequality for the Sub-Cauchy projection lead to a Cheeger constant $\Phi=\Omega(\sqrt d)$. The coupling argument above remains unchanged with $\delta\asymp d^{-1}$ and $\rho=\frac{1}{4}$. Moreover, the same exponential bound on $\pi(\mathcal C_{\rm SCP})$ verifies Assumption~\ref{ass:cap-bound}, since $\delta\Phi=\Omega(d^{-1/2})$ and $M=\mathcal O(d^q)$. Applying \Cref{prop:lbCs} and \Cref{lem:nonlazy_mixing} as above therefore yields
\[
n_{\rm mix}(\varepsilon,\bar\pi_0)=\mathcal O(d\log d).
\]
This proves the extension of both \Cref{cor:SPS_mixing_kappa_concave,cor:SPS_mixing_CD} to SCPS.
\end{proof}
\subsection{Technical results}
\label{sec:technical-results}

This section contains technical lemmas that are used to prove the results in the previous sections.

\subsubsection{Computation of differentials}

The following is an auxiliary result for the subsequent \Cref{lem:det_J_normalization_from_tangent_space}.
\begin{lemma}[Differential of the radial projection onto the sphere]
\label{lem:dn_along_eta}
    Let $\mathsf{n}:\mathbb{R}^{d+1}\setminus\{0\}\to \mathbb{S}^d $, $\mathsf{n}(x)=\tfrac{x}{|x|}$. The derivative of $\mathsf{n}$ at $x$ along a direction $\eta\in\mathbb{R}^{d+1}$ is given by
    \begin{align}
    \label{dn_along_eta}
        \partial_\eta \mathsf{n}(x)=\frac{1}{|x|}\left(\eta-\frac{\eta^Tx}{|x|^2}x\right).
    \end{align}
\end{lemma}

\begin{proof}
    The direction $\eta$ can be decomposed as $\eta=\eta_x \,x + w$, with $\eta_x=\frac{1}{|x|^2}\eta^Tx$, and $w = \eta - \eta_x\,x$. Note that $w^Tx = 0$.

    By linearity of the directional derivative, 
    \begin{align}
    \label{decompose_dn}
        \partial_\eta \mathsf{n}(x)=\eta_x\,\partial_x\mathsf{n}(x)+\partial_w \mathsf{n}(x).
    \end{align}

    To compute $\partial_x\mathsf{n}(x)$, observe that for $t\in \mathbb{R}$, $$\mathsf{n}(x+tx) = \frac{x+tx}{|x+tx|}=\frac{1+t}{|1+t|}\frac{x}{|x|}=\frac{(1+t)}{|1+t|}\mathsf{n}(x).$$ Therefore, $\partial_x\mathsf{n}(x) = \lim_{t\to 0 }\left[\mathsf{n}(x+tx)-\mathsf{n}(x)\right]/t=0$.

    Because $w^Tx=0$, by  the Pythagorean theorem, for $t\in\mathbb{R}$,
        $\mathsf{n}(x+tw)=\left(x+tw\right)/\sqrt{|x|^2+t^2|w|^2}$, and as $t\to 0$,
    \begin{align*}
        \mathsf{n}(x+tw)=\frac{x+tw}{|x|}+O(t^2).
    \end{align*}
    It follows that 
    \begin{align*}
        \partial_w\mathsf{n}(x) = \lim_{t\to 0 }\frac{\mathsf{n}(x+tw)-\mathsf{n}(x)}{t}=\frac{w}{|x|},
    \end{align*}
    and recalling \eqref{decompose_dn} and the definitions of $w$ and $\eta_x$, we conclude that
    \begin{align*}
        \partial_\eta \mathsf{n}(x)=\eta_x\, 0 + \frac{w}{|x|}=\frac{1}{|x|}\left(\eta-\eta_x\,x\right)=\frac{1}{|x|}\left(\eta - \frac{\eta^Tx}{|x|^2}x\right).
    \end{align*}
\end{proof}

The following lemma is used in \Cref{lem:monotonicity_Q} to prove the monotonicity of the random walk proposal density on the sphere.

\begin{lemma}[Jacobian determinant of the radial projection from the tangent plane onto the sphere]
\label{lem:det_J_normalization_from_tangent_space}
Let $u\in\mathbb{S}^d$ and let $H=\{w\in\mathbb{R}^{d+1}\,:\, w^Tu=0\}\subset\mathbb{R}^{d+1}$ be the tangent space to $\mathbb{S}^d$ at $u$. Define $f_u:H\to\mathbb{S}^d$ by $f_u(w)\coloneqq\frac{u+w}{|u+w|}$.
Then, the Jacobian of $f_u$ with respect to the Euclidean volume measure on $H$ and the Riemannian volume measure on $\mathbb{S}^d$ is given by
\begin{align*}
Jf_u(w)=(1+|w|^2)^{-\frac{d+1}{2}}.
\end{align*}
\end{lemma}

\begin{proof}
Define $\mathsf{n}:\mathbb{R}^{d+1}\setminus\{0\}\to \mathbb{S}^d$ by
$\mathsf{n}(x)\coloneqq\tfrac{x}{|x|}$ and note that
$f_u(w)=\mathsf{n}(u+w)$ for all $w\in H$.
Given a direction $\eta\in H$, the derivative of $f_u$ at $w\in H$ along $\eta$ is given by
\begin{equation}
\label{eq:partial_eta_f_w}
\begin{split}
    \partial_\eta f_u(w)
    &=
    \lim_{t\to0}
    \frac{f_u(w+t\eta)-f_u(w)}{t}\\
    &=
    \lim_{t\to0}
    \frac{\mathsf{n}(u+w+t\eta)-\mathsf{n}(u+w)}{t}
    =
    \partial_\eta \mathsf{n}(u+w)\\
    &\stackrel{(i)}{=}
    \frac{1}{|u+w|}
    \left(
    \eta-\frac{\eta^T(u+w)}{|u+w|^2}(u+w)
    \right)\\
    &\stackrel{(ii)}{=}
    \frac{1}{\sqrt{1+|w|^2}}
    \left(
    \eta-\frac{\eta^T(u+w)}{1+|w|^2}(u+w)
    \right),
\end{split}
\end{equation}
where in $(i)$ we used the result from
\Cref{lem:dn_along_eta}, and in $(ii)$ we used that
$w^Tu=0$ and $|u|=1$.

We first consider the case $w=0$. The identity~\eqref{eq:partial_eta_f_w}
holds for every $w\in H$ and every $\eta\in H$. In particular, taking
$w=0$ gives, for all $\eta\in H$,
\begin{align*}
    \partial_\eta f_u(0)
    =
    \eta-(\eta^Tu)\,u
    =
    \eta,
\end{align*}
where the last equality holds because $\eta\in H$ implies
$\eta^Tu=0$. Hence the differential $Df_u(0):H\to T_u\mathbb S^d$
coincides, under the canonical identification
$T_u\mathbb S^d=H$, with the identity map on $H$. Consequently, for any
orthonormal basis $\{\tilde e_1,\dots,\tilde e_d\}$ of $H$, the matrix
representation of $Df_u(0)$ is the $d\times d$ identity matrix, therefore
\begin{align*}
    Jf_u(0) =1 =(1+|0|^2)^{-\frac{d+1}{2}},
\end{align*}
which proves the claim when $w=0$.

For the remainder of the proof, we may therefore assume $w\neq0$.

Choose an orthonormal basis
$\{\tilde e_1,\dots,\tilde e_d\}$ of $H$ such that $\tilde e_d=\frac{w}{|w|}$.
Observe that $\tilde e_i^Tw=0$ for all $i=1,\dots,d-1$, and, by
definition of $H$, $\tilde e_i^Tu=0$ for all $i=1,\dots,d$. Therefore, by~\eqref{eq:partial_eta_f_w},
\begin{align*}
    \partial_{\tilde e_i}f_u(w)
    =
    \frac{\tilde e_i}{\sqrt{1+|w|^2}},
    \qquad
    i=1,\dots,d-1,
\end{align*}
and
\begin{equation}
\label{partial_ed_fu}
\begin{split}
    \partial_{\tilde e_d}f_u(w)
    &=
    \frac{1}{\sqrt{1+|w|^2}}
    \left(
    \tilde e_d
    -
    \frac{\tilde e_d^Tw}{1+|w|^2}(u+w)
    \right)=
    \frac{1}{\sqrt{1+|w|^2}}
    \left(
    \frac{w}{|w|}
    -
    \frac{|w|}{1+|w|^2}(u+w)
    \right)\\
    &=
    \frac{1}{(1+|w|^2)^{3/2}}
    \frac{1}{|w|}
    \left(
    w-|w|^2u
    \right).
\end{split}
\end{equation}

Because
$\partial_{\tilde e_d}f_u(w)\in\operatorname{span}(u,w)$, we have
\[
    \tilde e_i^T\partial_{\tilde e_d}f_u(w)=0,
    \qquad \text{for }\quad i=1,\dots,d-1.
\]
Moreover, using $w^Tu=0$ and $|u|=1$, Equation~\eqref{partial_ed_fu}
yields
\begin{align*}
    \left|
    \partial_{\tilde e_d}f_u(w)
    \right|^2=
    \frac{|w-|w|^2u|^2}
    {|w|^2(1+|w|^2)^3}=
    \frac{|w|^2+|w|^4}
    {|w|^2(1+|w|^2)^3}=
    \frac{1}{(1+|w|^2)^2}.
\end{align*}
Hence, $\left|\partial_{\tilde e_d}f_u(w)\right|=(1+|w|^2)^{-1}$.

Consequently,
$\{\hat e_1,\dots,\hat e_d\}$, with
\begin{align*}
    \begin{cases}
        \hat e_i=\tilde e_i,
        & \text{for }\quad i=1,\dots,d-1,\\[0.2cm]
        \hat e_d
        =
        \dfrac{\partial_{\tilde e_d}f_u(w)}
        {|\partial_{\tilde e_d}f_u(w)|},
    \end{cases}
\end{align*}
is an orthonormal basis for the tangent space to $\mathbb{S}^d$ at
$f_u(w)$. In this basis,
\begin{align*}
    \partial_{\tilde e_i}f_u(w)=\frac{\hat e_i}{\sqrt{1+|w|^2}},
    \qquad
    i=1,\dots,d-1,
\end{align*}
and
\begin{align*}
    \partial_{\tilde e_d}f_u(w)=
    \left|
    \partial_{\tilde e_d}f_u(w)
    \right|
    \hat e_d=
    (1+|w|^2)^{-1}\hat e_d.
\end{align*}

Thus, the matrix representation of $Df_u(w):H\to T_{f_u(w)}\mathbb{S}^d$
using the orthonormal bases
$\{\tilde e_1,\dots,\tilde e_d\}$ for $H$ and
$\{\hat e_1,\dots,\hat e_d\}$ for
$T_{f_u(w)}\mathbb{S}^d$ is given by
\begin{align*}
    \begin{pmatrix}
        (1+|w|^2)^{-1/2} & 0 & \dots & 0 & 0\\
        0 & (1+|w|^2)^{-1/2} & \dots & 0 & 0\\
        \vdots & \vdots & \ddots & \vdots & \vdots\\
        0 & 0 & \cdots & (1+|w|^2)^{-1/2} & 0\\
        0 & 0 & \dots & 0 & (1+|w|^2)^{-1}
    \end{pmatrix}.
\end{align*}
Therefore, its absolute determinant, equivalently the Jacobian of
$f_u$ with respect to the Euclidean volume measure on $H$ and the
Riemannian volume measure on $\mathbb{S}^d$, is
\begin{align*}
    Jf_u(w)=
    (1+|w|^2)^{-\frac{d-1}{2}}
    (1+|w|^2)^{-1}=
    (1+|w|^2)^{-\frac{d+1}{2}},
\end{align*}
showing the desired result.

\end{proof}

To translate the weighted functional isoperimetric inequalities on $\mathbb{R}^d$ into unweighted inequalities on the sphere $\mathcal{S}^d$ or its subset $\mathcal{S}^d_o$, we need bounds on the differentials of the inverse projection maps.  The following two lemmas provide these bounds, together with the corresponding forward-projection bounds used in the coupling proofs.

\begin{lemma}[Operator norms of the stereographic projection
and its inverse]
\label{lem:singular_values_SP}
Let $R>0$. For every $x\in\mathbb{R}^d$,
\begin{align}
\label{eq:op_norm_SP_inverse}
\|D({\rm SP}^{-1})_x\|_{\rm op}=\frac{4R}{|x|^2+4R^2}.
\end{align}
Moreover, for every
$u\in\mathcal{S}^d\setminus\{2e_{d+1}\}$,
\begin{align}
\label{eq:op_norm_SP}
\|D({\rm SP})_u\|_{\rm op}=\frac{4R^2+|{\rm SP}(u)|^2}{4R}=\frac{2R}{2-u_{d+1}}.
\end{align}
\end{lemma}

\begin{proof}
Recall from~\eqref{eq:SP_inverse} that
\begin{align*}
{\rm SP}^{-1}(x)=\begin{pmatrix}
\dfrac{4R}{|x|^2+4R^2}x\\[0.8em]
\dfrac{2|x|^2}{|x|^2+4R^2}
\end{pmatrix}.
\end{align*}
Set $q(x)=|x|^2+4R^2$. Then, for every $v\in\mathbb{R}^d$,
\begin{align*}
D({\rm SP}^{-1})_x[v]=\begin{pmatrix}
\dfrac{4R}{q(x)}v-\dfrac{8R\langle x,v\rangle}{q(x)^2}x\\[1em]
\dfrac{16R^2\langle x,v\rangle}{q(x)^2}
\end{pmatrix}.
\end{align*}
Therefore, expanding the squared norm of $D({\rm SP}^{-1})_x[v]$ yields
\begin{align*}
\left|D({\rm SP}^{-1})_x[v]\right|^2=\left|\dfrac{4R}{q(x)}v
-\dfrac{8R\langle x,v\rangle}{q(x)^2}x\right|^2
+\frac{256R^4\langle x,v\rangle^2}{q(x)^4}=
\frac{16R^2}{q(x)^2}|v|^2.
\end{align*}
Hence,
\begin{align*}
\left|D({\rm SP}^{-1})_x[v]\right|=\frac{4R}{|x|^2+4R^2}|v|.
\end{align*}

Thus, the differential of the inverse stereographic projection multiplies the length of every vector
by $4R/q(x)$. Since
\[
D({\rm SP})_u=\left(D({\rm SP}^{-1})_x\right)^{-1},\qquad x={\rm SP}(u),
\]
the differential of the forward stereographic projection multiplies lengths by
$q(x)/(4R)$. Hence
\[
\|D({\rm SP})_u\|_{\rm op}=\frac{q(x)}{4R}=\frac{4R^2+|{\rm SP}(u)|^2}{4R}.
\]
This gives the first equality in~\eqref{eq:op_norm_SP}. The second follows from
\[
|{\rm SP}(u)|^2=4R^2\frac{u_{d+1}}{2-u_{d+1}},
\]
as stated in~\eqref{eq:SP_radius_height_relation}.
\end{proof}

\begin{lemma}[Operator norms of the Sub-Cauchy projection and its inverse]
\label{lem:singular_values_inverse_SCP}
Let $R>0$ and $l_o\in[1,2)$. Define $S(x)$ and $L(x)$ as in~\eqref{def:SCP_D-S-L}. For every $x\in\mathbb R^d$,
\begin{align}
\label{eq:op_norm_SCP_inverse}
\|D({\rm SCP}^{-1})_x\|_{\rm op}\le \frac{L(x)}{R},
\end{align}
with equality when $d\ge2$. Moreover, for
$u={\rm SCP}^{-1}(x)$,
\begin{align}
\label{eq:op_norm_SCP}
\|D({\rm SCP})_u\|_{\rm op}=\frac{R\,S(x)}{l_oL(x)}.
\end{align}
\end{lemma}

\begin{proof}
From the definitions of $S(x)$ and $L(x)$ in~\eqref{def:SCP_D-S-L}, it follows that
\[
S(x)^2=l_o^2+l_o(2-l_o)\frac{|x|^2}{R^2},\qquad
L(x)\left(S(x)-l_o(l_o-1)\right)=l_o(2-l_o),\qquad\text{and}\qquad
\nabla L(x)=-\frac{L(x)^2}{S(x)\cdot R^2}x.
\]
Differentiating~\eqref{eq:SCP_inverse}, we obtain
\[
D({\rm SCP}^{-1})_x[v]=
\begin{pmatrix}
\dfrac{L(x)}{R}v-\dfrac{L(x)^2\langle x,v\rangle}{S(x)\cdot R^3}x\\[0.8em]
\dfrac{l_oL(x)^2\langle x,v\rangle}{S(x)\cdot R^2}
\end{pmatrix}.
\]
Using the preceding identities to expand the squared norm of
$D({\rm SCP}^{-1})_x[v]$, we obtain
\begin{align*}
\left|D({\rm SCP}^{-1})_x[v]\right|^2&=\frac{L(x)^2}{R^2}\left(|v|^2+\frac{-2L(x)S(x)+L(x)^2\left(\frac{|x|^2}{R^2}+l_o^2\right)}{R^2S(x)^2}\langle x,v\rangle^2\right)\\
&=\frac{L(x)^2}{R^2}\left(|v|^2-\frac{l_o(2-l_o)}{R^2S(x)^2}\langle x,v\rangle^2\right).
\end{align*}
Since $l_o(2-l_o)\ge0$, the upper bound~\eqref{eq:op_norm_SCP_inverse} follows immediately. 

To identify the smallest singular value of $D({\rm SCP}^{-1})_x$, we use the Cauchy-Schwarz inequality
\[
\langle x,v\rangle^2\le |x|^2|v|^2,
\]
together with the identity
\[
1-\frac{l_o(2-l_o)|x|^2}{R^2S(x)^2}=\frac{l_o^2}{S(x)^2},
\]
which follows from the definition of $S(x)$. This yields
\[
\frac{l_o^2L(x)^2}{R^2S(x)^2}|v|^2\le\left|D({\rm SCP}^{-1})_x[v]\right|^2\le\frac{L(x)^2}{R^2}|v|^2,\]
which shows that the smallest singular value of $D({\rm SCP}^{-1})_x$ satisfies
\[
\sigma_{\min}\left(D({\rm SCP}^{-1})_x\right)
\ge \frac{l_oL(x)}{R\,S(x)}.
\]

When $x\neq0$, the lower bound is attained by vectors parallel to $x$. Hence,
\[
\sigma_{\min}\left(D({\rm SCP}^{-1})_x\right)
=\frac{l_oL(x)}{R\,S(x)}.
\]
If $d\ge2$, the upper bound is attained by any nonzero vector perpendicular to $x$, which gives equality in~\eqref{eq:op_norm_SCP_inverse}.

Since
\[
D({\rm SCP})_u=\left(D({\rm SCP}^{-1})_x\right)^{-1},
\]
we obtain
\[
\|D({\rm SCP})_u\|_{\rm op}=\frac{R\,S(x)}{l_oL(x)},
\]
proving~\eqref{eq:op_norm_SCP}.

At $x=0$, we have $L(0)=1$ and $S(0)=l_o$, and
\[
D({\rm SCP}^{-1})_0[v]=\left(\frac{v}{R},0\right).
\]
Hence
\[
\|D({\rm SCP}^{-1})_0\|_{\rm op}=\frac{1}{R}=\frac{L(0)}{R},\qquad
\|D({\rm SCP})_u\|_{\rm op}=R=\frac{R\,S(0)}{l_oL(0)},
\]
so both conclusions also hold at $x=0$.
\end{proof}

\subsubsection{Other proofs}

We now prove the mixing bound stated in~\eqref{eq:lovasz1993-laziness-removed}, using the $s$-conductance estimate of~\cite[Lemma~1]{ascolani2026scalability}, which builds on results from~\citep{lovasz1993random}. This estimate requires the transition kernel to be reversible and positive semidefinite. While reversibility holds in our setting, positive semidefiniteness is not assumed. As discussed in~\citep{ascolani2026scalability}, the latter property can be ensured by replacing the kernel with its lazy version. However, this would give a mixing bound for the lazy chain rather than the original chain. We instead apply the estimate to the two-step kernel $T^2$, which is reversible and positive semidefinite whenever $T$ is reversible. Since each transition of $T^2$ corresponds to two transitions of $T$, this yields a mixing bound for the original chain. The following lemma also shows that $T^2$ satisfies the same conductance lower bound provided by \Cref{prop:lbCs} for $T$.

\begin{lemma}[Mixing without a laziness assumption]
\label{lem:nonlazy_mixing}
Suppose that the hypotheses of \Cref{prop:lbCs} hold for the Markov kernel $T$ and $s\in(0,\frac{1}{2})$. Then, for every $M$-warm initial distribution $\pi_0$ and every integer $n\ge1$,
\[
\|\pi_0T^n-\pi\|_{\rm TV}\le Ms+M\exp\left(-\frac{C_s(T^2)^2}{2}\left\lfloor\frac{n}{2}\right\rfloor\right),
\]
where
\[
C_s(T^2)\ge\frac{\rho}{4}\min\left\{1,\frac{\delta\Phi}{64}\right\}.
\]
\end{lemma}

\begin{proof}
By contraction of total variation under a Markov kernel,
\[
\|T^2(u,\cdot)-T^2(v,\cdot)\|_{\rm TV}\le\|T(u,\cdot)-T(v,\cdot)\|_{\rm TV}.
\]
Consequently, $T^2$ satisfies the same $(\delta,\rho)$-coupling condition as $T$ on $\mathcal X\setminus\mathcal C$.

Reversibility of $T$ implies that its Markov operator is self-adjoint on $L^2(\pi)$. Hence, $T^2$ is also reversible and, for every $g\in L^2(\pi)$,
\[
\langle g,T^2g\rangle_{L^2(\pi)}=\langle Tg,Tg\rangle_{L^2(\pi)}=\|Tg\|_{L^2(\pi)}^2\ge0.
\]
Thus, $T^2$ is positive semidefinite. The isoperimetric assumption and the bound on $\pi(\mathcal C)$ are unchanged, thus all hypotheses of \Cref{prop:lbCs} hold for $T^2$. Applying that proposition gives
\[
C_s(T^2)\ge\frac{\rho}{4}\min\left\{1,\frac{\delta\Phi}{64}\right\}.
\]

Applying~\cite[Lemma~1]{ascolani2026scalability} to $T^2$ and using $1-x\le e^{-x}$ yields,
for every integer $m\ge0$,
\[
\|\pi_0T^{2m}-\pi\|_{\rm TV}=\|\pi_0(T^2)^m-\pi\|_{\rm TV}\le Ms+M\exp\left(-\frac{C_s(T^2)^2}{2}m\right).
\]
For even $n=2m$, the preceding estimate applies directly. For odd $n=2m+1$, stationarity of $\pi$ and contraction of total variation give
\[
\|\pi_0T^{2m+1}-\pi\|_{\rm TV}=\|(\pi_0T^{2m})T-\pi T\|_{\rm TV}\le\|\pi_0T^{2m}-\pi\|_{\rm TV}.
\]
Hence, in both cases, the claimed bound follows with $m=\lfloor n/2\rfloor$.
\end{proof}

The result below is used in the proof of \Cref{lem:couplingTQ_SPS_bound_g',lem:couplingTQ_SCPS_bound_g'} on the closeness of the SPS and SCPS transition kernels to the proposal distribution.

\begin{lemma}[Concentration of a uniform random point on the sphere along fixed directions]
\label{lem:U_scalar_ei}
    Let $\{\tilde e_i\}_{i=1}^{d+1}$ be an orthonormal basis of $\mathbb{R}^{d+1}$. Define $S$ to be the $(d-1)$-dimensional sphere of unit radius on the Euclidean hyperplane spanned by $\tilde e_1,\dots,\tilde e_d$, that is, $S\coloneqq\{x\in\mathbb{R}^{d+1}\,:\,\sum_{i=1}^d\langle x,\tilde e_i\rangle^2=1,\text{ and }\langle x,\tilde e_{d+1}\rangle=0\}$. Let $U$ be a random vector, uniformly distributed on $S$. Then,
    \begin{align}
    \label{eq:U_scalar_ei}
        \langle U, \tilde e_i\rangle =\mathcal{O}_\mathbb{P}\left(\frac{1}{\sqrt{d}}\right),\qquad  \text{ for }i=1,\dots,d+1.
    \end{align}
    In particular, 
    \begin{align}
    \label{eq:U_scalar_w}
          \langle U, W\rangle =\mathcal{O}_\mathbb{P}\left(\frac{1}{\sqrt{d}}\right),
    \end{align}
    for any random vector $W\in\mathbb{R}^{d+1}$ independent of $U$ such that $|W|=1$ almost surely.
\end{lemma}

\begin{proof}
We start by proving the statement for $i=d+1$. By definition of $U$, the random vector belongs to the set $S$ almost surely, which is a subset of $\text{span}\{\tilde e_1,\dots,\tilde e_{d}\}$. In particular, $\langle U,\tilde e_{d+1}\rangle=0$. Therefore, $\langle U, \tilde e_{d+1}\rangle =\mathcal{O}_\mathbb{P}\left(\frac{1}{\sqrt{d}}\right)$ holds trivially.

Now consider $i=1,\dots,d$. Since $U\in S$ almost surely, then by definition of $S$, $\sum_{i=1}^d\langle U,\tilde e_i\rangle^2=1$. Consequently, $\sum_{i=1}^d\mathbb{E}[\langle U,\tilde e_i\rangle^2]=1$. Given that $U$ is uniformly distributed on $S$, it follows by symmetry that $\mathbb{E}[\langle U,\tilde e_i\rangle^2]=\tfrac{1}{d}$ for all $i=1,\dots,d$. Again by symmetry, $\mathbb{E}[\langle U,\tilde e_i\rangle]=0$. Thus, $\operatorname{Var}(\langle U,\tilde e_i\rangle)=\mathbb{E}[\langle U,\tilde e_i\rangle^2]-\mathbb{E}[\langle U,\tilde e_i\rangle]^2=\tfrac{1}{d}$. By Chebyshev's inequality,~\eqref{eq:U_scalar_ei} follows.

To prove~\eqref{eq:U_scalar_w}, observe that
\begin{align*}
     \langle U, W\rangle=\sum_{i=1}^{d+1}\langle U,\tilde e_i\rangle\cdot\langle W,\tilde e_i\rangle = \sum_{i=1}^{d}\langle U,\tilde e_i\rangle\cdot\langle W,\tilde e_i\rangle,
\end{align*}
since $\langle U,\tilde e_{d+1}\rangle=0$.

Conditionally on $W$, using the independence of $U$ and $W$ and the fact that $\mathbb{E}[\langle U,\tilde e_i\rangle]=0$, we have $\mathbb{E}[\langle U,W\rangle\mid W]=0$. Taking the expectation with respect to $W$ yields $\mathbb{E}[\langle U,W\rangle]=0$. Moreover,
\begin{align*}
\mathbb{E}[\langle U,W\rangle^2\mid W] &= \sum_{i=1}^{d}\sum_{j=1}^{d}\langle W,\tilde e_i\rangle\langle W,\tilde e_j\rangle\mathbb{E}[\langle U,\tilde e_i\rangle\langle U,\tilde e_j\rangle].
\end{align*}
By symmetry, the distribution of $U$ is invariant under sign flips of any coordinate $\tilde e_i$. 
Therefore, for $i\neq j$, $\mathbb{E}[\langle U,\tilde e_i\rangle\langle U,\tilde e_j\rangle]=0$. For $i=j$ we recover $\mathbb{E}[\langle U,\tilde e_i\rangle^2]=\frac{1}{d}$, as proved above. Then,
\begin{align*}
\mathbb{E}[\langle U,w\rangle^2\mid W]=\frac{1}{d}\sum_{i=1}^{d}\langle W,\tilde e_i\rangle^2\le\frac{1}{d}|W|^2=\frac{1}{d}.
\end{align*}

Taking the expectation with respect to $W$ gives $\mathbb{E}[\langle U,W\rangle^2]\le\frac{1}{d}$. Since $\mathbb{E}[\langle U,W\rangle]=0$, it follows that $\operatorname{Var}(\langle U,W\rangle)\le\frac{1}{d}$. Chebyshev's inequality then yields $\langle U,W\rangle=\mathcal{O}_{\mathbb{P}}\left(d^{-\frac{1}{2}}\right)$, concluding the proof.
\end{proof}

\subsection{Proofs for the examples in \Cref{sec:applications-results}}
\label{sec:proofs-examples}

In this section, we prove the mixing-time corollaries for the three examples introduced in \Cref{sec:applications-results}. Since the $\kappa$-concavity of each target has already been established, it remains to verify the additional assumptions of \Cref{cor:SPS_mixing_kappa_concave,cor:SCPS_mixing_kappa_concave}, namely the required covariance or norm-quantile bounds and the corresponding gradient and Hessian bounds. In the moderately heavy-tailed regime, \Cref{rmk:vanishing-kappa-SCPS} extends the SPS mixing-time result to SCPS under the same target assumptions. The curvature-improved Student's $t$ result in \Cref{rmk:student-curvature} follows from the discussion in \Cref{sec:shaper_iso}.

\subsubsection{Proof for Student's $t$ distributions}
\label{sec:proofs-examples-students-t}
\begin{proof}[Proof of \Cref{cor:student-applications}]
The $\kappa$-concavity of the target was established in \Cref{sec:example-student}. It remains to verify the assumptions of the corresponding mixing-time results.

Consider first a centered Student's $t$ distribution with scale matrix $S$ and $\nu$ degrees of freedom. Its log-density satisfies
\[
\nabla\log\bar\pi(x)=-\frac{(d+\nu)S^{-1}x}{\nu+x^\top S^{-1}x}.
\]
Since
\[
|S^{-1}x|\le\lambda_{\min}(S)^{-1/2}\sqrt{x^\top S^{-1}x},
\]
and
\[
\sup_{r\ge0}\frac{\sqrt r}{\nu+r}=\frac{1}{2\sqrt\nu},
\]
we obtain
\begin{equation}
\label{eq:student-gradient-bound}
\sup_x|\nabla\log\bar\pi(x)|\le\frac{d+\nu}{2\sqrt{\nu\lambda_{\min}(S)}}.
\end{equation}
Moreover,
\[
\nabla^2\log\bar\pi(x)=-(d+\nu)\left[\frac{S^{-1}}{\nu+x^\top S^{-1}x}-\frac{2(S^{-1}x)(S^{-1}x)^\top}{(\nu+x^\top S^{-1}x)^2}\right].
\]
Writing $y=S^{-1/2}x$, the matrix inside the brackets can be expressed as
\[
S^{-1/2}\left[\frac{I_d}{\nu+|y|^2}-\frac{2yy^\top}{(\nu+|y|^2)^2}\right]S^{-1/2}.
\]
The operator norm of the middle matrix is bounded by $1/\nu$. Hence,
\begin{equation}
\label{eq:student-hessian-bound}
\sup_x\|\nabla^2\log\bar\pi(x)\|_{\rm op}\le\frac{d+\nu}{\nu\lambda_{\min}(S)}.
\end{equation}

Suppose first that $\nu_d\asymp d$. Under the assumed uniform bounds on the eigenvalues of $S_d$, \eqref{eq:student-gradient-bound} and \eqref{eq:student-hessian-bound} are respectively $\mathcal O(\sqrt d)$ and $\mathcal O(1)$. Moreover, for sufficiently large $d$, the covariance matrix exists and is given by
\[
\operatorname{Cov}(X)=\frac{\nu_d}{\nu_d-2}S_d,
\]
whose eigenvalues are uniformly bounded above and away from zero. The assumptions of \Cref{cor:SPS_mixing_kappa_concave} are therefore satisfied, yielding the stated mixing-time bound for SPS and, by \Cref{rmk:vanishing-kappa-SCPS}, for SCPS.

Now suppose that $\nu>1$ is fixed. Using the representation $X=\sqrt{{\nu}/{V}}\,S_d^{1/2}Z$, where $Z\sim\mathcal N_d(0,I_d)$ and $V\sim\chi^2_\nu$ are independent, we obtain
\[
\mathbb E[|X|]\le\sqrt{\nu}\,\|S_d^{1/2}\|_{\rm op}\,\mathbb E[V^{-1/2}]\,\mathbb E[|Z|]=\mathcal O(\sqrt d),
\]
since $\nu>1$ implies $\mathbb E[V^{-1/2}]<\infty$. Markov's inequality then shows that the $2/3$-quantile of $|X|$ is $\mathcal O(\sqrt d)$. When $\nu$ is fixed, the bounds in \eqref{eq:student-gradient-bound} and \eqref{eq:student-hessian-bound} are both $\mathcal O(d)$. Thus, the assumptions of \Cref{cor:SCPS_mixing_kappa_concave} hold with $\kappa=-1/\nu$, which gives the stated condition on $\beta$ and the corresponding mixing-time bound.
\end{proof}

\subsubsection{Proof for exponential-power regression with an unknown scale}
\label{sec:proofs-examples-exponential-power}

\begin{proof}[Proof of \Cref{cor:quartic-applications}]
The $\kappa$-concavity of the target was established in \Cref{sec:example-quartic}. It remains to verify the quantile and derivative assumptions of \Cref{cor:SCPS_mixing_kappa_concave}.

Recall that $\theta_0=D_0^{-1}y_0$ and that we recenter the distribution by setting $x=\theta-\theta_0$. Define
\begin{equation}
\label{eq:quartic-epsilon-H}
\epsilon=y_1-D_1\theta_0,\qquad H=D_1D_0^{-1}.
\end{equation}
We first show that the $2/3$-quantile of $|x|$ is $\mathcal O(\sqrt d)$.

To establish this quantile bound, we introduce an auxiliary distribution chosen so that, after recentering at $\theta_0$, it reproduces exactly the contribution of the first $d$ residuals to the joint posterior. The contribution of the remaining $k$ residuals is then incorporated through a multiplicative weight. Define the auxiliary distribution by taking
\begin{equation}
\label{eq:quartic-auxiliary-distribution}
\tau\sim\operatorname{Gamma}(a+k/4,b),
\end{equation}
and, independently of $\tau$, letting $Z=(Z_1,\ldots,Z_d)^\top$ have independent coordinates with density
\[
z\mapsto\frac{2}{\Gamma(1/4)}e^{-z^4}.
\]
The shape parameter $a+k/4$ in~\eqref{eq:quartic-auxiliary-distribution} arises from the change of variables
\begin{equation}
\label{eq:quartic-change-variables}
\theta=\theta_0+\tau^{-1/4}D_0^{-1}Z.
\end{equation}
Indeed, since $n=d+k$, the factor $\tau^{a+n/4-1}$ in the joint posterior combines with the Jacobian factor $\tau^{-d/4}$ to give $\tau^{a+k/4-1}$.

Since $D_0\theta_0=y_0$, Equation~\eqref{eq:quartic-change-variables} gives
\[
y_0-D_0\theta=-\tau^{-1/4}Z,
\]
and hence
\begin{equation}
\label{eq:quartic-first-residuals}
\tau\|y_0-D_0\theta\|_4^4=\|Z\|_4^4.
\end{equation}
Thus, the auxiliary distribution accounts exactly for the contribution of the first $d$ residuals. The remaining $k$ residuals contribute the multiplicative weight
\begin{equation}
\label{eq:quartic-weight}
W=\exp\left\{-\tau\|\epsilon-D_1(\theta-\theta_0)\|_4^4\right\}
=\exp\left\{-\|\tau^{1/4}\epsilon-HZ\|_4^4\right\},
\end{equation}
where the second equality follows from~\eqref{eq:quartic-epsilon-H} and~\eqref{eq:quartic-change-variables}. Therefore, the joint posterior of $(\theta,\tau)$ is obtained from the auxiliary distribution by weighting with $W$ and renormalizing.

Let $\mathbb E_0$ denote expectation under the auxiliary distribution. By~\eqref{eq:control-norms-quartic-appl} and~\eqref{eq:quartic-epsilon-H},
\begin{equation}
\label{eq:quartic-epsilon-H-bounds}
|\epsilon|=\mathcal O(1),\qquad \|H\|_{\rm op}\le\|D_1\|_{\rm op}\|D_0^{-1}\|_{\rm op}=\mathcal O(1).
\end{equation}
Since $k$ is fixed and the coordinates of $Z$ are centered with finite fourth moments, Equation~\eqref{eq:quartic-weight} gives
\begin{equation}
\label{eq:quartic-log-weight-bound}
\mathbb E_0[-\log W]
=\mathbb E_0\left[\|\tau^{1/4}\epsilon-HZ\|_4^4\right]
\le8\mathbb E_0[\tau]\|\epsilon\|_4^4+8\mathbb E_0\left[\|HZ\|_4^4\right]
\le C,
\end{equation}
where $C>0$ is independent of $d$. To verify the last inequality, let $h_j^\top$ denote the $j$th row of $H$. Then
\[
\mathbb E_0\left[\|HZ\|_4^4\right]
=\sum_{j=1}^k\mathbb E_0\left[|h_j^\top Z|^4\right]
\le C_0\sum_{j=1}^k|h_j|^4
\le C_0k\|H\|_{\rm op}^4
=\mathcal O(1),
\]
where we used that $k$ is fixed and $\|H\|_{\rm op}=\mathcal O(1)$ by~\eqref{eq:quartic-epsilon-H-bounds}. Jensen's inequality and~\eqref{eq:quartic-log-weight-bound} therefore imply
\begin{equation}
\label{eq:quartic-weight-normalizer}
\mathbb E_0[W]\ge\exp\{-\mathbb E_0[-\log W]\}\ge e^{-C}.
\end{equation}

Since the posterior is obtained from the auxiliary distribution by weighting with $W$ and renormalizing, for any integrable function $g$,
\begin{equation}
\label{eq:quartic-reweighting-identity}
\mathbb E_{\bar\pi_d}[g]=\frac{\mathbb E_0[gW]}{\mathbb E_0[W]}.
\end{equation}
Taking $g(\theta)=|\theta-\theta_0|$, using $W\le1$, and applying~\eqref{eq:quartic-weight-normalizer}, we obtain
\begin{equation}
\label{eq:quartic-posterior-first-moment}
\mathbb E_{\bar\pi_d}[|\theta-\theta_0|]
\le e^C\mathbb E_0[|\theta-\theta_0|].
\end{equation}
The shape parameter in~\eqref{eq:quartic-auxiliary-distribution} satisfies
\[
a+\frac{k}{4}=\frac{\nu}{4}>\frac14,
\]
so $\mathbb E_0[\tau^{-1/4}]$ is finite and independent of $d$. Moreover, since the coordinates of $Z$ have finite second moments,
\[
\mathbb E_0[|Z|]=\mathcal O(\sqrt d).
\]
Using~\eqref{eq:quartic-change-variables} and the independence of $\tau$ and $Z$ gives
\begin{equation}
\label{eq:quartic-auxiliary-first-moment}
\mathbb E_0[|\theta-\theta_0|]
\le\|D_0^{-1}\|_{\rm op}\mathbb E_0[\tau^{-1/4}]\mathbb E_0[|Z|]
=\mathcal O(\sqrt d).
\end{equation}
Combining~\eqref{eq:quartic-posterior-first-moment} and~\eqref{eq:quartic-auxiliary-first-moment} yields
\begin{equation*}
\mathbb E_{\bar\pi_d}[|\theta-\theta_0|]=\mathcal O(\sqrt d).
\end{equation*}
Markov's inequality then shows that the $2/3$-quantile of $|x|=|\theta-\theta_0|$ is $\mathcal O(\sqrt d)$.

It remains to verify the derivative bounds. Let
\begin{equation}
\label{eq:quartic-q-definitions}
\alpha=a+\frac{n}{4},\qquad q(\theta)=b+\|y-D\theta\|_4^4,\qquad S=\|y-D\theta\|_4^4.
\end{equation}
Since $D=(D_0^\top,D_1^\top)^\top$, the assumptions in~\eqref{eq:control-norms-quartic-appl} imply
\begin{equation}
\label{eq:quartic-D-bound}
L:=\|D\|_{\rm op}
\le\left(\|D_0\|_{\rm op}^2+\|D_1\|_{\rm op}^2\right)^{1/2}
=\mathcal O(1).
\end{equation}
Differentiating $q(\theta)$, defined in~\eqref{eq:quartic-q-definitions}, gives
\begin{equation}
\label{eq:quartic-q-derivative-bounds}
|\nabla q(\theta)|\le4LS^{3/4},\qquad 0\preceq\nabla^2q(\theta)\preceq12L^2S^{1/2}I_d.
\end{equation}
Since $\log\bar\pi_d(\theta)=-\alpha\log q(\theta)$ up to an additive constant,
\[
\nabla\log\bar\pi_d(\theta)=-\alpha\frac{\nabla q(\theta)}{q(\theta)}.
\]
Using~\eqref{eq:quartic-q-derivative-bounds} and maximizing $S^{3/4}/(b+S)$ over $S\ge0$ gives
\begin{equation}
\label{eq:quartic-gradient-bound}
\sup_\theta|\nabla\log\bar\pi_d(\theta)|
\le3^{3/4}\alpha Lb^{-1/4}.
\end{equation}
Similarly,
\[
\nabla^2\log\bar\pi_d(\theta)
=\alpha\left(\frac{\nabla q(\theta)\nabla q(\theta)^\top}{q(\theta)^2}-\frac{\nabla^2q(\theta)}{q(\theta)}\right).
\]
The two matrices inside the difference are positive semidefinite, so the operator norm of their difference is bounded by the larger of their operator norms. Using~\eqref{eq:quartic-q-derivative-bounds} and maximizing the corresponding scalar factors over $S\ge0$ yields
\begin{equation}
\label{eq:quartic-hessian-bound}
\sup_\theta\|\nabla^2\log\bar\pi_d(\theta)\|_{\rm op}
\le6\alpha L^2b^{-1/2}.
\end{equation}
Since $n=d+k$ with $a,b,k$ fixed, Equation~\eqref{eq:quartic-q-definitions} gives $\alpha=\mathcal O(d)$. Together with~\eqref{eq:quartic-D-bound}, the bounds in~\eqref{eq:quartic-gradient-bound} and~\eqref{eq:quartic-hessian-bound} are therefore both $\mathcal O(d)$. These bounds hold globally and are unchanged by translation, so they imply the derivative conditions of \Cref{cor:SCPS_mixing_kappa_concave} for every fixed $C>0$.

Together with the $2/3$-quantile bound established above, all assumptions of \Cref{cor:SCPS_mixing_kappa_concave} are satisfied. Since $\kappa=-1/\nu$, its condition on $\beta$ becomes
\[
\beta>\frac{1+2/\nu}{2(1-1/\nu)}=\frac{\nu+2}{2(\nu-1)},
\]
which gives the stated mixing-time bound.
\end{proof}

\subsubsection{Proof for Gaussian regression with censored responses}
\label{sec:proofs-examples-censored}

\begin{proof}[Proof of \Cref{cor:censored-applications}]
The $\kappa$-concavity of the target was established in \Cref{sec:example-censored}. It remains to verify the covariance and derivative assumptions of \Cref{cor:SPS_mixing_kappa_concave} when $\nu_d\asymp d$, and the quantile and derivative assumptions of \Cref{cor:SCPS_mixing_kappa_concave} when $\nu_d=\nu>1$ is fixed. We suppress the index $d$ when no confusion arises.

Before conditioning on the censoring event $E$, the marginal distribution of $\theta$ given $y_o$ is Student's $t$ with $\nu=2a$ degrees of freedom, location $m$, and scale matrix
\[
S=\frac{b}{a}V.
\]
Let $f_t$ denote its density and let $f_C$ denote the coefficient posterior density after conditioning on $E$. The predictive distribution of $Z$ given $y_o$ is Student's $t$ in dimension $n_c$, with location $D_cm$ and scale matrix
\[
S_Z=\frac{b}{a}\left(I_{n_c}+D_cVD_c^\top\right).
\]
By the assumptions of the corollary, the eigenvalues of $S_Z$ are uniformly bounded above and away from zero, while $|c-D_cm|=\mathcal O(1)$. Since $n_c$ is fixed, $c-D_cm$ remains in a bounded set, and the eigenvalues of $S_Z$ remain in a fixed interval bounded away from zero. Hence the probability of the censoring event is uniformly bounded away from zero: for some constant $p_*>0$ independent of $d$,
\begin{equation*}
p_E:=\mathbb P(E\mid y_o)\ge p_*.
\end{equation*}
Bayes' formula therefore gives
\begin{equation}
\label{eq:censored-student-domination}
f_C(\theta)
=
\frac{f_t(\theta)\mathbb P(E\mid\theta,y_o)}{p_E}
\le
p_*^{-1}f_t(\theta).
\end{equation}

We next verify the derivative bounds. Define
\begin{equation}
\label{eq:censored-q-definitions}
\alpha=\frac{d+n_c+\nu}{2},\qquad q(\theta,z)=2b+(\theta-m)^\top V^{-1}(\theta-m)+|z-D_c\theta|^2,
\end{equation}
and
\begin{equation}
\label{eq:censored-L-definition}
L^2=\|V^{-1}+D_c^\top D_c\|_{\rm op}.
\end{equation}
Before conditioning on $E$, the joint density of $(\theta,Z)$ given $y_o$ has the form
\[
f_J(\theta,z)\propto q(\theta,z)^{-\alpha}.
\]
Differentiating $q(\theta,z)$ gives
\[
|\nabla_\theta q(\theta,z)|\le2L\sqrt{q(\theta,z)-2b},
\qquad
\nabla_\theta^2q(\theta,z)=2(V^{-1}+D_c^\top D_c).
\]
Consequently,
\begin{equation}
\label{eq:censored-joint-derivative-bounds}
|\nabla_\theta\log f_J(\theta,z)|
\le
\frac{\alpha L}{\sqrt{2b}},
\qquad
\nabla_\theta^2\log f_J(\theta,z)
\succeq
-\frac{\alpha L^2}{b}I_d.
\end{equation}

Since $f_C(\theta)$ is obtained by integrating the joint density $f_J(\theta,z)$ over the censoring region $\{z\le c\}$, differentiating with respect to $\theta$ under this integral gives
\[
\nabla\log f_C(\theta)
=
\mathbb E\left[\nabla_\theta\log f_J(\theta,Z)\mid\theta,E,y_o\right],
\]
and
\[
\nabla^2\log f_C(\theta)
=
\mathbb E\left[\nabla_\theta^2\log f_J(\theta,Z)\mid\theta,E,y_o\right]
+
\operatorname{Cov}\left(\nabla_\theta\log f_J(\theta,Z)\mid\theta,E,y_o\right).
\]
Since $f_C$ is $(-1/\nu)$-concave, the function $f_C^{-1/(d+\nu)}$ is convex, which implies
\[
\nabla^2\log f_C(\theta)
\preceq
\frac{1}{d+\nu}
\nabla\log f_C(\theta)\nabla\log f_C(\theta)^\top.
\]
Combining these identities with~\eqref{eq:censored-joint-derivative-bounds} yields
\begin{equation}
\label{eq:censored-derivative-bounds}
\sup_\theta|\nabla\log f_C(\theta)|
\le
\frac{\alpha L}{\sqrt{2b}},
\qquad
\sup_\theta\|\nabla^2\log f_C(\theta)\|_{\rm op}
\le
\max\left\{
\frac{\alpha L^2}{b},
\frac{\alpha^2L^2}{2b(d+\nu)}
\right\}.
\end{equation}

Suppose first that $\nu_d\asymp d$. Then $a_d\asymp d$, and the bounds on $b_d/a_d$ imply $b_d\asymp d$. Moreover, the assumptions on $V_d$ and $D_{c,d}$ imply $L=\mathcal O(1)$, while $\alpha=\mathcal O(d)$. Therefore,~\eqref{eq:censored-derivative-bounds} gives
\[
\sup_\theta|\nabla\log f_C(\theta)|=\mathcal O(\sqrt d),
\qquad
\sup_\theta\|\nabla^2\log f_C(\theta)\|_{\rm op}=\mathcal O(1).
\]
These bounds hold globally and are unchanged by translation.

It remains to verify the covariance bounds. The covariance matrix of the underlying Student's $t$ distribution is
\[
\operatorname{Cov}_{f_t}(\theta)
=
\frac{\nu_d}{\nu_d-2}\frac{b_d}{a_d}V_d,
\]
whose eigenvalues are uniformly bounded above and away from zero for sufficiently large $d$. For any unit vector $u$, the bound~\eqref{eq:censored-student-domination} gives
\[
\operatorname{Var}_{f_C}(u^\top\theta)
\le
\mathbb E_{f_C}\left[(u^\top(\theta-m))^2\right]
\le
p_*^{-1}
\mathbb E_{f_t}\left[(u^\top(\theta-m))^2\right]
=
\mathcal O(1).
\]
For the lower bound, fix a unit vector $u\in\mathbb R^d$, and consider the scalar projection $u^\top\theta$ along the direction $u$. Let $g_{C,u}$ and $g_{t,u}$ denote its densities under $f_C$ and $f_t$, respectively. Marginalizing~\eqref{eq:censored-student-domination} gives $g_{C,u}\le p_*^{-1}g_{t,u}$. Under $f_t$, $u^\top\theta$ is univariate Student's $t$ with scale parameter $\sqrt{u^\top S u}$, where $S=(b/a)V$. Since the eigenvalues of $S$ are uniformly bounded away from zero, there exists a constant $B>0$, independent of $d$ and $u$, such that
\[
\sup_{s\in\mathbb R}g_{C,u}(s)\le B
\]
for all sufficiently large $d$ and all $|u|=1$. Hence, with $r=1/(4B)$,
\[
\mathbb P_{f_C}\left(|u^\top\theta-\mathbb E_{f_C}[u^\top\theta]|\le r\right)\le 2rB=\frac12,
\]
so
\[
\mathbb P_{f_C}\left(|u^\top\theta-\mathbb E_{f_C}[u^\top\theta]|\ge r\right)\ge\frac12.
\]
Therefore,
\[
\operatorname{Var}_{f_C}(u^\top\theta)
\ge r^2\mathbb P_{f_C}\left(|u^\top\theta-\mathbb E_{f_C}[u^\top\theta]|\ge r\right)
\ge\frac{1}{32B^2}.
\]
Since this holds uniformly for all $|u|=1$, the smallest eigenvalue of $\operatorname{Cov}_{f_C}(\theta)$ is uniformly bounded away from zero.

Since $\kappa=-1/\nu_d=\mathcal O(d^{-1})$, all assumptions of \Cref{cor:SPS_mixing_kappa_concave} are satisfied after recentering at the mean of $f_C$, which gives the first mixing-time bound.

Now suppose that $\nu_d=\nu>1$ is fixed. Then $a=\nu/2$ is fixed, and the bounds on $b/a$ imply that $b$ is bounded above and away from zero. Again $L=\mathcal O(1)$ and $\alpha=\mathcal O(d)$, so~\eqref{eq:censored-derivative-bounds} gives
\[
\sup_\theta|\nabla\log f_C(\theta)|=\mathcal O(d),
\qquad
\sup_\theta\|\nabla^2\log f_C(\theta)\|_{\rm op}=\mathcal O(d).
\]
These bounds are again global and unchanged by translation.

It remains to verify the $2/3$-quantile bound. Since $\nu>1$, the usual Student's $t$ representation gives
\[
\mathbb E_{f_t}[|\theta-m|]=\mathcal O(\sqrt d).
\]
Using~\eqref{eq:censored-student-domination},
\[
\mathbb E_{f_C}[|\theta-m|]
\le
p_*^{-1}\mathbb E_{f_t}[|\theta-m|]
=
\mathcal O(\sqrt d).
\]
Thus, the mean $\mu_C=\mathbb E_{f_C}[\theta]$ exists. Since
\[
|\mu_C-m|
\le
\mathbb E_{f_C}[|\theta-m|],
\]
we obtain
\[
\mathbb E_{f_C}[|\theta-\mu_C|]
\le
2\mathbb E_{f_C}[|\theta-m|]
=
\mathcal O(\sqrt d).
\]
Markov's inequality then shows that the $2/3$-quantile of $|\theta-\mu_C|$ is $\mathcal O(\sqrt d)$.

Together with the derivative bounds above, all assumptions of \Cref{cor:SCPS_mixing_kappa_concave} are satisfied. Since $\kappa=-1/\nu$, its condition on $\beta$ becomes
\[
\beta>\frac{1+2/\nu}{2(1-1/\nu)}
=
\frac{\nu+2}{2(\nu-1)},
\]
which gives the second mixing-time bound.
\end{proof}
\section{Weighted Isoperimetric Inequalities}
\label{sec:cheeger}
In this section, we obtain tighter Cheeger inequalities for a smaller class of \(\kappa\)-concave measures. We do this by combining a weighted Poincar\'e inequality with a \textit{curvature--dimension condition}, both read through the conformal change of metric induced by the stereographic projection. An analogous technique yields tighter isoperimetric inequalities in the log-concave setting (e.g., \cite[Section 2.4.4]{Chewi2023-hc}).

Throughout this section, we restrict to the radially symmetric case where
\begin{equation*}
    \bar{\pi}(dx) \propto \exp(-f(|x|)) dx,
\end{equation*}
and we consider the weighted Poincar\'e inequalities of the form,
\begin{equation}\label{eq:weighted_pi}
    \int ( \varphi(x) - \bar{\pi}(\varphi))^2 \bar{\pi}(dx) \leq C_{\text{PI}} \int w(x)^2 |\nabla \varphi(x)|^2 \bar{\pi}(dx), \qquad \forall \varphi \in \mathcal{C}_\infty(\mathbb{R}^d).
\end{equation}

\subsection{Curvature--dimension conditions on Riemannian manifolds}

In this section, we show that the weighted Cheeger inequality can be obtained from the weighted Poincar\'e inequality in \eqref{eq:weighted_pi} under an additional curvature condition.

We begin by viewing the weighted Poincar\'e inequality through a geometric lens. The inequality in \eqref{eq:weighted_pi} can be viewed as the \emph{standard} Poincar\'e associated with Riemannian manifold \((\mathbb{R}^d, g)\) with \(g_{ij} = (w)^{-2} \delta_{ij}\) where \(\delta\) denotes the Euclidean metric. This is known as a conformal transformation of the Euclidean metric, or a conformal manifold. By standard results, we obtain the associated volume measure \(\lambda_g(dx) = w(x)^{-d} dx\) and metric \(d_g(\cdot, \cdot)\) \citep{lee2018introduction}. The fact that \eqref{eq:weighted_pi} is natural for this setting follows from how the Riemannian gradient satisfies \(\nabla_g f = w^2 \nabla f\) and the Riemannian inner product satisfies \(|\cdot|_g^2 = (w)^{-2} |\cdot|^2\) and hence \(|\nabla_g f|_g^2 = w^2 |\nabla f|^2\).

Similarly, for sets with smooth boundary, the weighted Minkowski content \(\bar{\pi}^+_w\) coincides with the Minkowski content of \(\bar{\pi}\) on \((\mathbb{R}^d, g)\), so that weighted isoperimetric inequalities correspond to standard isoperimetric inequalities on \((\mathbb{R}^d, g)\). In the following lemma, we use a curvature--dimension condition on \((\mathbb{R}^d, g)\) to convert the Poincar\'e inequality into an isoperimetric inequality.

\begin{lemma}\label{lem:cheeger_poincare}
Consider the Riemannian metric \(g = (\tilde{w})^{-2} \delta\), suppose that \(g\) is complete, and suppose \(\bar{\pi}(dx) = \exp(-V(x)) \lambda_g(dx)\) satisfies \eqref{eq:weighted_pi} with weighting \(\tilde{w}\) and constant \(C_{\text{PI}}\) and that the curvature--dimension condition \(\operatorname{Ric}_g + \operatorname{Hess}_g V \succcurlyeq - K g\) holds for some \(K \geq 0\). Then, for every Borel set \(A\) with \(\bar{\pi}_{\tilde{w}}(A) < \infty\), we have the Cheeger inequality,
\begin{equation}\label{eq:cheeger_poincare_1}
    \bar{\pi}^+_{\tilde{w}}(A) \geq \frac{1}{6} \min \bigg \{ \frac{1}{C_{\text{PI}}\sqrt{K}}, \frac{1}{\sqrt{C_{\text{PI}}}} \bigg \} \bar{\pi}(A) (1-\bar{\pi}(A)).
\end{equation}
\end{lemma}
\begin{proof}
The proof proceeds by showing that the weighted Poincar\'{e} inequality \eqref{eq:weighted_pi} is equivalent to the standard Poincar\'{e} inequality with metric $g$. We then apply a Buser-type inequality by \citet{Ledoux2004-ry} to obtain a Cheeger inequality with respect to the geodesic distance \(d_g\), and finally transfer it to the weighted boundary measure \(\bar{\pi}^+_{\tilde{w}}\).

First, since \(|\nabla_g \varphi|_g^2 = \tilde{w}^2 |\nabla \varphi|^2\), the weighted Poincar\'e inequality \eqref{eq:weighted_pi} with weighting \(\tilde{w}\) is precisely the standard Poincar\'e inequality for \(\bar{\pi}\) on \((\mathbb{R}^d, g)\), with the same constant \(C_{\text{PI}}\).

Let \(\lambda_1\) denote the spectral gap of \(\bar{\pi}\) on \((\mathbb{R}^d, g)\), and let,
\begin{equation*}
    \Phi_g = \inf_{A} \frac{\bar{\pi}^+_{(g)}(A)}{\min \big \{ \bar{\pi}(A), 1 - \bar{\pi}(A) \big \}},
\end{equation*}
be the Cheeger constant associated with \(d_g\), where \(\bar{\pi}^+_{(g)}\) denotes the Minkowski content with respect to \(d_g\). Since \(g\) is complete and \(\operatorname{Ric}_g + \operatorname{Hess}_g V \succcurlyeq - K g\), \citet[Theorem 5.2]{Ledoux2004-ry} gives \(\lambda_1 \leq \max \{ 6 \sqrt{K} \Phi_g, 36 \Phi_g^2 \}\), and so, in either case, \(\Phi_g \geq \frac{1}{6} \min \{ \lambda_1 / \sqrt{K}, \sqrt{\lambda_1} \}\). By \eqref{eq:weighted_pi} we have \(\lambda_1 \geq C_{\text{PI}}^{-1}\), and hence,
\begin{equation*}
    \Phi_g \geq \Phi, \qquad \text{where} \qquad \Phi = \frac{1}{6} \min \bigg \{ \frac{1}{C_{\text{PI}}\sqrt{K}}, \frac{1}{\sqrt{C_{\text{PI}}}} \bigg \}.
\end{equation*}

It remains to transfer this to the weighted boundary measure. By the co-area formula, the Cheeger inequality for \(d_g\) is equivalent to the functional inequality,
\begin{equation*}
    \int |\nabla_g \varphi|_g \, d\bar{\pi} \geq \Phi_g \int |\varphi - m_\varphi| \, d\bar{\pi},
\end{equation*}
for every locally Lipschitz \(\varphi\), where \(m_\varphi\) denotes a median of \(\varphi\) under \(\bar{\pi}\) (see e.g., \citep{bobkov1997some}). Since \(|\nabla_g \varphi|_g = \tilde{w} |\nabla \varphi|\) and \(\iint |\varphi(x) - \varphi(y)| \bar{\pi}(dx) \bar{\pi}(dy) \leq 2 \int |\varphi - m_\varphi| \, d\bar{\pi}\), it follows that,
\begin{equation*}
    \int |\nabla \varphi(x)| \, \tilde{w}(x) \, \bar{\pi}(dx) \geq \frac{\Phi}{2} \iint |\varphi(x) - \varphi(y)| \, \bar{\pi}(dx) \bar{\pi}(dy),
\end{equation*}
for every locally Lipschitz \(\varphi\). Finally, \Cref{lem:BL09prop6.1}, applied with the weight \(\tilde{w}/\Phi\), together with \Cref{rmk:extendLemmaBK09}, since \(\tilde{w} > 0\), yields \(\bar{\pi}^+_{\tilde{w}}(A) \geq \Phi \, \bar{\pi}(A)(1 - \bar{\pi}(A))\) for every Borel set \(A\) with \(\bar{\pi}_{\tilde{w}}(A) < \infty\), which is the claim.
\end{proof}

Since the curvature dimension condition is defined in the weighted metric and is defined for a density with respect to the natural volume measure, we must do some calculations to move between the two. Luckily, the relationship between the two metrics is one that is familiar to the Riemannian geometry literature. Two metrics that are related by the scaling of the metric are said to be conformally related. In the following lemma, we collect some known calculations that relate Ricci curvature and the Hessian between conformally related metrics and apply them to the present setting to relate these objects to Euclidean notions of Hessian and gradient.

\begin{lemma}\label{lem:conformal_metric}
For a conformal metric \(g = e^{2v} \delta\) on \(\mathbb{R}^d\), we have the formulas,
\begin{gather*}
    \operatorname{Ric}_g = - (d-2) D^2 v + (d-2) \nabla v \otimes \nabla v - (\Delta v + (d-2) |\nabla v|^2) \delta,\\
    \operatorname{Hess}_g V = D^2 V - (\nabla v \otimes \nabla V) - (\nabla V \otimes \nabla v) + \langle \nabla v, \nabla V \rangle \delta.
\end{gather*}
Furthermore, we have
\begin{equation*}
    \lambda_g(dx) = e^{d v} dx.
\end{equation*}
\end{lemma}
This can be derived easily from \cite[Section 7]{lee2018introduction}.

\begin{lemma}\label{lem:cd_radial_weight}
Suppose that \(d \geq 2\) and both \(w\) and \(f\) are twice continuously differentiable and let \(V\) be defined by \(\bar{\pi}(dx) = \exp(-V(x)) \lambda_g(dx)\). For \(K \geq 0\) the curvature--dimension condition \(\operatorname{Ric}_g + \operatorname{Hess}_g V \succcurlyeq - K g\) holds if and only if, for all \(r > 0\), both of the conditions
\begin{align*}
    w^2 f'' + w w' f' - w w'' + (d-1) \frac{w w'}{r} - (d-1) (w')^2 &\geq - K,\\
    w^2 \frac{f'}{r} - w w' f' + w w'' + (w')^2 + (d-3) \frac{w w'}{r} &\geq - K,
\end{align*}
are satisfied.
\end{lemma}
\begin{proof}
Suppose that \(\bar{\pi}(dx) = \exp(-W(x)) dx, W(x) = f(|x|)\), and set \(V = W + dv\) so that \(\bar{\pi}\) is as in \Cref{lem:cheeger_poincare} (note that we have used \Cref{lem:conformal_metric} for this argument). It follows from \Cref{lem:conformal_metric} that the left-hand side of the condition can be written in coordinate form:
\begin{align*}
    \operatorname{Ric}_g + \operatorname{Hess}_g V &= D^2 W - (\nabla v \otimes \nabla W) - (\nabla W \otimes \nabla v) + \langle \nabla v, \nabla W \rangle I_d\\
    & \qquad + 2 D^2 v - (d + 2) \nabla v \otimes \nabla v - (\Delta v - 2 |\nabla v|^2) I_d\\
    &=: A + B
\end{align*}
Set \(v(x) = - \log(w(|x|))\) so that \(g = e^{2v} \delta\). Writing \(P = x \otimes x/|x|^2\), we have for any radial \(u(x) = u(r)\) that,
\begin{gather*}
    \nabla u = u'(r) \frac{x}{r}, \qquad D^2 u = u''(r) P + \frac{u'(r)}{r} \bigg ( I_d - P \bigg ),\\
    |\nabla u|^2 = u'(r)^2, \qquad \Delta u = u''(r) + (d-1) \frac{u'(r)}{r}.
\end{gather*}
Since \(v\) and \(W\) are both radial, it follows that \(\nabla v \otimes \nabla W = \nabla W \otimes \nabla v = v' f' P\) and \(\langle \nabla v, \nabla W \rangle = v' f'\). Thus, we compute,
\begin{equation*}
    A = \bigg ( f'' - \frac{f'}{r} - 2 v' f' \bigg ) P + \bigg ( \frac{f'}{r} + v' f' \bigg ) I_d,
\end{equation*}
and, using \(\Delta v - 2 |\nabla v|^2 = v'' + (d-1) v'/r - 2 (v')^2\),
\begin{equation*}
    B = \bigg ( 2 v'' - 2 \frac{v'}{r} - (d+2) (v')^2 \bigg ) P + \bigg ( - v'' + (3-d) \frac{v'}{r} + 2 (v')^2 \bigg ) I_d.
\end{equation*}
Since \(P\) is rank-one, it follows that \(A + B\) has two distinct eigenvalues:
\begin{align*}
    \lambda_{\text{rad}}(r) &= f'' - v' f' + v'' - (d-1) \frac{v'}{r} - d (v')^2,\\
    \lambda_{\text{tan}}(r) &= \frac{f'}{r} + v' f' - v'' + (3-d) \frac{v'}{r} + 2 (v')^2.
\end{align*}
Thus, the curvature dimension condition \(\operatorname{Ric}_g + \operatorname{Hess}_g V \succcurlyeq - K g\) is equivalent to \(\lambda_{\text{rad}}(r) \geq - w(r)^{-2} K, \lambda_{\text{tan}}(r) \geq - w(r)^{-2} K\). It remains to substitute \(v' = - w'/w\) and \(v'' = - w''/w + (w'/w)^2\) which yields the two conditions in the statement.
\end{proof}

We can now state the result that connects weighted Poincar\'{e} inequalities to weighted Cheeger inequalities under an additional curvature assumption. We do this by applying the previous lemmas to the setting where the weighting is comparable with the weighting \(\tilde{w}(x) = (R^2 + |x|^2)^{1/2}\).

\begin{proposition}\label{prop:wp_to_wc2}
Suppose that \(d \geq 2\) and that \(\bar{\pi}(dx) \propto \exp(-f(|x|)) dx\) where \(f\) is twice continuously differentiable on \([0, \infty)\), and that the weighted Poincaré inequality in \eqref{eq:weighted_pi} is satisfied, with weighting \(\tilde{w}(x) = (R^2 + |x|^2)^{1/2}\) and constant \(C_{\text{PI}} > 0\). Suppose further that the function
\begin{equation*}
    q(r) = \frac{(R^2 + r^2) f'(r)}{r}
\end{equation*}
is non-decreasing on \((0, \infty)\) and satisfies \(q(r) + (d - 2) \geq - K(1 + \frac{r^2}{R^2})\) for some \(K \geq 0\). Then, for any weighting \(w\) satisfying \(w(x) \geq C \tilde{w}(x)\), and every Borel set \(A\) with \(\bar{\pi}_w(A) < \infty\), \(\bar{\pi}\) satisfies the weighted isoperimetric inequality,
\begin{equation}
\label{eq:weighted_isoper_ineq}
    \bar{\pi}^+_w( A) \geq \frac{C}{6} \min \bigg \{ \frac{1}{C_{\text{PI}}\sqrt{K}}, \frac{1}{\sqrt{C_{\text{PI}}}} \bigg \} \bar{\pi}(A)(1 - \bar{\pi}(A)).
\end{equation}
\end{proposition}
\begin{proof}
Let \(g = \tilde{w}^{-2} \delta\), so that \(\tilde{w}(r) = (R^2 + r^2)^{1/2}\). We compute,
\begin{equation*}
    \tilde{w}' = \frac{r}{(R^2 + r^2)^{1/2}}, \qquad \tilde{w}'' = \frac{R^2}{(R^2 + r^2)^{3/2}},
\end{equation*}
and hence,
\begin{gather*}
    \tilde{w}^2 = R^2 + r^2, \qquad \tilde{w} \tilde{w}' = r, \qquad \frac{\tilde{w} \tilde{w}'}{r} = 1, \qquad \tilde{w} \tilde{w}'' = \frac{R^2}{R^2 + r^2}, \qquad (\tilde{w}')^2 = \frac{r^2}{R^2 + r^2}.
\end{gather*}
Substituting these into \Cref{lem:cd_radial_weight}, the left-hand side of the first condition equals,
\begin{equation*}
    (R^2 + r^2) f''(r) + r f'(r) - \frac{R^2}{R^2 + r^2} + (d-1) - \frac{(d-1) r^2}{R^2 + r^2},
\end{equation*}
and since \(- R^2 + (d-1)(R^2 + r^2) - (d-1) r^2 = (d-2) R^2\), this is equal to \((R^2 + r^2) f''(r) + r f'(r) + (d-2) R^2/(R^2 + r^2)\). Similarly, the left-hand side of the second condition equals
\begin{equation*}
    \frac{(R^2 + r^2) f'(r)}{r} - r f'(r) + \frac{R^2}{R^2 + r^2} + \frac{r^2}{R^2 + r^2} + d - 3 = \frac{R^2 f'(r)}{r} + d - 2.
\end{equation*}
Since \(f'(r) = r q(r)/(R^2 + r^2)\) and \(d \geq 2\), the latter is at least \(R^2 (q(r) + d - 2)/(R^2 + r^2) \geq -K\), by assumption. For the former, let \(\eta(r) = (R^2 + r^2) f'(r) = r q(r)\). Since \(q' \geq 0\), we have that,
\begin{equation*}
    \eta'(r) = q(r) + r q'(r) \geq q(r) = \frac{\eta(r)}{r}, \qquad \forall r > 0.
\end{equation*}
Since \(\eta' = 2 r f' + (R^2 + r^2) f''\), it follows that,
\begin{equation*}
    (R^2 + r^2) f''(r) + r f'(r) = \eta'(r) - \frac{r \eta(r)}{R^2 + r^2} \geq \frac{\eta(r)}{r} - \frac{r \eta(r)}{R^2 + r^2} = \frac{R^2 q(r)}{R^2 + r^2},
\end{equation*}
and so, by assumption, the former satisfies,
\begin{equation}\label{eq:wp_to_wc}
    (R^2 + r^2) f''(r) + r f'(r) + \frac{(d-2) R^2}{R^2 + r^2} \geq \frac{R^2 (q(r) + d - 2)}{R^2 + r^2} \geq -K, \qquad \forall r > 0.
\end{equation}
Thus, by \Cref{lem:cd_radial_weight}, the curvature--dimension condition \(\operatorname{Ric}_g + \operatorname{Hess}_g V \succcurlyeq - K g\) holds.

Moreover, \(g\) is complete. Indeed, since \(\tilde{w}\) is radial, the length with respect to \(g\) of any curve from \(0\) to \(x\) is at least,
\begin{equation*}
    \int_0^{|x|} \frac{dr}{\tilde{w}(r)} = \int_0^{|x|} \frac{dr}{\sqrt{R^2 + r^2}} = \operatorname{arsinh}(|x|/R),
\end{equation*}
which tends to infinity as \(|x| \to \infty\). Hence, sets that are bounded with respect to \(d_g\) are bounded in Euclidean space, and so \(g\) is complete by the Hopf--Rinow theorem.
\Cref{lem:cheeger_poincare} therefore applies and gives, for every Borel set \(B\) with \(\bar{\pi}_{\tilde{w}}(B) < \infty\),
\begin{equation*}
    \bar{\pi}^+_{\tilde{w}}( B) \geq \frac{1}{6} \min \bigg \{ \frac{1}{C_{\text{PI}}\sqrt{K}}, \frac{1}{\sqrt{C_{\text{PI}}}} \bigg \} \bar{\pi}(B)(1 - \bar{\pi}(B)).
\end{equation*}
Finally, if \(w \geq C \tilde{w}\) then \(\bar{\pi}_{\tilde{w}}(B) \leq \bar{\pi}_w(B)/C\), and \(\bar{\pi}_w \geq C \bar{\pi}_{\tilde{w}}\) as measures, and hence \(\bar{\pi}^+_w(B) \geq C \bar{\pi}^+_{\tilde{w}}(B)\), which yields \eqref{eq:weighted_isoper_ineq}.
\end{proof}

\subsection{Weighted Poincar\'{e} inequalities for \(\kappa\)-concave measures}

To obtain a Cheeger constant of order \(\sqrt{d}\), we require \(C_{\text{PI}} = \mathcal{O}(1/d)\) and \(K = \mathcal{O}(d)\). In this section, we give a simple condition on \(f\) that implies both, with \(K = 0\). To control \(C_{\text{PI}}\), we use a result of \citet{Bonnefont2016-if} which reduces the spectral gap of a radially symmetric measure to that of its radial part. We then bound the spectral gap of the radial part with a variational argument based on \citet{Chen1997-it}.

Throughout this section, let \(\tilde{w}(x) = (R^2 + |x|^2)^{1/2}\) as in \Cref{prop:wp_to_wc2}, and let \(\nu\) denote the law of \(|X|\) for \(X \sim \bar{\pi}\) which has density \(\rho(r) \propto r^{d-1} \exp(-f(r))\). The canonical operator associated with the \(\tilde{w}\)--weighted Poincar\'{e} inequality is given by,
\begin{equation*}
    L_{\bar{\pi}} \varphi(x) = \tilde{w}(x)^2 \Delta \varphi(x) + \bigg ( 2x - \tilde{w}(x)^2 f'(|x|) \frac{x}{|x|} \bigg ) \cdot \nabla \varphi(x).
\end{equation*}
The weighted Poincar\'{e} inequality with constant \(C_{\text{PI}}\) is then equivalent to a lower bound on the spectral gap of \(-L_{\bar{\pi}}\) by \(1/C_{\text{PI}}\) \citep{Bakry2014-ut}. The following lemma, that we borrow from \citet{Bonnefont2016-if}, shows that in the radial case, the Poincar\'{e} constant can be approximated using the Poincar\'{e} constant associated with the operator on the half-line \((0, \infty)\):
\begin{equation*}
    L_\nu u(r) = (R^2 + r^2) u''(r) + b(r) u'(r), \qquad b(r) = 2r - (R^2 + r^2) \bigg ( f'(r) - \frac{d-1}{r} \bigg ).
\end{equation*}

\begin{lemma}[{\cite[Theorem 5.1]{Bonnefont2016-if}}]\label{lem:bjm_radial_reduction}
Suppose that \(d \geq 2\) and that \(\int |x|^2 \bar{\pi}(dx) < \infty\). Then,
\begin{equation*}
    \lambda_1(-L_{\bar{\pi}}) \geq \min \big \{ \lambda_1(-L_\nu), d-1 \big \}.
\end{equation*}
\end{lemma}

To lower bound \(\lambda_1(-L_\nu)\), we begin with a method that uses that, when the conditions on \(q\) in \Cref{prop:wp_to_wc2} hold with \(K = 0\), a simple transformation makes \(\nu\) log-concave. In the proof, we use the variable
\begin{equation*}
    T = \operatorname{arsinh}(|X|/R),
\end{equation*}
which is the length with respect to the metric \(g = \tilde{w}^{-2} \delta\) of the radial segment from \(0\) to \(X\), since \(\int_0^{|x|} (R^2 + r^2)^{-1/2} dr = \operatorname{arsinh}(|x|/R)\).

\begin{proposition}\label{prop:wpi_thin_shell}
Suppose that \(d \geq 2\) and that \(\bar{\pi}(dx) \propto \exp(-f(|x|)) dx\) where \(f\) is twice continuously differentiable on \([0, \infty)\), and that the function \(q\) of \Cref{prop:wp_to_wc2} satisfies the conditions there with \(K = 0\). Suppose further that \(\int |x|^2 \bar{\pi}(dx) < \infty\). Then the weighted Poincar\'e inequality \eqref{eq:weighted_pi} is satisfied with weighting \(\tilde{w}\) and constant,
\begin{equation*}
    C_{\text{PI}} = \max \bigg \{ \frac{12 \operatorname{Var}(|X|)}{R^2}, \frac{1}{d-1} \bigg \}.
\end{equation*}
\end{proposition}
\begin{proof}
By \Cref{lem:bjm_radial_reduction}, it suffices to show that \(\lambda_1(-L_\nu) \geq R^2/(12 \operatorname{Var}(|X|))\). For this, we first perform a change of variables. Given a locally absolutely continuous \(u : (0, \infty) \to \mathbb{R}\), let \(v(t) = u(R \sinh t)\), so that \(u(|X|) = v(T)\) and \((R^2 + r^2) u'(r)^2 = v'(t)^2\) for \(r = R \sinh t\). Hence,
\begin{equation*}
    \operatorname{Var}(u(|X|)) = \operatorname{Var}(v(T)), \qquad \int (R^2 + r^2) u'(r)^2 \nu(dr) = \mathbb{E}[v'(T)^2].
\end{equation*}
Thus, \(\lambda_1(-L_\nu)\) is the spectral gap of the law of \(T\), with respect to the un-weighted Dirichlet form \(\mathbb{E}[v'(T)^2]\).

Next, the law of \(T\) has density \(\rho(R \sinh t) R \cosh t\), which is proportional to \(\exp(-U(t))\) on \((0, \infty)\), where,
\begin{equation*}
    U(t) = f(R \sinh t) - (d-1) \log \sinh t - \log \cosh t.
\end{equation*}
Using \(dr/dt = (R^2 + r^2)^{1/2}\), \(1/\sinh^2 t = R^2/r^2\) and \(1/\cosh^2 t = R^2/(R^2 + r^2)\), we have that,
\begin{equation*}
    U''(t) = (R^2 + r^2) f''(r) + r f'(r) + \frac{(d-1) R^2}{r^2} - \frac{R^2}{R^2 + r^2} \geq (R^2 + r^2) f''(r) + r f'(r) + \frac{(d-2) R^2}{R^2 + r^2} \geq 0,
\end{equation*}
where the first inequality uses \(R^2/r^2 \geq R^2/(R^2 + r^2)\), and the second is \eqref{eq:wp_to_wc} with \(K = 0\). Thus, the law of \(T\) is log-concave.

By \cite[Corollary 4.3]{Bobkov1999-vu}, the spectral gap of a log-concave probability measure on the real line is at least \((12 \sigma^2)^{-1}\), where \(\sigma^2\) is its variance. Hence, the law of \(T\) has spectral gap at least \((12 \operatorname{Var}(T))^{-1}\).

Finally, since \(r \mapsto \operatorname{arsinh}(r/R)\) is \((1/R)\)-Lipschitz, letting \(X'\) denote an independent copy of \(X\),
\begin{equation*}
    \operatorname{Var}(T) = \frac{1}{2} \mathbb{E} \Big [ \big ( \operatorname{arsinh}(|X|/R) - \operatorname{arsinh}(|X'|/R) \big )^2 \Big ] \leq \frac{1}{2 R^2} \mathbb{E} \big [ (|X| - |X'|)^2 \big ] = \frac{\operatorname{Var}(|X|)}{R^2},
\end{equation*}
which completes the proof.
\end{proof}

For an alternative approach to lower bounding \(\lambda_1(-L_\nu)\), we extend the variational formula of \cite{Chen1997-it}.

\begin{lemma}\label{lem:variational_gap}
Suppose that \(f\) is twice continuously differentiable and let \(\psi \in \mathcal{C}^3((0,\infty))\) satisfy \(\psi' > 0\), \(\int |\psi| d\nu < \infty\), and
\begin{equation}\label{eq:variational_gap_boundary}
    \lim_{r \to 0} \rho(r) (R^2 + r^2) \psi'(r) = \lim_{r \to \infty} \rho(r) (R^2 + r^2) \psi'(r) = 0.
\end{equation}
If, for some \(\lambda > 0\),
\begin{equation*}
    - (L_\nu \psi)'(r) \geq \lambda \psi'(r), \qquad \forall r > 0,
\end{equation*}
then \(\lambda_1(-L_\nu) \geq \lambda\).
\end{lemma}
\begin{proof}
Let \(u : (0, \infty) \to \mathbb{R}\) be locally absolutely continuous. For \(0 < s < t\), the Cauchy--Schwarz inequality gives,
\begin{equation*}
    (u(t) - u(s))^2 = \bigg ( \int_s^t u'(r) dr \bigg )^2 \leq \int_s^t \frac{u'(r)^2}{\psi'(r)} dr \int_s^t \psi'(r) dr = (\psi(t) - \psi(s)) \int_s^t \frac{u'(r)^2}{\psi'(r)} dr.
\end{equation*}
Since \(\nu\) has a density, the diagonal \(\{s = t\}\) has \(\nu \otimes \nu\)-measure zero, and so by symmetry and the previous display,
\begin{align*}
    \frac{1}{2} \iint (u(t) - u(s))^2 \nu(ds) \nu(dt) &= \iint_{s < t} (u(t) - u(s))^2 \nu(ds) \nu(dt)\\
    &\leq \iint_{s < t} (\psi(t) - \psi(s)) \int_s^t \frac{u'(r)^2}{\psi'(r)} dr \, \nu(ds) \nu(dt).
\end{align*}
We now exchange the order of integration on the right-hand side. Writing \(\int_s^t = \int_0^\infty \mathbf{1}\{s < r < t\} \, dr\), and noting that the integrand is non-negative since \(\psi\) is increasing, Tonelli's theorem gives,
\begin{gather*}
    \iint_{s < t} (\psi(t) - \psi(s)) \int_s^t \frac{u'(r)^2}{\psi'(r)} dr \, \nu(ds) \nu(dt) = \int_0^\infty \frac{u'(r)^2}{\psi'(r)} H(r) dr, \\
    H(r) = \iint \mathbf{1}\{s < r < t\} (\psi(t) - \psi(s)) \, \nu(ds) \nu(dt).
\end{gather*}
We then note the factorisation \(\mathbf{1}\{s < r < t\} = \mathbf{1}\{s < r\} \mathbf{1}\{t > r\}\), so that each of the two terms in \(H(r)\) simplifies as,
\begin{gather*}
    \iint \mathbf{1}\{s < r < t\} \, \psi(t) \, \nu(ds) \nu(dt) = \nu((0, r)) \int_r^\infty \psi \, d\nu,\\
    \iint \mathbf{1}\{s < r < t\} \, \psi(s) \, \nu(ds) \nu(dt) = \nu((r, \infty)) \int_0^r \psi \, d\nu.
\end{gather*}
Hence, letting \(\bar{\psi} = \int \psi \, d\nu\), we have,
\begin{equation*}
    H(r) = \nu((0, r)) \bigg ( \bar{\psi} - \int_0^r \psi \, d\nu \bigg ) - \big ( 1 - \nu((0, r)) \big ) \int_0^r \psi \, d\nu = \int_0^r (\bar{\psi} - \psi) \, d\nu.
\end{equation*}
Next, let \(G(r) = \rho(r) (R^2 + r^2) \psi'(r)\). With this, it suffices to show that \(\lambda H \leq G\) on \((0,\infty)\), since then it would hold that,
\begin{equation*}
    \int_0^\infty \frac{u'(r)^2}{\psi'(r)} H(r) dr \leq \frac{1}{\lambda} \int_0^\infty u'(r)^2 (R^2 + r^2) \rho(r) dr = \frac{1}{\lambda} \int (R^2 + r^2) u'(r)^2 \nu(dr).
\end{equation*}
To show this, we first express \(G\) as an integral. Since \(\rho'/\rho = (d-1)/r - f'\), we have that,
\begin{equation}\label{eq:radial_divergence_form}
    \rho(r) L_\nu u(r) = \big ( \rho(r) (R^2 + r^2) u'(r) \big )', \qquad \forall u \in \mathcal{C}^2((0,\infty)).
\end{equation}
and hence \(G' = \rho L_\nu \psi\). Since \(G(r) \to 0\) as \(r \to 0\) by \eqref{eq:variational_gap_boundary}, it follows that \(G(r) = \int_0^r L_\nu \psi \, d\nu\). Combining this with the expression for \(H\) above, we have,
\begin{equation*}
    G(r) - \lambda H(r) = \int_0^r k \, d\nu, \qquad \text{where} \qquad k = L_\nu \psi + \lambda (\psi - \bar{\psi}).
\end{equation*}
By assumption, \(k' = (L_\nu \psi)' + \lambda \psi' \leq 0\), so that \(k\) is non-increasing. Moreover, since \(G(r), H(r) \to 0\) as \(r \to \infty\), we have \(\int_0^\infty k \, d\nu = 0\). Hence, it must hold that
\begin{equation*}
    G(r) - \lambda H(r) = \int_0^r k \, d\nu \geq 0.
\end{equation*}
\end{proof}

With this, we obtain the following sufficient conditions for the Poincar\'{e} inequality.
\begin{proposition}\label{prop:wpi_q_lower_bound}
Suppose that \(d \geq 2\) and that \(\bar{\pi}(dx) \propto \exp(-f(|x|)) dx\) where \(f\) is twice continuously differentiable on \([0, \infty)\). Suppose further that the function \(q\) of \Cref{prop:wp_to_wc2} is non-negative and non-decreasing on \((0, \infty)\), and satisfies \(\inf_{r > 0} q(r) \geq m + d + 2\) for some \(m > 0\). Then the weighted Poincar\'{e} inequality in \eqref{eq:weighted_pi} is satisfied with weighting \(\tilde{w}\) and constant,
\begin{equation*}
    C_{\text{PI}} = \max \bigg \{ \frac{1}{2m}, \frac{1}{d-1} \bigg \}.
\end{equation*}
\end{proposition}
\begin{proof}
Let \(\eta(r) = (R^2 + r^2) f'(r) = r q(r)\). Since \(q' \geq 0\), we have that,
\begin{equation*}
    \eta'(r) = q(r) + r q'(r) \geq q(r) = \frac{\eta(r)}{r}, \qquad \forall r > 0.
\end{equation*}
We first verify the integrability conditions. Since \(f'(r) \geq (m + d + 2) r/(R^2 + r^2)\), integrating gives \(f(r) \geq f(0) + \frac{m + d + 2}{2} \log(1 + r^2/R^2)\), and so,
\begin{equation*}
    \rho(r) \leq c \, r^{d-1} \bigg ( 1 + \frac{r^2}{R^2} \bigg )^{-(m + d + 2)/2},
\end{equation*}
for some constant \(c > 0\). Since \(m > 0\), it follows that \(\int |x|^2 \bar{\pi}(dx) = \int r^2 \rho(r) dr < \infty\). Similarly, with \(\psi(r) = r^2\), the quantity \(\rho(r)(R^2 + r^2) \psi'(r)\) is at most a constant multiple of \(r^d (1 + r^2/R^2)^{-(m + d)/2}\), which tends to \(0\) as \(r \to 0\) and, since \(m > 0\), as \(r \to \infty\). Thus \eqref{eq:variational_gap_boundary} holds and \(\int |\psi| d\nu < \infty\).

Next, with \(\psi(r) = r^2\), we have \(\psi' = 2r\) and \(\psi'' = 2\), and since \(b(r) = 2r - \eta(r) + (d-1)(R^2 + r^2)/r\),
\begin{equation*}
    L_\nu \psi(r) = 2(R^2 + r^2) + 2 r b(r) = 2 d R^2 + 2(d+2) r^2 - 2 r \eta(r).
\end{equation*}
Hence,
\begin{equation*}
    - \frac{(L_\nu \psi)'(r)}{\psi'(r)} = \frac{2 (r \eta(r))' - 4(d+2) r}{2r} = \eta'(r) + \frac{\eta(r)}{r} - 2(d+2) \geq 2 q(r) - 2(d+2) \geq 2m.
\end{equation*}
\Cref{lem:variational_gap}, applied with \(\lambda = 2m\), gives \(\lambda_1(-L_\nu) \geq 2m\), and then \Cref{lem:bjm_radial_reduction} gives \(\lambda_1(-L_{\bar{\pi}}) \geq \min \{ 2m, d - 1 \}\).
\end{proof}

\end{document}